\documentclass[nocoverpage,a4paper,10pt,final,openright,swedish]{thesis}
\usepackage{graphicx}
\usepackage{calsymbols}
\usepackage{wasysym}
\usepackage{threeparttable}
\usepackage{eufrak}

\title{Gamma Rays from Primordial Black Holes in Supersymmetric Scenarios}
\subtitle{}
\author{Martin Sahl\'{e}n}
\date{April 2003}
\shortdate{2003}
\type{Master's Thesis}
\department{Department of Physics}
\address{SE-106 91 Stockholm, Sweden}
\city{Stockholm}
\country{Sweden}
\publisher{}
\copyrightline{\copyright\ Martin Sahl\'en, April 2003}
\isrn{KTH/FYS/-{}-03:11-{}-SE}
\issn{0280-316X}
\trita{FYS-2003:11}
\dedication{To my Mother.}
\comment{\emph{Thesis on the subject of Physics for the degree of Master of Science in Engineering from the School of Engineering
         Physics ($\mathfrak{F}$) at KTH.}}
\foregincomment{\emph{Examensarbete inom \"amnet fysik f\"or avl\"aggande av civilingenj\"orsexamen inom\newline utbildningsprogrammet
         Teknisk Fysik ($\mathfrak{F}$) vid KTH.}}
\division{Mathematical Physics}
\centercomment{\centerline{Typeset in \LaTeX}  \newline \newline \newline \newline \newline \newline
This work was performed in the Field and Particle Theory Group, \newline Department of Physics,
    Stockholm University.}

\cplogo{\includegraphics[height=2.5cm]{kthtext.eps}}
\innerlogo{\includegraphics[height=6cm]{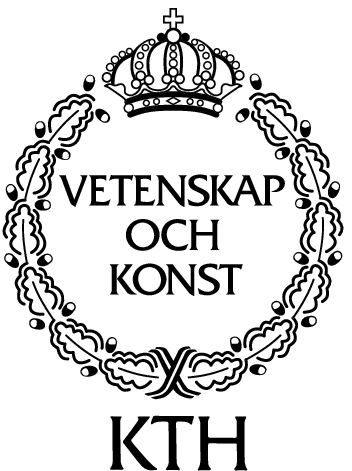}}

\extrainstitute{Stockholm University}
\extradepartment{Department of Physics}
\extradivision{Field and Particle Theory}
\extraaddress{SE-106 91 Stockholm, Sweden}
\extracplogo{\includegraphics[height=2.5cm]{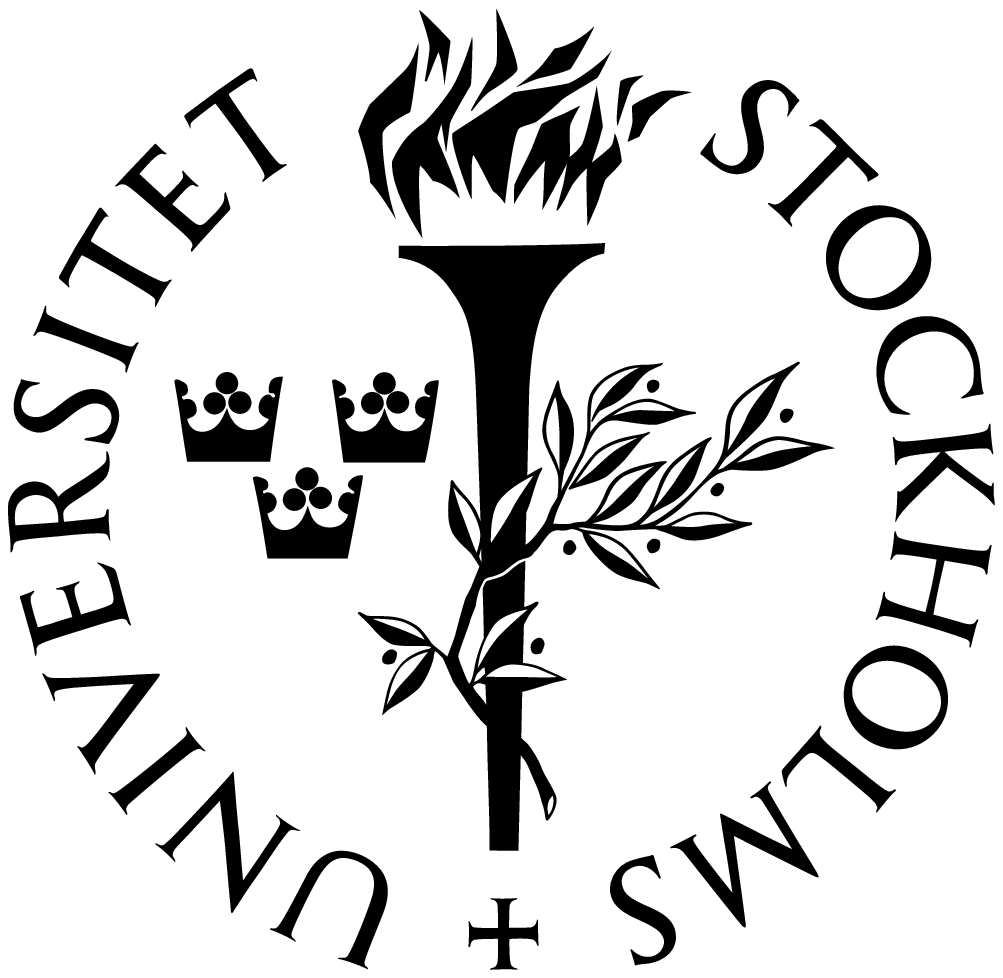}}
\extrainnerlogo{\includegraphics[height=6cm]{suorg_sv.eps}}

\begin{document}

\begin{abstract}
Particle emission from primordial black holes (PBHs) has been studied during the past three decades, and remains an important
testing ground for early-Universe and high-energy physics. This thesis gives a review of PBH formation and distribution, the
Hawking emission process and its later revisions and some of the cosmological consequences associated with PBHs.
The concept of supersymmetry in quantum mechanics and particle physics is also introduced.

Using the MacGibbon-Webber model, the instantaneous photon (gamma-ray) emission spectra from PBHs in the Standard Model and mSUGRA
benchmark models were investigated using the particle physics event generator Pythia.
The point source and diffuse gamma-ray spectra for a range of PBH models were also
found numerically.

It was found that the instantaneous photon flux from PBHs in mSUGRA benchmark models is generally less than approximately
four times greater than in
the Standard Model. The point source flux from a PBH in its final stages of emission
will in mSUGRA benchmark models both quantitatively and qualitatively differ
from that in the Standard Model. The dominant part of the diffuse gamma-ray
flux should not be affected by any supersymmetric particles. However,
for PBHs with initial mass approximately in the range $[4\times 10^{13}\,{\rm g}, 1\times 10^{14}\,{\rm g}]$, the diffuse
Standard Model flux 
at photon energies $E_0 \apprge 1 \,\rm GeV$
will generally be less than roughly $50\%$ greater than for mSUGRA benchmark models. The results generally show good agreement with previous
works, although model differences give some disagreement.

It is concluded that it is unlikely that the effects found could be detected with the upcoming gamma-ray satellite GLAST, due for
launch in 2006.

 \noindent \strut 

\end{abstract}

\tableofcontents

\mainmatter

\chapter{Introduction}

Emission from Primordial Black Holes (PBHs) has been investigated during the past 
three decades, following the works of Hawking \,\cite{Hawking:1974rv,Hawking:1975sw} which predicted black holes 
should emit quantum particles with a thermal spectrum.

PBH emission should have a range of effects on cosmological processes, such as entropy production, nucleosynthesis and
the cosmic microwave background. Photon emission from PBHs will contribute to the diffuse gamma-ray background observed.
This has been used to place constraints on early-Universe conditions.

The gamma-ray constraints will most certainly improve as the new gamma-ray satellite GLAST \,\cite{GLAST:www}
is launched in 2006.
Therefore, it is timely to revise these estimates, and also include a possible supersymmetric contribution to the
gamma-ray emission spectra. 

MacGibbon and Webber modified the emission picture in 1990-1991 \,\cite{MacGibbon:1990zk,MacGibbon:1991tj,Halzen:1991uw}
following the 
work by Oliensis and Hill \,\cite{Oliensis:1984ih}, by assuming only elementary particles are directly 
emitted (rather than composite particles), later fragmenting into hadrons, leptons
 and gauge bosons. This alters the spectrum away from thermality. 
In this work, we will adopt the standard approach of MacGibbon and Webber \,\cite{MacGibbon:1990zk} and 
apply it to supersymmetric extensions of the Standard Model of elementary particle
 physics. We will find emission spectra for a set of mSUGRA benchmark models \,\cite{Ellis:2001hv}.

In Chapter 2, the formation mechanisms and distribution models for PBHs are reviewed.
In Chapter 3, the emission process is reviewed along with lifetime approximations.
Chapter 4 gives an account of cosmological consequences and constraints,
specifically on primordial density perturbations, from PBHs.
Chapter 5 gives a brief introduction to supersymmetry in quantum mechanics and reviews
some of the reasons why supersymmetry is interesting in particle models.
In Chapter 6, expressions for point source and diffuse photon flux from PBHs are derived,
the particulars of photon interactions are presented, an estimate of effects from supersymmetric
particles on the diffuse flux from PBHs is given and the model used presented.
The simulation of PBH emission is covered in Chapter 7. The results are presented and discussed in Chapter 8,
and final conclusions are given in Chapter 9.

Where convenient, natural units with $\hbar = k = c = G = 1$ will be used.

\chapter{PBH Formation and Distribution}

\section{Formation Mechanisms}

The original analysis of PBH formation and distribution was made by Carr and Hawking \,\cite{Carr:1974nx, Carr:1975qj}.
They investigated the formation of PBHs by the
gravitational collapse of primordial density perturbations \,\cite{Hawking:1971}. 
This remains one of the
most interesting formation mechanisms, especially since it is closely related to
very small scales in the early universe.

However, there are of course also other possibilities for PBH formation, for example
phase transitions \,\cite{Hawking:1982ga, La:1989st, Crawford:1982yz},
collapse of cosmic strings \,\cite{Hawking:1989bb, Polnarev:1991bb} and
softening of the equation of state \,\cite{Canuto:1978bb}.

In our further discussion, we will limit ourselves to PBH formation from the
collapse of density perturbations.

\subsection{Power Spectrum $P(\mathbf{k})$}

Let $\delta(\mathbf{x})$ be the density contrast, i.e. the deviation in energy density 
relative to the background $\rho_b$, at $\mathbf{x}$,
\begin{equation}
\delta(\mathbf{x}) = 
\frac{\rho\left(\mathbf{x}\right)-\rho_{b}}{\rho_{b}} \, .
\end{equation}
Looking at the Fourier transform of $\delta(\mathbf{x})$, denoted $\delta(\mathbf{k})$,
we can write
\,\cite{Bergstrom:1999bk}
\begin{equation}
\langle \delta^{*}(\mathbf{k})\delta(\mathbf{k}^{'}) \rangle = 
(2\pi)^3\delta^{(3)}(\mathbf{k}-\mathbf{k}^{'})P(\mathbf{k}) \, ,
\end{equation}
where $\delta^{(3)}$ is the 3D Dirac function and $P(\mathbf{k})$ the
 \emph{power spectrum} of the density fluctuations. This expression assumes the phases 
are uncorrelated, and hence the Fourier transforms are uncorrelated. 
The standard model for $P$ is a power law, i.e.
\begin{equation}
P(\mathbf{k}) \propto |\mathbf{k}|^n \, .
\end{equation}
Here, $n$ is the so called \emph{spectral index}. The case $n=1$ is the
scale-invariant so called \emph{Harrison-Zel'dovich} spectrum. Inflation
\cite{Bergstrom:1999bk} produces a power spectrum that is close to
scale-invariant.

\subsection{Hawking-Carr Model}
\label{sec:hcmod}
Hawking and Carr imposed limits on $\delta$ for black hole formation to be
possible. They found, using analytical arguments, that a spatial region will form a
black hole if the density contrast
$\delta$ in the region fulfills \,\cite{Carr:1974nx, Carr:1975qj}
\begin{equation}
\label{eq:hcdelta}
1/3 \leq \delta \leq 1 \, .
\end{equation}
The lower constraint comes from requiring that the size of the region must be
greater than the Jeans length at the time of collapse, so that gravity can overcome pressure.
The upper bound comes from
requiring that the region not form a disconnected topology, i.e. a separate closed
universe. The black holes formed in this picture are assumed to have an initial
mass
\begin{equation}
M_{i} = \gamma^{3/2}M_{H} \, ,
\end{equation}
where
\begin{equation}
M_H = \frac{4\pi H^{-3}\rho}{3}
\end{equation}
is the horizon mass when the scale of interest crossed the horizon and
$\gamma$ determines the equation of state $p=\gamma\rho$. For radiation
domination, we have $\gamma = 1/3$.

\subsection{Niemeyer-Jedamzik Model}
A recent development with interesting consequences for PBH formation is the realisation
that \emph{near the threshold of black hole formation},
 gravitational collapse behaves
as a critical phenomenon.
Numerical simulations by Choptuik \,\cite{Choptuik:1993jv}
and Evans and Coleman \,\cite{Evans:1994pj} suggest a
scaling relation for black hole formation of the form
\begin{equation}
M_{i}(\delta) = kM_{H}(\delta-\delta_{c})^{\eta} \, .
\end{equation}
Here, $k$ is a constant dependent on the shape of the fluctuation,
$\delta_{c}$ the (shape-dependent) critical density, $\eta$ the universal
shape-independent scaling exponent and $M_H$ the horizon mass at PBH
formation.

A valid question to ask is whether this phenomenon will be relevant for black
hole formation. In general, it will not be relevant for astrophysical black
holes, since along with other inhibiting effects it would require significant
fine tuning of initial conditions.
However, the effect on PBH formation
was further investigated by Niemeyer and Jedamzik, who found that this 
scaling should be relevant for PBH formation \,\cite{Niemeyer:1998mt}. 
This is essentially due to the fact that for PBH formation, one expects
most regions collapsing to black holes to have an over-density close to
$\delta_{c}$. Further
numerical simulations \,\cite{Niemeyer:1999ak} showed that the scaling relation
holds for PBH formation with nearly-critical initial conditions. They found
\begin{equation}
\eta \approx 0.37
\end{equation}
and, significantly, for many fluctuation shapes 
\begin{equation}
\delta_{c} \approx 0.7\, ,
\end{equation}
roughly twice the analytically predicted value of 1/3 (Eq. (\ref{eq:hcdelta})).

\subsection{Epoch of Formation}
\label{sec:formation}
Most authors, as will we, 
today assume an inflationary scenario, since this solves several problems
with the standard Big Bang model \cite{Bergstrom:1999bk}.
Hence any PBHs formed
before the period of inflation will be diluted away. We consider therefore
PBHs formed after inflation, typically immediately after the reheating epoch.
The time of formation is then related to the reheating temperature $T_{RH}$ as \cite{Kolb:1990bk}
\begin{equation}
t_{iRH} = 0.301g_*^{-1/2}\frac{m_{Pl}}{T_{RH}^2} \, ,
\end{equation}
where $g_*$ is the degrees of freedom of the constituents in the early Universe ($g_* \sim 100$ in the
Standard Model and $g_* \sim 200$ in the MSSM) and $m_{Pl}$ the Planck mass. 
This time is easily transformed to the redshift $z_{iRH}$ using \cite{Bergstrom:1999bk}
\begin{equation}
\label{eq:zi}
1+z_{iRH} \approx \left(\frac{3\Omega_M^{1/2}t_{iRH}H_0}{2}\right)^{-2/3} \, .
\end{equation}
In the Hawking-Carr model, we can also relate the minimum initial mass at a certain reheating temperature as \cite{Kim:1999iv}
\begin{equation}
\label{eq:mrh}
M_{RH} \approx \frac{1}{8}\gamma^{3/2}m_{Pl}\left(\frac{T_{RH}}{T_{Pl}}\right)^{-2} \, ,
\end{equation}
where $T_{Pl}$ is the Planck temperature.

\section{Spatial Distributions}

The distribution of PBHs is of course closely related to the primordial density 
perturbations (we consider only formation from the collapse of such 
perturbations). This Section will present some models for the PBH mass spectrum, although in the subsequent
simulations and calculations, these models will not be used (see Section \ref{sec:model}).

The present comoving number density of black holes, can be expressed as 
\begin{equation}
\frac{dn}{dM} = \frac{dn}{dM_{i}}\frac{dM_{i}}{dM}=\frac{dn}{dM_{i}}\frac{dM_{i}}{dt}\frac{dt}{dM} = \frac{dn}{dM_{i}}(1-\frac{dt}{dM}\frac{d}{dt}\int_{0}^{t}\frac{dM}{dt^{'}}dt^{'}) \, .
\end{equation}
The two last steps are in practice not very useful, but at least clearly show the dependence on $dM/dt$. In practice, $dM/dt$ most likely yields a separable ODE, from which one easily finds $M_{i}$ as a function of $M$ and $t$ (c.f. Eq. 
(\ref{eq:m})).

\subsection{Carr Mass Spectrum}

The function $dn/dM_{i}$, called the \emph{initial mass function} or sometimes the mass spectrum
describes the initial number density of the PBHs. 
The classical result due to Carr is \,\cite{Carr:1975qj}
\begin{equation}
\frac{dn_{C}}{dM_{i}} \propto M_{i}^{-5/2} \, ,
\end{equation}
which is found using the Hawking-Carr model (see Section \ref{sec:hcmod}), assuming
the spectral index $n = 1$.
This type of initial mass function is cosmologically irrelevant (see next Section).

\subsection{Kim-Lee Mass Spectrum}

The initial mass function for a general power law power spectrum (in the
Hawking-Carr model), i.e.
 $P(k) \propto k^{n}$, was given by Kim-Lee as \,\cite{Kim:1996hr}
\begin{equation}
\frac{dn_{KL}}{dM_{i}} = \frac{n+3}{4} \sqrt{\frac{2}{\pi}} \gamma^{7/4}\rho_{i}M_{H,i}^{1/2}M_{i}^{-5/2}\sigma_{H}^{-1}\times 
\exp\left(-\frac{\gamma^2}{2\sigma_{H}^2}\right) \, ,
\end{equation}
where $\gamma$ determines the equation of state ($p=\gamma \rho$), $\rho_i$ is the energy density and $M_{H,i}$ the horizon mass
when the PBHs form. This was derived
following Carr \,\cite{Carr:1975qj}, using the Press-Schechter formalism 
\,\cite{Press:1974iz}. The mass variance $\sigma_{H}^2$, which is roughly
the mean square size of the density perturbations, will be treated in a
following Section.

The case $n = 1$ corresponds to a scale-invariant
(Harrison-Zel'dovich) spectrum which yields a Carr initial mass function,
 $dn/dM_{i} \propto M_{i}^{-5/2}$. 
As some authors realised, the $n = 1$ spectrum does not yield a significant PBH
abundance when normalised to COBE observations \,\cite{Carr:1993aq, Green:1999xm}.
Green estimates an initial energy fraction $\sim 10^{-6\times10^6}$ in PBHs \,\cite{Green:2001kw}.

\subsection{Niemeyer-Jedamzik Mass Spectrum}

It was found by Green and Liddle \,\cite{Green:1999xm} that for both power law spectra and flat spectra with a spike on a certain 
scale, in the limit where the number of PBHs formed is small enough to satisfy constraints on their abundance, one can assume that all the 
PBHs form at a single horizon mass. 

Using that assumption, one can calculate analytically the initial mass function
in the Niemeyer-Jedamzik model \,\cite{Niemeyer:1998mt} 
\begin{eqnarray}
\nonumber
\frac{dn_{NJ}}{dM_{i}} & = & \frac{\rho_{i}}{\sqrt{2\pi}\eta\sigma_{H}M_{i}M_{H,i}}\left(\frac{M_{i}}{kM_{H,i}}\right)^{1/\eta} \times \\
&& \times
\exp\left(-\frac{[\delta_{c}+(M_{i}/kM_{H,i})^{1/\eta}]^2}{2\sigma_{H}^2}\right) \, ,
\end{eqnarray}
where $k$, $\delta_c$ and $\eta$ are the critical parameters, $\rho_i$ is the energy density and $M_{H,i}$ the horizon mass
when the PBHs form.
The mass variance $\sigma_{H}^2$, which is roughly
the mean square size of the density perturbations, is treated in the following Section.

\subsection{Mass Variance}
The horizon mass variance $\sigma(M_H)$, roughly
the mean square size of the density perturbations,
 is related to the power spectrum as \,\cite{Green:1997sz}
\begin{equation}
\sigma_{H}^2 \equiv \sigma^2(M_{H}) = \frac{1}{2\pi^2}\int_0^{\infty} W(kR)P(k) k^2 dk \, ,
\end{equation}
with $R=a^{-1}H^{-1}$. $W(kR)$ is a window function picking out scales around $k \sim 1/R$.
Using the well-known time evolution of cosmological quantities during radiation and matter domination,
one finds for power law power spectra \,\cite{Green:1997sz}
\begin{equation}
\sigma_H = \sigma(M_{H,0})\left(\frac{M_{H,eq}}{M_{H,0}}\right)^{(1-n)/6}
\left(\frac{M_{H}}{M_{H,eq}}\right)^{(1-n)/4} \, ,
\end{equation}
with $\sigma(M_{H,0})=9.5\times10^{-5}$ from COBE data \,\cite{Green:1997sz, 
Bunn:1996py}. The subscript ``eq'' refers to values at matter-radiation equality, and the subscript ``0'' refers to
values at the present epoch.

The expression for the mass variance has been reconsidered by Blais, Bringmann,
Kiefer and Polarski \,\cite{Bringmann:2001yp, Blais:2002gw}, who give a more
accurate result, which will not be reproduced here.

\chapter{Emission Process}
\section{Direct Emission}
\subsection{Hawking Spectrum}

Using semi-classical arguments, Hawking found that black holes should emit
quantum mechanical particles \,\cite{Hawking:1974rv, Hawking:1975sw}. 
Hawking's expression for the emission spectrum from a black hole with 
angular velocity $\Omega$, electric potential $\Phi$ and surface gravity
$\kappa$, for each degree 
of freedom of particles with energy $Q$ per unit time $t$ is
\begin{equation}
\frac{d^2N}{dQdt} = \frac{\Gamma_{s}}{2\pi\hbar\left(\exp\left(
\frac{Q-n\hbar\Omega-q\Phi}{\hbar\kappa/2\pi c}\right)-(-1)^{2s}\right)} \, .
\end{equation}
Here, $\Gamma_{s}$ is the absorption probability for the emitted particle,
$n\hbar$ the angular momentum and $q$ the particle charge.
A black hole is generally characterised by these three basic quantities -
its mass, charge and
angular momentum. This is the essence of the so called 
\emph{no hair theorem} \,\cite{Hawking:1977wz}.
For $\Omega = \Phi = 0$, the surface gravity $\kappa$ can be written
\begin{equation}
\kappa = \frac{c^4}{4GM} \, .
\end{equation}
If we thus define a temperature $T$ of the black hole,
\begin{equation}
T=\frac{\hbar c{^3}}{8\pi kGM} = 1.06\left(\frac{10^{13}\,{\rm g}}{M}\right) \,{\rm GeV} \, ,
\end{equation}
we see that when $\Omega = \Phi = 0$ the Hawking emission formula mimics
a black-body spectrum,
\begin{equation}
\label{eq:hawking}
\frac{d^2N}{dQdt} = \frac{\Gamma_{s}}{2\pi\hbar\left(\exp\left(
\frac{Q}{kT}\right)-(-1)^{2s}\right)} \, .
\end{equation}

In the following Sections, we will limit ourselves to non-rotating, uncharged 
(i.e. Schwarzschild) black holes. Page \,\cite{Page:1977um} and Gibbons \,\cite{Gibbons:1975kk} have shown that black holes with spin and/or 
charge will quickly (compared to the times we are interested in) emit particles 
which carry away the spin and charge. Fluctuations will of course still exist, but these have quite a negligible effect
on the emission rates \,\cite{MacGibbon:1990zk}.

\begin{figure}
  \begin{center}
    \includegraphics[width=0.75\textwidth]{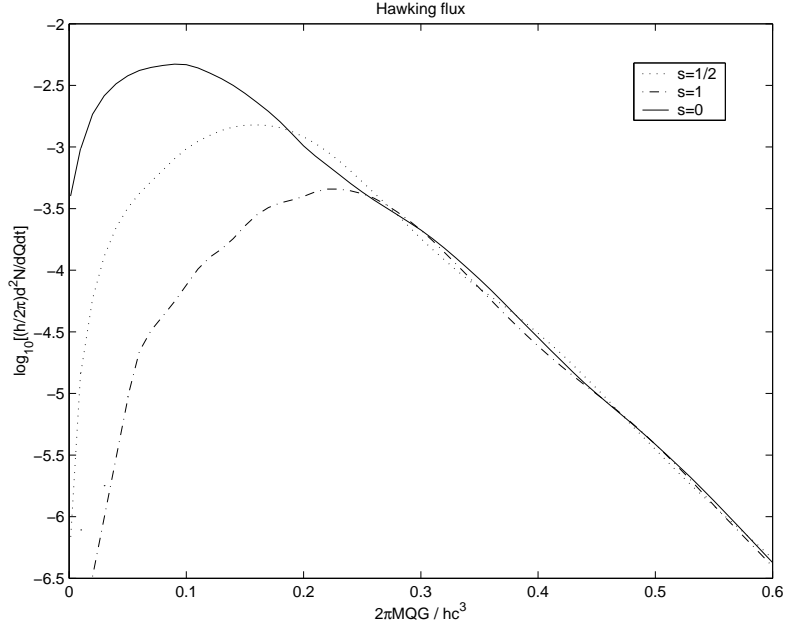}
  \end{center}
  \caption{Hawking flux for $s=0,1/2,1$ in the relativistic limit.}
  \label{figure:flux}
\end{figure}

A plot of the direct emission spectra for $s=0,1/2,1$ in the relativistic
limit can be seen in Figure \,\ref{figure:flux}.

\begin{figure}
  \begin{center}
    \includegraphics[width=0.75\textwidth]{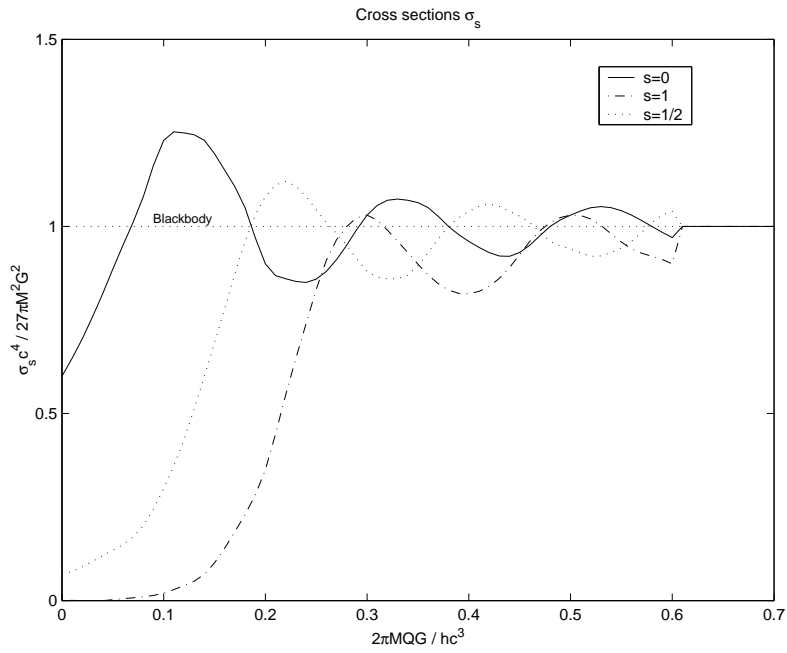}
  \end{center}
  \caption{Cross sections for $s=0, 1/2, 1$. Fitted using cubic splines from
  data in \,\cite{MacGibbon:1990zk}.
  The straight line corresponds to the
  geometric limit blackbody cross section $\sigma_g=27\pi M^2G^2/c^4$.
  The behaviour is similar to an 
  over-damped oscillation around the blackbody limit. For parameter values
  $>0.6$ we then approximate with that limit, due to lack of data. The accuracy
  is generally within $5\%$, increasing with the parameter value. However, as can be seen, for parameter values
  slightly greater than $0.6$ the accuracy for $s=1$ is only within $\sim 10\%$.}
  \label{figure:csec}
\end{figure}

The absorption probability $\Gamma_{s}$ can be further expressed as
\,\cite{MacGibbon:1990zk, Barrau:2001ev}
\begin{equation}
\Gamma_{s} = \frac{\sigma_{s}(Q,M,\mu)}{\pi \hbar^2c^2}(Q^2-\mu^2) \, ,
\end{equation}
where $\mu$ is the (rest) mass of the emitted particle and $\sigma_{s}$ the 
absorption cross-section. The absorption cross-sections have to be numerically
computed by solving the Teukolsky equation \,\cite{Teukolsky:1973ha, 
Teukolsky:1974ab}.
 This was done by Page for $s=1/2, 1, 2$ in 1976-77 \,\cite{Page:1976phd,
Page:1976df, Page:1976ki, Page:1977um} (for several different scenarios)
 and for $s=0$ by Simkins in 1986 \,\cite{Simkins:1986phd}. 
A plot of the cross sections, produced from data in 
Ref. \,\cite{MacGibbon:1990zk}, is shown in Figure \ref{figure:csec}.

\subsection{Modified Hawking Spectra}

Hawking's semi-classical derivation has been modified by different authors.
We will specifically look at a recent derivation by Parikh and 
Wilczek \,\cite{Parikh:1999mf}. In contrast to Hawking, the authors assume a dynamical geometry, 
and modeling emission as a tunneling process, arrive at 
\begin{equation}
\frac{d^2N}{dQdt} = \frac{\Gamma_{s}}{2\pi\hbar\left(\exp\left(8\pi Q(M-Q/2)\right)-(-1)^{2s}\right)} \, .
\end{equation}
We see that the quadratic correction in the exponential becomes important only 
when $Q$ is comparable to $M$, i.e. in the late stages of evaporation. We then 
recover the Hawking spectrum in the case $Q\ll M$. These ideas have also recently 
been used in the context of string gravity to arrive at similar corrections \,\cite{Alexeyev:2002}.

\section{Fragmentation and Hadronisation}

In the model proposed by Oliensis and Hill \,\cite{Oliensis:1984ih} and expanded upon by MacGibbon and 
Webber \,\cite{MacGibbon:1990zk,MacGibbon:1991tj}, all fundamental particles with rest mass less than or equal to the 
black hole temperature are emitted (of course, occasionally heavier particles are emitted, but as seen in Hawking's expression they are exponentially damped). 
When the black hole temperature reaches the 
QCD confinement scale $\Lambda_{QCD}=250-300$ MeV, quark and gluon jets are
emitted. These fragment and eventually produce hadrons and other particles. 

The authors argue that almost all emitted particles are much smaller than 
the black hole, since the de Broglie wavelength of most emitted particles satisfies
\begin{eqnarray}
\nonumber
  \lambda_{dB} \apprle 0.04 \times 10^{-13} \left(\frac{T}{1 \,\rm GeV}\right)^{-1} \,\,\rm cm \ll r_{bh} = 
\frac{2GM}{c^2} = \\
 = 1.57 \times 10^{-13} \left(\frac{T}{1 \,\rm GeV}\right)^{-1} \,\,\rm cm \, .
\end{eqnarray}

Furthermore, successive emissions should not interact, since for most particles
\begin{equation}
\lambda_{dB} = \frac{1}{Q} \ll \Delta t \approx \frac{20}{Q} \, ,
\end{equation}  
where $Q$ is the particle energy and $\Delta t$ is the time between successive emissions.
Oliensis finds numerically that $99.9\%$ of the emitted particles should satisfy this. 

Likewise, the previously emitted particles should not affect the emission process itself,
since
\begin{equation}
r_{bh} \approx \frac{\Delta t}{50} \ll \Delta t \, .
\end{equation}  

There should also be no strong interactions between successively emitted particles.
Lastly, gravitational effects on lifetimes, etc.
should not have a significant effect on the final spectrum.

In summary, no short or long range forces should significantly affect the emission 
process before fragmentation and hadronisation, which means the emission is similar
to that of $e^+e^-$-annihilation events.

However, it should be noted that these ideas have been challenged by Heckler.
He proposed in 1997 that QED and QCD photospheres (or perhaps better termed photosphere and chromosphere 
as for the Sun) 
may form around the black hole above a certain black hole 
temperature when bremsstrahlung and pair-production becomes non-negligible \,\cite{Heckler:1997jv,Heckler:1997ab}. 
These ideas remain controversial, but seem at least partly supported by numerical 
simulations by Cline, Mostoslavsky and Servant \,\cite{Cline:1998xk} and Daghigh and Kapusta \,\cite{Daghigh:2001gy}.
In light of the uncertainties associated with photosphere effects, we will neglect photosphere effects in this work.
Barrau et al. \cite{Barrau:2001ev} use a parameterisation based on Ref. \cite{Cline:1998xk} to account for photosphere effects,
but this method seems dubious.

To find the actual emitted spectra after fragmentation, one has to convolve the 
direct (i.e. Hawking) emission spectrum with a function describing what final 
particles are created by the fragmentation of an initial particle. For a particle 
species $X$ we can write
\begin{equation}
\label{eq:frag}
\frac{d^2N_{X}}{dEdt} = \sum_{j} \int_{Q=E}^{\infty} \alpha_{j}\frac{\Gamma_{s}}{2\pi 
\hbar \left(\exp\left(\frac{Q}{kT}\right)-(-1)^{2s}\right)}\times \frac{dg_{jX}(Q,E)}{dE} dQ \, ,
\end{equation}
where we sum over all particle species $j$ and weight with their degrees of 
freedom $\alpha_{j}$. The \emph{fragmentation function} $dg_{jX}(Q,E)/dE$ is the 
number of $X$ particles with energy between $E$ and $E+dE$ created by the 
particle (jet) $j$ with energy Q. Here we can (and should) include any decay 
products, so for most simulation purposes one may readily decay all but particles 
stable on astrophysical time scales, i.e. $\gamma$, $p\overline{p}$, $e^{\pm}$, $\nu \overline{\nu}$. This list should also contain the lightest supersymmetric 
particle (LSP) in the case of a supersymmetric model (with R-parity conserved).

One should specifically note that since we have for pions $m_{\pi} \approx 
135-140$ MeV $< \Lambda_{QCD}$, in the (approximate) energy interval [$m_{\pi}, \Lambda_{QCD}$] 
the black hole will directly emit pions.

\section{Mass Loss Rate and Lifetime}

From the emission rate we can find the black hole mass loss rate as \,\cite{MacGibbon:1991tj}
\begin{equation}
\label{eq:dmexact}
\frac{dM}{dt}=-\sum_{j} \alpha_j \int_{\mu_{j}}^{\infty} \frac{d^2N_{j}}{dQdt} \times \frac{Q}{c^2} dQ \, ,
\end{equation}
where we sum the direct emission rate, Eq. (\ref{eq:hawking}),
over all particle species $j$
with rest mass $\mu_{j}$ and $\alpha_j$ degrees of freedom.
Using the mass loss rate, the lifetime $\tau$ 
of a black hole is
\begin{equation}
\label{eq:tauexact}
\tau = \int_{M_{i}}^{m_{Pl}} (\frac{dM}{dt})^{-1} dM \, ,
\end{equation}
where the inverse mass loss rate is integrated from the initial black hole mass 
$M_{i}$ to the Planck mass $m_{Pl}$. The latter limit is an approximation to deal 
with the fact that our derivations, particularly Hawking's,
break down close to the Planck scale. Uncertainties in the final mass will have a
negligible effect on the lifetime (see e.g. Eq. (\ref{eq:m})).

Similarly, we find the initial mass $M_{i}$ of a black hole formed at $t=0$, that 
would just have evaporated at time $t_{f}$ to be
\begin{equation}
M_{i} = \int_{t_{f}}^{0} \frac{dM}{dt} dt \, .
\end{equation}
If we assume all PBHs are formed very close to the Big Bang, we can replace 
$t_{f}$ by $t_{u}$, the age of the universe, to get the initial mass $M_{*}$ 
of PBHs that would just have evaporated today. The generally used value is 
$M_{*} = 5 \times 10^{14}$ g \,\cite{Page:1976ab}.

\begin{figure}
  \begin{center}
    \includegraphics[width=0.75\textwidth]{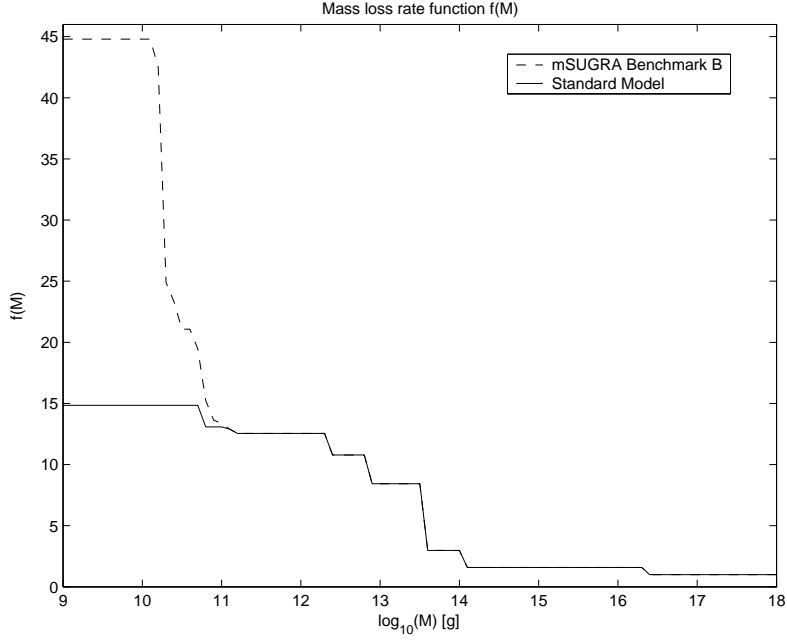}
  \end{center}
  \caption{Mass loss rate coefficient function $f(M)$.}
  \label{figure:f}
\end{figure}

Far from mass thresholds, one can assume that \,\cite{Page:1976ab,MacGibbon:1991tj}
\begin{equation}
\label{eq:dmapprox}
\frac{dM}{dt} \approx -\Phi(M) M^{-2} \, ,
\end{equation}
where $\Phi(M)$ then is a function of the number of emitted particle species. 
This function generally has to be calculated numerically, and the values reported by MacGibbon and Webber \,\cite{MacGibbon:1990zk,MacGibbon:1991tj} are
\begin{equation}
\label{eq:macf}
\frac{dM}{dt} = -5.34\times 10^{25}f(M)\left(\frac{M}{1\,\rm g}\right)^{-2} {\,\,\rm g\, s^{-1}} \, ,
\end{equation}
where $f(M)$ in the relativistic regimes, i.e. far from mass thresholds, is
\begin{eqnarray}
\label{eq:macff}
\nonumber
f(M) & = & 0.267d_{s=0} + 0.060d_{s=1} + 0.007d_{s=2} + 0.020d_{s=3/2} +\\
&& + 0.147d_{s=1/2}^{q=\pm e} + 0.142d_{s=1/2}^{q=0} \, ,
\end{eqnarray}
where the factors $d$ count the degrees of freedom for particles with spin and 
charge according to the sub- and superscripts, that are being emitted by the 
black hole (this is how the $M$-dependence comes in). 
The $s=3/2$ contribution is for hypothetical gravitino emission.
 The error in counting quarks as having 
charge $\pm e$ or charged elementary particles as uncharged is $<5\%$ \,\cite{MacGibbon:1990zk}. This function is plotted in
Figure \ref{figure:f} for the Standard Model and mSUGRA Benchmark B. 
MacGibbon also made approximations to account for the effect of mass thresholds
 \,\cite{MacGibbon:1991tj}.

\begin{figure}
  \begin{center}
    \includegraphics[width=0.75\textwidth]{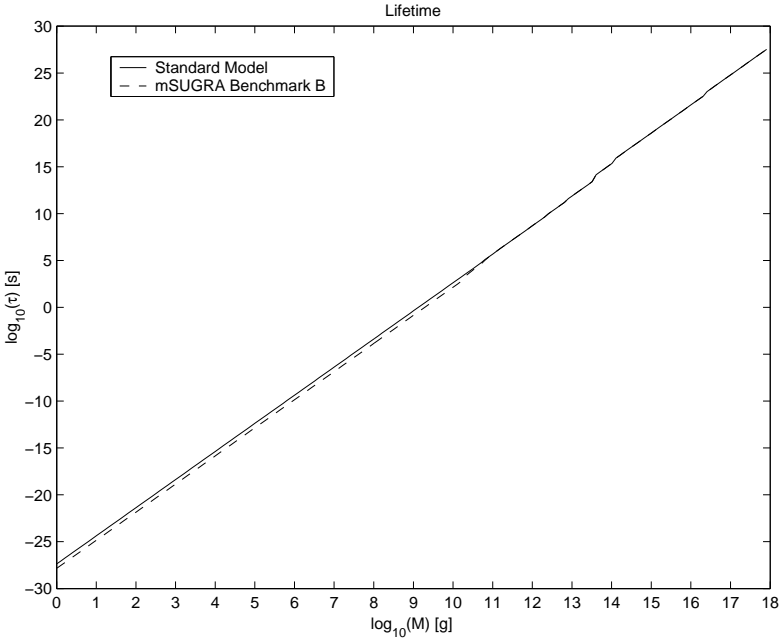}
  \end{center}
  \caption{Black hole lifetime as a function of mass}
  \label{figure:lifetime}
\end{figure}

Using the assumptions (\ref{eq:dmapprox})-(\ref{eq:macff}), Eqs. (\ref{eq:dmexact}) and (\ref{eq:tauexact}) can be solved
for $M$ and $\tau$ respectively, to find
\begin{equation}
\label{eq:m}
M = (M_{i}^3-3\Phi t)^{1/3}
\end{equation}
and
\begin{equation}
\tau = \frac{M_{i}^3}{3\Phi} \, ,
\end{equation}
when $\Phi(M)$ is roughly constant. 
MacGibbon estimates the black hole lifetime to be \,\cite{MacGibbon:1991tj}
\begin{equation}
  \tau \approx 6.24 \times 10^{-27} \left(\frac{M_{i}}{1 \,\rm g}\right)^3f(M_{i})^{-1} {\,\,\rm s} \, ,
\end{equation}
with $f(M)$ given by Eq. (\ref{eq:macf}). This is naturally a slight overestimate of the lifetime, since
the increase in degrees of freedom as the temperature rises is neglected. However, since the black hole
spends most of its time near $M_i$ the relative error is small.
This estimate of the PBH lifetime is shown in Figure \ref{figure:lifetime}, where the effect of the aforementioned error can be
seen in that the mSUGRA Benchmark B lifetimes start to differ from the Standard Model only at masses
lower than approximately $10^{11} \,\rm g$.

These expressions for the mass and lifetime hint at why PBHs are particularly interesting, i.e. why we do not consider
emission from black holes formed from the collapse of stars at later epochs. A black hole with $M = M_{\astrosun} \approx
2 \times 10^{33} \,\rm g$ would have an initial temperature $T_i \sim 10^{-21}\,\rm GeV$ and
a lifetime that is $\sim 10^{56}$ times the age of the universe. For all practical purposes, these black holes will thus emit
nothing.

\chapter{Cosmological Consequences}
\section{Effects on Cosmological Phenomena}
Hawking emission provides a mechanism for PBHs to affect the early Universe. 
The emitted particles can interact with radiation and matter in the Universe and thus affect 
many of the early-Universe phenomena: the CMBR \,\cite{Naselsky:1978, Zeldovich:1976},
entropy production \,\cite{Zeldovich:1976},
baryogenesis \,\cite{Barrow:1980hh} and nucleosynthesis \,\cite{Lindley:1980b, Miyama:1978, Novikov:1979a, Rothman:1981}
to name some. Recently is was also realised that PBHs are potentially a great probe of the early Universe
with a varying gravitational constant \,\cite{Carr:2000my}.

\section{Physics Constraints}
PBHs can be used to probe the primordial mass distribution, hence
the spectral index $n$ in a power law power spectrum. Page and Hawking required that the emitted gamma rays from PBHs should 
not exceed the observed diffuse gamma-ray background (DGB), and thus found a constraint 
on the abundance of PBHs \,\cite{Page:1976ab}. This is known as the Page-Hawking bound.
This constraint has later been updated by MacGibbon and 
Carr \,\cite{MacGibbon:1991ca, Carr:1998mc} using data from EGRET \,\cite{Sreekumar:1998un} and COMPTEL \,\cite{Kribs:1997ac}.
Kim, Lee and  MacGibbon \,\cite{Kim:1999iv} also use the normalisation to the DGB to constrain the primordial density fluctuations
in the case of a Kim-Lee initial mass function. They find $n \apprle 1.23 - 1.25$ for the spectral index.

Another popular approach is to require that some critical energy fraction $\Omega < 1$, typically the total energy in PBHs
$\Omega_{PBH}$. This kind of approach was used by Carr, Gilbert and Lidsey to find constraints on the spectral index $n$
\cite{Carr:1994ar}. They find $n \apprle 1.3-1.4$. Green and Liddle have corrected errors in their results and find 
$n \apprle 1.25-1.3$ \cite{Green:1997sz}. Kim and Lee performed an analysis using the Kim-Lee
mass spectrum and
obtained $n \apprle 1.43-1.48$ \cite{Kim:1996hr}. Kribs, Leibovich and Rothstein used the Niemeyer-Jedamzik mass
spectrum to obtain $n \apprle 1.2-1.45$ \cite{Kribs:1999bs} (they also use the DGB bound).
All these calculations have been done using normalisation to COBE anisotropy measurements
\cite{Bunn:1996py, Bunn:1997da}.

Other methods of constraining $n$ are also used in the aforementioned and other articles.
Reviews of these and other cosmological constraints from PBHs are given in Refs. \cite{Liddle:1998nt} and \cite{Carr:1998mc}.

Currently, the best constraints on $n$ come from CMBR measurements such as
COBE \,\cite{Bunn:1996py, Bunn:1997da} and WMAP \,\cite{Bennett:2003bz}. However, it is important to note
that $n$ may vary with $k$, and PBH constraints probe much smaller scales than CMBR measurements do.
Consequently, the PBH constraints provide information that CMBR measurements do not, especially in the case
that $P(k)$ is not a power law but a spike or step \cite{Blais:2002gw}.

\chapter{Supersymmetry}

\section{Supersymmetry in Quantum Mechanics}
To understand the basic concept behind supersymmetry (SUSY), let us start from the quantum 
mechanical Hamiltonian for a single particle:
\begin{equation}
H_{1} = -\frac{\hbar^2}{2m}\frac{d^2}{dx^2} + V_{1}(x) \, .
\end{equation}
We now assume that the potential $V_{1}(x)$ is unknown, but we know the ground 
state wave function $\psi_{0}(x)$ which we will assign the energy eigenvalue 0 
(no loss of generality). Inserting the ground state wave function into the Schr\"{o}dinger equation for the above Hamiltonian we find 
\begin{equation}
V_{1}(x) = \frac{\hbar^2}{2m}\frac{\psi_{0}^{''}(x)}{\psi_{0}(x)} \, .
\end{equation}
If we now set out to write the Hamiltonian $H_{1}$ in a standard form, i. e. $H_{1}=A^\dag A$, we can write
\begin{equation}
A=\frac{\hbar}{\sqrt{2m}}\frac{d}{dx} + W(x),\,\, A^\dag = -\frac{\hbar}{\sqrt{2m}}\frac{d}{dx} + W(x) \, ,
\end{equation}
where $W(x)$ is the so called \emph{superpotential}. Looking at the original expression, we can now write
\begin{equation}
V_{1}(x) = W^2(x) - \frac{\hbar}{\sqrt{2m}}W^{'}(x) \, ,
\end{equation}
which is called the \emph{Riccati equation}. Using the fact that $A\psi_{0} \Rightarrow H_{1}\psi_{0} = 0$ we find the solution for $W$ in terms of the ground state wave function
\begin{equation}
W(x) = - \frac{\hbar}{\sqrt{2m}} \frac{\psi_{0}^{'}(x)}{\psi_{0}(x)} \, .
\end{equation}

Let us now define the operator $H_{2} = AA^{\dag}$. We then find 
\begin{eqnarray}
\nonumber
H_{2} =  -\frac{\hbar^2}{2m}\frac{d^2}{dx^2} + V_{2}(x) \, , \\
V_{2}(x) = W^2(x) + \frac{\hbar}{\sqrt{2m}}W^{'}(x) \, .
\end{eqnarray}
We call $V_{1}(x)$ and $V_{2}(x)$ supersymmetric partner potentials.
Using the Schr\"{o}dinger equation for the two Hamiltonians, one finds that the eigenvalues and eigenfunctions are related as
\begin{equation}
E_{n}^{(2)} = E_{n+1}^{(1)},\,\, E_{0}^{(1)}=0 \, ,
\end{equation}
\begin{equation}
\psi_{n}^{(2)} = (E_{n+1}^{(1)})^{-1/2}A\psi_{n+1}^{(1)} \, ,
\end{equation}
\begin{equation}
\psi_{n+1}^{(1)} = (E_{n}^{(2)})^{-1/2}A^{\dag}\psi_{n}^{(2)} \, .
\end{equation}
Both sets of eigenvalues are positive semi-definite, and the energy levels are degenerate, except for $H_{1}$ which has an ``extra'' level at zero energy. We also see that one effect of the operators $A$ and $A^{\dag}$ is to ``jump between the potentials''.
One can now define a SUSY matrix Hamiltonian
\begin{equation}
H = \left( \begin{array}{ccc}
H_{1} & 0 \\
0 & H_{2} \\
\end{array} \right)
\end{equation}
and the operators
\begin{equation}
Q = \left( \begin{array}{ccc}
0 & 0 \\
A & 0 \\
\end{array} \right), \,\,
Q^{\dag} = \left( \begin{array}{ccc}
0 & A^{\dag} \\
0 & 0 \\
\end{array} \right) \, .
\end{equation}
These three operators now form the closed superalgebra $sl$(1/1) with commutation and anticommutation relations
\begin{eqnarray}
\nonumber
[H,Q] & = & [H,Q^{\dag}] = 0 \\
\nonumber
\{Q,Q^{\dag}\} & = & H \\
 \{Q,Q\} & = & \{Q^{\dag},Q^{\dag}\} = 0
\end{eqnarray}
The so called \emph{supercharges} $Q$, $Q^{\dag}$ commute with $H$, hence the degeneracy in the spectra of $H_{1}$ and $H_{2}$. The supercharges can be interpreted as operators that exchange bosonic and fermionic degrees of freedom.

In broader terms, what we have seen is that given a particle (ground state wave function), we may construct another particle such that they are both described by the same superpotential, and bosonic degrees of freedom in one particle correspond to fermionic degrees of freedom in the other. In a sense they are both ``components'' of a ``superparticle''. 

\section{Supersymmetry and Particle Physics}
\subsection{Background}
The ideas laid out in the preceding Section form the basis of supersymmetric extensions
of the Standard Model. The main motivations for considering supersymmetry are:
\begin{itemize}
\item solution of hierarchy and naturalness problems
\item connection between gravity and other fundamental interactions
\item can learn about high-energy physics in general
\item rich experimental consequences
\end{itemize}
The hierarchy and naturalness problems are intimately related. Suppose we wish to write down a Lagrangian which accounts
correctly for physics up to some energy scale $\Lambda$ much larger than the electro-weak scale $E_{weak} \approx 200
\,\rm GeV$. Choosing to break electro-weak symmetry by the Higgs mechanism, one finds that the masses of scalar excitations
of the field will be of order $\Lambda$. This is due to radiative corrections, the mass corrections being 
$\delta m_{scalar}^2 \sim g^2\Lambda^2$. Specifically, the Higgs mass would become huge, destroying
the hierarchy between the electroweak and $\Lambda$ scales,
giving us the hierarchy problem.
The simplest way to avoid this problem is to fine-tune the coupling constants to an accuracy of $E_{weak}^2/\Lambda^2
\apprle 10^{-28}$ (in the case $\Lambda = E_{GUT} \approx 10^{16} \,\rm GeV$).
This is the naturalness problem. Supersymmetry removes or significantly improves on
these problems by introducing new corrections from the new particles, which reduces the fine-tuning requirement to
at most $\log(\Lambda/E_{weak})$ and completely removes it in some cases. Since we have not observed supersymmetric
particles with the same masses as their Standard Model partners, supersymmetry can not be an exact symmetry. We thus need
a mechanism for spontaneously breaking supersymmetry. The symmetry breaking imposes approximate limits on the mass 
splitting between a particle and its supersymmetric partner: if the Higgs boson mass correction $\delta m_H^2 \sim 
g^2(m_B^2-m_F^2)$, where $m_B$ is the boson mass and $m_F$ the fermion mass of a particle-supersymmetric partner pair,
is to be of the order of the electroweak gauge-boson masses or less,
the mass splittings should not be much greater than the electroweak scale. 

On a side note, there are other models for
dealing with the naturalness problem, for instance the Technicolor approach where fundamental scalars are composites
of new fermions. This approach is however technically difficult.

Local supersymmetry, i.e. where transformations are allowed to depend on position and time, and gravity are closely related. 
Loosely speaking, local supersymmetry gives local Poincar\'e symmetry, which is the basis for general relativity. Because
of this relation, local supersymmetry is called supergravity (SUGRA). Initially, it was thought that this supergravity could
hold as a successful theory of everything. However, theoretical considerations have shown that it most likely would only
be an effective theory up to some energy scale. Nevertheless, it remains important since it turns out to be the proper
framework for spontaneous breaking of supersymmetry (which is required).

There are rich experimental consequences from supersymmetric particle models, so it is possible to test the ideas. One important
prediction is the existence of stable particles, which could explain at least part of the cold dark matter problem in cosmology.

Even if supersymmetry in the end turns out not to be a correct description of Nature, we can learn a lot about how
high-energy physics should behave from studying these theories.

There are also aesthetic and philosophical arguments for considering supersymmetry.
For instance, Nature has shown it ``likes'' gauge symmetries, and supersymmetry would be the next natural such symmetry. 

A review of supersymmetric matter and observational aspects can be found in ref. \,\cite{Jungman:1996df}. Ref. \,\cite{Cooper:2001bk} gives a general introduction to supersymmetry in quantum mechanics. A comprehensive introduction to supersymmetry and supergravity
is given in Ref. \,\cite{Wess:1992bk}. The discussion above is based on those references.

\subsection{Minimal Supersymmetric Standard Model and mSUGRA}
\subsubsection{Definition}
The Minimal Supersymmetric Standard Model (MSSM) is, as the name implies, the standard supersymmetric model.
It contains all the known particles in the Standard Model plus
an extra Higgs isospin doublet (that is theoretically required)
together with their supersymmetric partners. The interactions of the model are those allowed by
$SU(3) \times SU(2) \times U(1)$ symmetry and renormalisability.
The MSSM, as do most supersymmetric models, requires R-parity, defined as
\begin{equation}
R=(-1)^{3(B-L)+2S}
\end{equation}
to be multiplicatively conserved ($B$ is baryon number, $L$ lepton number and $S$ spin).
Standard Model particles have $R=1$ and
 supersymmetric particles have $R=-1$. This places constraints on the allowed processes.
 For instance, we see that there must be a lightest supersymmetric particle that is stable
 (the LSP), since it has no allowed decay mode. Also, supersymmetric particles produced
 from Standard Model particles will be pair-produced.

The most general MSSM contains 63 parameters, disregarding the usual Standard Model parameters.
For practical purposes, some form of restrictions are usually imposed, limiting the number
of free parameters.

We will use a parameterisation known as mSUGRA (also UHM, denoting ``Unification of Higgs Masses''
or CMSSM, ``Constrained MSSM'')
which consists of five parameters, the common scalar mass $m_0$, the common gaugino mass $m_{1/2}$,
the sign of the higgsino mass parameter sgn$(\mu)$, the ratio of Higgs vacuum expectation values $\tan \beta$ 
and the common trilinear coupling $A$ (see next Section for an introduction to these particles). This
model assumes that the supersymmetry breaking is gravity-mediated, and in this framework the masses $m_0$ and 
$m_{1/2}$ are assumed universal at some GUT scale. This parameter space has been investigated by Ellis et al.
\cite{Ellis:2001hv} applying various experimental and theoretical constraints.
They give a set of benchmark models (parameter sets) that represent parts of the parameter space that are 
particularly interesting in light of the constraints. We will use these benchmark models in this work, referring
to them as ``mSUGRA Benchmark A-M'' (see also Appendix \ref{apx:models}).

\subsubsection{Particle Content}
For each ordinary Standard Model particle we have a supersymmetric partner, 
or \emph{sparticle}. The naming convention is such that fermions gain a prefix 
\emph{s} and bosons a postfix \emph{ino}. For instance, quarks have \emph{squark} 
partners and photons \emph{photino} partners. Table \ref{table:part} should make the notation conventions clear.

  \begin{center}
    \begin{threeparttable}

    \begin{tabular}{|lcc|lcc|}
      \hline
      \sc{Particles} & & & \bf{\sc{Sparticles}} & & \\
      \hline
      \bf{Name} & \bf{Symbol} & \bf{Spin} &
      \bf{Name} & \bf{Symbol} & \bf{Spin} \\
      \hline
      Charged & $\ell^{\pm}$ & 1/2 & 
      Charged & $\tilde{\ell}_{L,R}^{\pm}$ & 0 \\
      leptons & & & sleptons & & \\
      \hline
      Neutrinos
      \tnote{a} 
      & $\nu_{L},\bar{\nu}_{R}$ & 1/2 &
      Sneutrinos & $\tilde{\nu}_{L},\bar{\tilde{\nu}}_{R}$ & 0\\
      \hline
      Up quarks    & $q^{u}, \bar{q}^{u}$ &  1/2 &
      Up squarks   & $\tilde{q}_{L,R}^{u}, \bar{\tilde{q}}_{L,R}^{u}$ & 0 \\
	\hline
      Down quarks    & $q^{d}, \bar{q}^{d}$ &  1/2 &
      Down squarks   & $\tilde{q}_{L,R}^{d}, \bar{\tilde{q}}_{L,R}^{d}$ & 0 \\
      \hline
      Gluon     & $g$ & 1 &
      Gluino    & $\tilde{g}$ & 1/2\\
      \hline
      W-bosons  & $W^{\pm}$ & 1 &
      & &\\
      Charged Higgs & $H^{\pm}$ & 0 & 
      Charginos & $\tilde{\chi}_{1,2}^{\pm}$ & 1/2\\
      boson & & & & &\\
      \hline
      Photon    & $\gamma$ & 1 & 
      & &\\
      Z-boson   & $Z^0$ & 1 &
      & &\\
      Light scalar & $h^0$ & 0 &
      & & \\
      Higgs boson & & &
      Neutralinos & $\tilde{\chi}_{1,2,3,4}^{0}$ & 1/2\\
      Heavy scalar & $H^0$ & 0 &
      & & \\
      Higgs boson & & & & &\\
      Pseudoscalar & $A^0$ & 0 &
      & &\\
      Higgs boson & & & & &\\
      \hline
      Graviton  & $G$ & 2 &
      Gravitino & $\tilde{G}$ & 3/2 \\
      \hline
    \end{tabular}
    \begin{tablenotes}
    \item[a] Since neutrinos actually have a small mass, we should
      expect there to be a right-handed version also, although this has never
      been observed. Consequently, the sparticle side should in principle also contain
      corresponding degrees of freedom.
    \end{tablenotes}
     \caption{Standard Model particles and their SUSY partner sparticles.}
     \label{table:part} 
    \end{threeparttable}
  \end{center}

Table \ref{table:part} lists charginos and neutralinos, which are the mass eigenstates one finds by
diagonalising the Hamiltonian. 
Hence these states are the natural ones to use in most physical contexts. 
They can be written as superpositions of photinos, zinos, winos and higgsinos, i.e.
\begin{eqnarray}
  \tilde{\chi}_i^{\pm} & = & M_{i1}^{\pm} \tilde{W}^{\pm}+ M_{i3}^{\pm} \tilde{H}^{\pm} \, , \\
  \tilde{\chi}_i^{0} & = & N_{i1} \tilde{\gamma}+ N_{i2} \tilde{Z}^0 + N_{i3} \tilde{H}^{0}_1 
+ N_{i4} \tilde{H}^{0}_2 \, ,
\end{eqnarray}
where the $M$s and $N$s are the usual expansion coefficients.

In this work, we use two different meanings when referring to ``the Standard Model''.
When considering ``the Standard Model'' in its own right, e.g. in contrast to a supersymmetric
particle model, ``the Standard Model''
will be understood \emph{not to contain any Higgs particles}. 
When considering a supersymmetric particle model in its own right, Higgs bosons will be looked upon as 
``Standard Model'' particles, although formally not the same ``Standard Model'' as mentioned in the previous
sentence (one should perhaps rather use ``the Extended Standard Model'' in this case).
The notation of Table \ref{table:part} might confuse this issue, so the reader is advised to take special note
of these conventions.

\chapter{Photon Spectra from PBHs}

\section{Point Source Flux}
Choosing our position in the universe as the origin of a spherical coordinate system $(r, \theta
, \phi)$, we can express the (unattenuated) photon flux at Earth from a single PBH at $r$ as
\begin{equation}
\label{eq:pointflux}
\frac{dN_{1}^{\oplus}(r,M,E)}{dEdtdA} = 
\frac{1}{4\pi r^2} \frac{d^2N_{\gamma}(M,E)}{dEdt} \, .
\end{equation}
The symbol $\oplus$ is the astronomical symbol for Earth.
In principle, we should of course include redshift dependence and effects from interactions.
However, since we can only expect to detect evaporating PBHs within $\sim 1\,\rm pc$ \cite{Green:2001kw} we neglect those effects.

\section{Diffuse Flux}
The diffuse gamma-ray spectrum reaching Earth will in general be dependent on $dM/dt$, $dn/dM_{i}$, the cosmological model used and the interactions of the gamma rays while traveling towards Earth. These are suitably described using the optical depth $\tau(E, t)$, which is the effective number of interactions.

If we consider an isotropic distribution $dn/dM$ of PBHs, one can easily
write down the flux of photons with energy between $E$ and $E+dE$
within the solid angle $\Omega_{0}$ from PBHs with mass between $M$ and $M+dM$ as
\begin{eqnarray}
\frac{dN_{n}^{\oplus}(M,E,\Omega_{0})}{dEdtdAdM} =
 \int_{0}^{\Omega_{0}}\int_{0}^{r_{max}} \frac{dN_{1}^{\oplus}(r,M,E)}{dEdtdA}
 \frac{dn(M)}{dM} r^2drd\Omega \, .
\end{eqnarray}
Since we assume an
isotropic PBH distribution we can perform the solid angle integral to obtain
\begin{equation}
\frac{dN_{n}^{\oplus}(M,E,\Omega_{0})}{dEdtdAdM} =
\frac{\Omega_{0}}{4\pi} 
\int_{0}^{r_{max}} \frac{d^2N_{\gamma}(M,E)}{dEdt}\frac{dn(M)}{dM} dr \, ,
\end{equation}
where we have made use of the expression for the flux from a single PBH, Eq. (\ref{eq:pointflux}).
To turn this into a useful expression, we note that since we are dealing with photons, we have
$dr = -c dt$ (the minus sign since the photons are traveling inwards in the chosen coordinate system).
We can thus write
\begin{equation}
\frac{dN_{n}^{\oplus}(M,E)}{dEdtdAdMd\Omega_{0}} =
\frac{c}{4\pi} \int_{t_{min}}^{t_{0}} f_{\gamma}(M,E) \frac{dn(M)}{dM} dt_{1} \, ,
\end{equation}
where for simplicity we introduce
\begin{equation}
f_{\gamma}(M,E') \equiv \frac{d^2N_{\gamma}(M,E)}{dEdt}\arrowvert_{E=E'} \, ,
\end{equation}
the instantaneous photon flux at energy $E'$ from a PBH with mass $M$.
We should also integrate over $M$ to get the total spectrum,
\begin{equation}
\frac{dN_{n}^{\oplus}(E)}{dEdtdAd\Omega_{0}} =
\frac{c}{4\pi} \int_{t_{min}}^{t_{0}} \int_{M_{*}(t_{1})}^{M_{max}(t_{1})}
 f_{\gamma}(M_{evap},E) 
 \frac{dn(M_{i})}{dM_{i}} dM_{i} dt_{1} \, ,
\end{equation}
where we have changed variables to the initial PBH mass $M_{i}$.
 The limits are given
by the range of masses that are emitting at time $t_{1}$, which will be $[M_{*}
(t_{1}), M_{max}(t_{1})]$ ($M_{max}(t_{1})$ is the heaviest existing PBH at time $t_{1}$).
The quantity $M_{evap}(M_i,t)$ is the mass at time $t$ of a PBH with initial mass $M_i$, evaluated
according to Eq. (\ref{eq:m}).
This expression neglects effects from the expansion of the universe and
 degradation due to interactions (the optical depth).
Let $a(t)$ be the FLRW scale
factor (denoting $a_{j} \equiv a(t_{j})$). The relevant expansion effects are photon redshift and the decrease
in volume density. Including these and the optical depth, we thus get
\begin{eqnarray}
\label{eq:djde}
\nonumber
\frac{dJ}{dE_0} \equiv \frac{dN_{n}^{\oplus}(E_0)}{dE_{0} dt d\Omega_{0} dA}
& = & \frac{c}{4\pi} \int_{t_{min}}^{t_{0}} \int_{M_{*}(t_{1})}^{M_{max}(t_{1})}
 \frac{a_{0}}{a_{1}} \left(\frac{a_{1}}{a_{i}}\right)^{-3} e^{-\tau(E_0,z_{1})} \times \\
&& \times
 f_{\gamma}(M_{evap},(1+z_1)E_0) 
 \frac{dn(M_{i})}{dM_{i}} dM_{i} dt_{1} \, ,
\end{eqnarray}
where $E=(1+z_1)E_0$ was used to rewrite in terms of the observed photon energy $E_0$ and
$z_1 = z(t_1)$. In keeping with convention, subscripts $i$ denote the time of PBH formation.
We may also rewrite this expression completely in terms of the redshift $z$ using the 
definition of redshift, $1+z_{j} = a_{0}/a_{j}$, and the following 
expression (for derivation, see for instance Ref. \,\cite{Bergstrom:1999bk}).
\begin{equation}
\frac{dt}{dz} = -H_{0}^{-1}(1+z)^{-1}[(1+z)^2(1+\Omega_{M}z)-z(2+z)\Omega_{\Lambda}+\Omega_{R}(1+z)^4]^{-1/2} \, .
\end{equation}
Here we clearly see the dependence on the cosmological parameters. The radiation
contribution $\Omega_{R}$ is only relevant for very high redshifts in the radiation-dominated epoch ($z \apprge 1100$). 

This derivation of the diffuse flux follows those of Hawking and Page \,\cite{Page:1976ab} and Kim, Lee and MacGibbon \,\cite{Kim:1999iv}.

\section{Optical Depth}
\label{sec:opdepth}
The optical depth $\tau(E_0,z)$ has been studied by Zdziarski and Svensson 
\,\cite{Zdziarski:1988ab}. We will look at the processes they consider in the following
subsections. Since we are considering a supersymmetric scenario, one also has to consider
potential contributions from such particles to the optical depth. However, any such sparticles
would quickly
decay to ordinary particles and LSPs. The LSP-photon cross sections are very small, so we should
not expect any difference in the optical depth due to sparticles.

For an introduction to the notion of optical depth, see e.g. Ref. \,\cite{Weinberg:1972bk}.

In the following, we will use the same parameterisation as Zdziarski and Svensson, i.e.
\begin{eqnarray}
\epsilon_0 & \equiv & \frac{E_0}{511\,{\rm keV}} \, ,\\
T_{2.7} & \equiv & \frac{T_0}{2.7\,{\rm K}} \, ,\\
h_{50} & \equiv & \frac{h_0}{0.5} \, , \\
\Omega_{0.1} & \equiv & \frac{\Omega_b}{0.1} \, , \\
\Theta_0 & \equiv & 4.55 \times 10^{-10} T_{2.7} \, .
\end{eqnarray}
Furthermore, the formulae given below assume $\Omega_{TOT}=1$.

\subsection{Photoionisation}
Neutral H and He atoms may become photoionised by photons with energies larger than the ionisation energies, 13.6 eV and 24.6 eV
respectively ($\gamma A \rightarrow A^+ e^-$). 

Below the reionisation redshift $z_{ri}$, the optical depth due to photoionisation is essentially zero.
With that assumption, the optical depth due to photoionisation, $\tau_{pi}$, takes the form
\begin{equation}
\tau_{pi} = 1.1\times10^{-10}\Omega_{0.1}h_{50}\epsilon_0^{-3.3}\left[ \left(1+z_{ri}\right)^{-1.8} - \left(1+z\right)^{-1.8} \right] \, .
\end{equation}
The expression for $\tau_{pi}$ is a numerical fit valid for photon energies $E \ge 250 \,\, {\rm eV}$. 

\subsection{Compton Scattering}
We divide the effects of Compton scattering ($\gamma e \rightarrow \gamma e$) in a scattering loss and an energy loss
component.
\subsubsection{Scattering Loss}
The optical depth $\tau_{cs}$ due to the scattering component is given exactly by
\begin{equation}
  \tau_{cs} = 2.27\times10^{-3}\Omega_{0.1}h_{50}\epsilon_0^{-3/2} \left\{ F_{cs}\left[ \epsilon_0\left(1+z\right) \right]
   - F_{cs}\left(\epsilon_0\right) \right\} \, ,
\end{equation}
where
\begin{eqnarray}
\nonumber
  F_{cs}(\epsilon) & = & \frac{(1+\epsilon)^2 - 1/3}{\epsilon^{3/2}}\ln(1+2\epsilon) - \frac{3\epsilon+7/4}{1+2\epsilon}\epsilon^{1/2} - \frac{4}{3\epsilon^{1/2}} - \\
& & - \frac{11 \arctan\left[(2\epsilon)^{1/2}\right]}{3\times2^{5/2}} \, .
\end{eqnarray}

\subsubsection{Energy Loss}
The optical depth $\tau_{ce}$ due to the energy loss component is given exactly by
\begin{equation}
  \tau_{ce} = 1.14\times10^{-3}\Omega_{0.1}h_{50}\epsilon_0^{-3/2} \left\{ F_{ce}\left[ \epsilon_0\left(1+z\right) \right]
   - F_{ce}\left(\epsilon_0\right) \right\} \, ,
\end{equation}
where
\begin{eqnarray}
\nonumber
  F_{ce}(\epsilon) & = & \frac{2(1+\epsilon)^2}{\epsilon^{3/2}}\ln(1+2\epsilon) - 
\frac{272\epsilon^3+470\epsilon^2+261\epsilon+48}{12\epsilon^{1/2}(1+2\epsilon)^2} + \\
&& + \frac{7 \arctan\left[(2\epsilon)^{1/2}\right]}{2^{5/2}} \, .
\end{eqnarray}

\subsection{Photon-Matter Pair Production}
Pair production from photon-matter interaction is primarily in the form of photon-atom pair production 
($\gamma A \rightarrow \gamma A e^+ e^-$)
for
$z<z_{rec}$ and photon-ionised matter pair production ($\gamma e \rightarrow e e^+, \gamma Z \rightarrow
Z e^+ e^-$)
for $z>z_{rec}$, where $z_{rec} \approx 1300$ is the redshift of
recombination.

\subsubsection{Atoms}
The optical depth $\tau_{app}$ due to photon-atom pair production is given by 
\begin{equation}
  \tau_{app} = 9.3\times10^{-6}\Omega_{0.1}h_{50} \epsilon_0^{-3/2} 
\left\{ F_{app}\left[ \epsilon_0\left(1+z\right) \right]
   - F_{app}\left(\epsilon_0\right) \right\} \, ,
\end{equation}
where
\begin{eqnarray}
\nonumber
 F_{app}(\epsilon) & = &
\epsilon^{3/2} \ln \left( \frac{513\epsilon}{825+\epsilon} \right) - 1650\epsilon^{1/2} + \\
&& + 2\times825^{3/2}
\arctan\left[\left(\frac{\epsilon}{825}\right)^{1/2}\right] \, .
\end{eqnarray}
The expression for $\tau_{app}$ is found using fitted numerical data, and is valid for $(1+z)\epsilon_0 > 6$.

\subsubsection{Ionised Matter}
The optical depth $\tau_{ipp}$ due to photon-ionised matter pair production is given by
\begin{equation}
\tau_{ipp} =
1.15\times10^{-5}\Omega_{0.1}h_{50}
\left\{ \left(1+z\right)^{3/2}F_{ipp}\left[ \epsilon_0\left(1+z\right) \right]
   - F_{ipp}\left(\epsilon_0\right) \right\} \, ,
\end{equation}
where
\begin{equation}
 F_{ipp}(\epsilon) =
 \ln(2\epsilon) - \frac{109}{42} - \frac{2}{3} \, .
\end{equation}
This expression for $\tau_{ipp}$ is valid for $(1+z)\epsilon_0 \gg 1$.

\subsection{Photon-Photon Scattering}
Photon-photon scattering ($\gamma \gamma \rightarrow \gamma \gamma$) gives an optical depth $\tau_{\gamma s}$
\begin{equation}
\tau_{\gamma s} = 2.44\times10^{-28}h_{50}^{-1}T_{2.7}^6 \epsilon_0^3
\left[ \left(1+z\right)^{15/2}  - 1 \right] \, .
\end{equation}
This expression is valid for $(1+z)^2\epsilon_0\Theta_0 \apprle 0.1$.

\subsection{Photon-Photon Pair Production}
\subsubsection{Single Photon Pair Production}
The optical depth $\tau_{\gamma spp}$ due to single photon pair production ($\gamma \gamma \rightarrow e^+ e^-$)
is divided in two limits. For $(1+z)^2\epsilon_0\Theta_0 \ll 1$ we have
\begin{eqnarray}
\nonumber
  \tau_{\gamma spp} & = & 3.83\times10^{5}h_{50}^{-1}T_{2.7}^{3} \left(\pi\epsilon_0\Theta_0\right)^{1/2}\times \\
 &&  \times \left\{ \left(1+z\right)^{5/2}F_{sppl}\left[ \epsilon_0\left(1+z\right), \Theta_0\left(1+z\right) \right]
   - F_{sppl}\left(\epsilon_0, \Theta_0\right) \right\} \, ,
\end{eqnarray}
where
\begin{equation}
 F_{sppl}(\epsilon,\Theta) =
\left(1+\frac{7\epsilon\Theta}{2}\right)e^{-1/\left(\epsilon\Theta\right)} \, ,
\end{equation}
and for $(1+z)^2\epsilon_0\Theta_0 \gg 1$ we have
\begin{eqnarray}
\nonumber
  \tau_{\gamma spp} & = & 3.83\times10^{5}h_{50}^{-1}T_{2.7}^{3} \frac{4\pi^2}{3\epsilon_0\Theta_0}\times \\
  && \times \left\{ F_{spph}\left( \epsilon_0, \Theta_0 \right)
   - \left(1+z\right)^{-1/2} F_{spph}\left[\epsilon_0\left(1+z\right), \Theta_0\right] \right\} \, ,
\end{eqnarray}
where
\begin{equation}
 F_{spph}(\epsilon,\Theta) =
\ln\left(25.5\epsilon\Theta\right) \, .
\end{equation}

\subsubsection{Double Photon Pair Production}
The optical depth $\tau_{\gamma dpp}$ due to double photon pair production ($\gamma \gamma \rightarrow e^+ e^- e^+ e^-$)
is given by
\begin{equation}
  \tau_{\gamma dpp} = 47.5h_{50}^{-1}T_{2.7}^{3}\left[\left(1+z\right)^{3/2}-1\right] \, .
\end{equation}
This expression is valid for $(1+z)^2\epsilon_0\Theta_0 \gg 1$.

\section{Sparticle Effects - An Estimate}
\label{sec:speff}
We can expect effects from sparticles to show up in the PBH emission above 
$T \sim 100$ GeV (if we have a fairly low sparticle mass spectrum).
To illustrate the behaviour of the diffuse flux from PBHs further,
especially how SUSY scenarios might affect it, we turn to studying the
diffuse flux integral \emph{at peak photon energies}.
We begin by considering the low-temperature part of the integral
(\ref{eq:djde})
\begin{eqnarray}
\label{eq:djltappr}
\nonumber
\frac{dJ}{dE_{0}} & \propto & \frac{c}{4\pi} \int_{t_{min}}^{t'}
(1+z_1)^4 f_{\gamma}(T(t_1),E) dt_1 \approx \\
\nonumber
& \approx & 10^{25} \,{\rm m^{-2} \, sr^{-1}}
\int_{z'}^{z_{max}} (1+z_1)^{3/2} f_{\gamma}(T_i,E) dz_1 \le \\
\nonumber
& \le & \frac{10^{25} \,{\rm m^{-2}\, sr^{-1}}}{5/2}
\left[(1+z_{max})^{5/2} - (1+z')^{5/2} \right] f_{\gamma}(T_i,E_{peak}) \approx \\
& \approx & \frac{10^{25} \,{\rm m^{-2}\, sr^{-1}}}{5/2}(1+z_{max})^{5/2}f_{\gamma}(T_i,E_{peak}) \, ,
\end{eqnarray}
where $t'$ is the time when the black hole temperature becomes some order of magnitude higher
than the initial temperature $T_i$, $z_{max}$ is an approximation for the highest contributing
redshift and $E_{peak} \approx 100 \,\rm MeV$ is the peak emission photon energy.
This approach was chosen since the black hole spends most of
its time near $M_i$, and during this time the flux does not change very much.
It was assumed that $\Omega_{TOT} = \Omega_M = 1$ for simplicity.
In the last step we assumed $z_{max} \gg z'$ which is valid for the range
of PBHs we consider.
The optical depth was also assumed to be significantly higher
than 1 for $z \apprge z_{max}$ and not to change significantly for lower redshifts. This should be
sufficient for an order of magnitude estimate (within a factor 10 or so)
in the observable photon energy range where one can see the furthest back in time.
It will become clear later that only an upper limit such as this will be sufficient for our purposes.

To get an estimate for the high temperature flux, the integral (\ref{eq:djde}) is rewritten in terms of the
black hole temperature $T$. First, we need the expression
\begin{equation}
1+z \approx \left[\left(\frac{\hbar c^3}{8 \pi kG}\right)^3
\frac{\Omega_M^{1/2} H_0}{2 \Phi} \left(T_i^{-3}-T^{-3} \right) + \left(1+z_i\right)^{-3/2}  \right]^{-2/3} \, .
\end{equation}
This expression neglects the dependence in $\Phi$ on $t$, i.e. we choose to evaluate $\Phi$ at some time,
typically the initial time $t_i$. The function $\Phi$ generally varies less than a factor 10 over the PBH
lifetime (at least for the PBHs we consider), so our assumption will underestimate $\Phi$ with at most a factor 10.
When $T \gg T_i$ and $z_i \sim 10^{28}$ the expression is simplified to
\begin{equation}
\label{eq:zsimple}
1+z \approx \left(\frac{\hbar c^3}{8 \pi kG}\right)^{-2} 
\left( \frac{H_0}{2 \Phi} \right)^{-2/3} \Omega_M^{-1/3} T_i^{2} \, .
\end{equation}
Using this expression for $1+z$, we can write the high-temperature part of the integral (\ref{eq:djde}) as
\begin{eqnarray}
\nonumber
\frac{dJ}{dE_{0}} & \propto & 5 \times 10^{33}\,{\rm m^{-2}\,  sr^{-1}}
\left(\frac{T_i}{1\,\rm GeV}\right)^{8} \times \\
&& \times
\int_{T'}^{T_{0}}f(T)^{5/3} 
 e^{-\tau(E_0, z_1(T))} f_{\gamma}(T,E) T^{-4} dT \, ,
\end{eqnarray}
where $T' = T(t')$ with $t'$ as earlier and $T_0$ is the temperature at the present time (or rather, the highest
temperature the PBH has had by today in the case that the PBH has completely evaporated by today).
Neglecting optical depth, and realising that the redshift at photon emission remains fairly constant, see Eq.
(\ref{eq:zsimple}), we may simplify the integral to
\begin{eqnarray}
\label{eq:speffhigh}
\nonumber
\frac{dJ}{dE_{0}} & \propto & 5 \times 10^{33} \,{\rm m^{-2}\, sr^{-1}} \left(\frac{T_i}{1\,\rm GeV}\right)^{8}
f_{\gamma}(T_{ref}, E_{peak}) T_{ref}^{-\beta}
\times \\
\nonumber
& & \times
\int_{T'}^{T_{0}}f(T)^{5/3} T^{\beta-4} dT \approx \\
\nonumber
& \approx &
\frac{5}{3-\beta} \times 10^{33} \,{\rm m^{-2}\,  sr^{-1}}\left(\frac{T_i}{1\,\rm GeV}\right)^{8} 
\frac{T'^{\beta-3}}{T_{ref}^{\beta}} f(T_i)^{5/3} \times \\
& & \times
f_{\gamma}(T_{ref}, E_{peak}) \, .
\end{eqnarray}
It was here assumed that the instantaneous flux $f_{\gamma} \propto T^{\beta}$. This is supported by simulation data, with
$1 \apprle \beta \apprle 2$ (see e.g. Figure \ref{figure:SMinst} or Ref. \cite{MacGibbon:1990zk}).
We thus evaluate $f_{\gamma}$ at some reference temperature $T_{ref}$ and photon energy
$E_{peak} \approx 100\,\rm MeV$ (since we are interested in the peak flux). We can choose $T_{ref} = T'$ to simplify
the expression somewhat.
The assumption of constant redshift of emission is further supported by the fact that around the instantaneous peak
energy, the flux does not vary significantly with photon energy. Hence, the flux is not very sensitive to variations
in $z$ at these photon energies. The last step assumes $T_0 \gg T'$ which will be true in the cases we consider.

We can now use these assumptions for the cases we consider. In these cases, the low-temperature
contribution (\ref{eq:djltappr}) will in fact be negligible compared to the contribution from higher temperatures, so
we need only consider the case $T \gg T_i$. 
For a black hole with mass $M_i \approx 5 \times 10^{14} \,\rm g$ we have
\begin{eqnarray}
\nonumber
T_i & \approx & 0.02 \,{\rm GeV} \, ,\\
\nonumber
T' & \approx & 0.1 \,{\rm GeV} \, ,\\
z_{max} & \sim & 10^3 \, ,
\end{eqnarray}
and inserting these in Eq. (\ref{eq:speffhigh}) with $T_{ref}=T'$ we obtain the requirement
\begin{equation}
 f_{\gamma}^{0.1\rightarrow 100} 0.1^{-3} \apprle
 f_{\gamma}^{100\rightarrow T_{Pl}} 100^{-3}
 \Rightarrow 
 f_{\gamma}^{100\rightarrow T_{Pl}} \apprge 10^9 f_{\gamma}^{0.1\rightarrow 100} \, ,
\end{equation}
for the flux from $T \apprge 100\,\rm GeV$ to be non-negligible. Here $f_{\gamma}^{T_1 \rightarrow T_2}$ symbolically 
denotes ``the flux between $T=T_1$ and $T=T_2$''.
In the case of a black hole with $M_i \approx 3 \times 10^{13} \,\rm g$ we use
\begin{eqnarray}
\nonumber
T_i & \approx & 0.35 \,{\rm GeV} \, , \\
\nonumber
T' & \approx & 1 \,{\rm GeV} \, ,\\
z_{max} & \sim & 2\times10^3 \, ,
\end{eqnarray}
and in the same manner as above, we obtain the requirement
\begin{equation}
 f_{\gamma}^{1\rightarrow 100} 1^{-3} \apprle
 f_{\gamma}^{100\rightarrow T_{Pl}} 100^{-3}
 \Rightarrow 
 f_{\gamma}^{100\rightarrow T_{Pl}} \apprge 10^6 f_{\gamma}^{1\rightarrow 100} \, .
\end{equation}
From simulation data (see Figure \ref{figure:SMinst} or Ref. \cite{MacGibbon:1990zk}) we know that the fluxes are approximately
\begin{eqnarray}
f_{\gamma}^{0.1\rightarrow 100} & \approx & 10^{23} \,{\rm GeV^{-1} s^{-1}} \, ,\\
 f_{\gamma}^{1\rightarrow 100} & \approx & 10^{25} \,\rm GeV^{-1} s^{-1} \, .
\end{eqnarray}
Inserting this above we find that the instantaneous flux from
temperatures above 100 GeV should fulfill
\begin{eqnarray}
 f_{\gamma}^{100\rightarrow T_{Pl}} & \apprge & 10^{32} \,{\rm GeV^{-1} s^{-1}}, M_i \approx 5\times 10^{14} \,{\rm g} \, , \\
f_{\gamma}^{100\rightarrow T_{Pl}} & \apprge & 10^{31} \,{\rm GeV^{-1} s^{-1}}, M_i \approx 3\times 10^{13} \,{\rm g} \, ,
\end{eqnarray}
for the observed flux at peak energies to be significantly affected.
These approximations should be valid to within about
one or two factors of ten, taking into account also optical depth effects.
The values agree well with MacGibbon's estimate that the increase in degrees of freedoms for
$T > 100 \,\rm GeV$ would have to be of the order $10^8$ to significantly affect the
lifetime emission of a PBH \cite{MacGibbon:1991tj}.

In the preceding, we have looked at the dominant peak emission, with emitted photon energy $\sim 100\,\rm MeV$.
There is of course the possibility that the diffuse flux coming from higher emitted energies may be
affected. This will be explored further in the Results Section.

\section{Applied Model} 
\label{sec:model}
\subsection{Initial Mass Function}
Green and Liddle have shown \,\cite{Green:1999xm} that a reasonable first approximation is to use
a PBH distribution created
at $t \sim 0$ with a single initial mass $M_i^0$ and volume number density $n_i$.
This gives the initial mass function
\begin{equation}
\frac{dn}{dM_i} = n_i\delta(M_i - M_i^0) =  n_0(1+z_i)^3 \delta(M_i - M_i^0) \, ,
\end{equation}
where $n_0$ is the present number density and $z_i$ is the redshift of PBH formation. 
An important note to make is that this definition of the present number density is just the diluted initial density,
hence the effect of evaporation is not considered (i.e. no PBHs disappear). The value of
$n_0$ can be constrained using the diffuse gamma-ray flux from PBHs. It can also be independently
constrained by requiring $\Omega_{PBH} < 1$. Using the formulae of Section \ref{sec:formation} and Ref.
\cite{Bergstrom:1999bk}, one finds that the value at the epoch of PBH formation is
\begin{equation}
\Omega_{PBH}^i \equiv \frac{\rho_{PBH}^i}{\rho_{c}^i} \approx 2 \times 10^{-26}\left(
\frac{n_0}{1 \, \rm pc^{-3}} \right) \left( \frac{M_i^0}{1 \,\rm g} \right) \, ,
\end{equation}
where $\rho_c = 3H^2 / (8\pi G)$ is the critical energy density.
The value of $\Omega_{PBH}^i$ is also an upper limit on $\Omega_{PBH}$ at later times. 
Apparently, the value of $n_0$ 
can be significantly greater than $1 \,\rm pc^{-3}$ for PBHs emitting particles in non-negligible amounts (i.e. with
sufficiently small initial masses) without the Universe being overclosed.

We will examine two ``limiting'' cases for $M_i^0$,
\begin{itemize}
\item a PBH expiring today
\item a PBH expiring when (the thinnest part of) the optical depth dropped below unity ($z \sim 900$)
\end{itemize}
These limits correspond to the range of PBHs contributing to the DGB today, which 
have initial masses approximately
in the range $3\times10^{13}$ - $5\times10^{14}\,\rm g$, corresponding
to initial temperatures approximately in the range $0.02 - 0.35\,\rm GeV$.
We will thus use $M_i^0 \approx 5 \times 10^{14} \,\rm g$ and 
$M_i^0 \approx 3 \times 10^{13} \,\rm g$ respectively for the two cases listed above. 

We will also look at some ``intermediate'' mass PBHs in the range $4 \times 10^{13}\,{\rm g} \apprle
M_i \apprle 1 \times 10^{14}\,{\rm g}$, since the flux from these might possibly show a
difference in the high-energy part of the diffuse photon flux.

\subsection{Cosmological Parameters}
We will use cosmological parameters according to WMAP results
 \,\cite{Bennett:2003bz}, except for $\Omega_{TOT} = 1.0$, $z_{rec} = 1300$ (this deviation from
WMAP results is still within uncertainties, and does not affect our conclusions). In summary, we use
\begin{eqnarray}
\nonumber
T_{0} & = & 2.725\,{\rm K} \\
\nonumber
h_0 & = & 0.71 \\
\nonumber
\Omega_{TOT} & = & 1.0 \\
\nonumber
\Omega_M & = & 0.28 \\
\nonumber
\Omega_{\Lambda} & = & 0.72 \\
\nonumber
\Omega_b & = & 0.044 \\
\nonumber
z_{ri} & = & 20 \\
z_{rec} & = & 1300
\end{eqnarray}
and thus we have the Zdziarski-Svensson parameters 
\begin{eqnarray}
\nonumber
T_{2.7} & \approx & 1.01  \\
\nonumber
h_{50} & = & 1.42 \\
\nonumber
\Omega_{0.1} & = & 0.44 \\
\Theta_0 & \approx & 4.60 \times 10^{-10}
\end{eqnarray}

\subsection{Optical Depth}
\begin{figure}
  \begin{center}
    \includegraphics[width=0.75\textwidth]{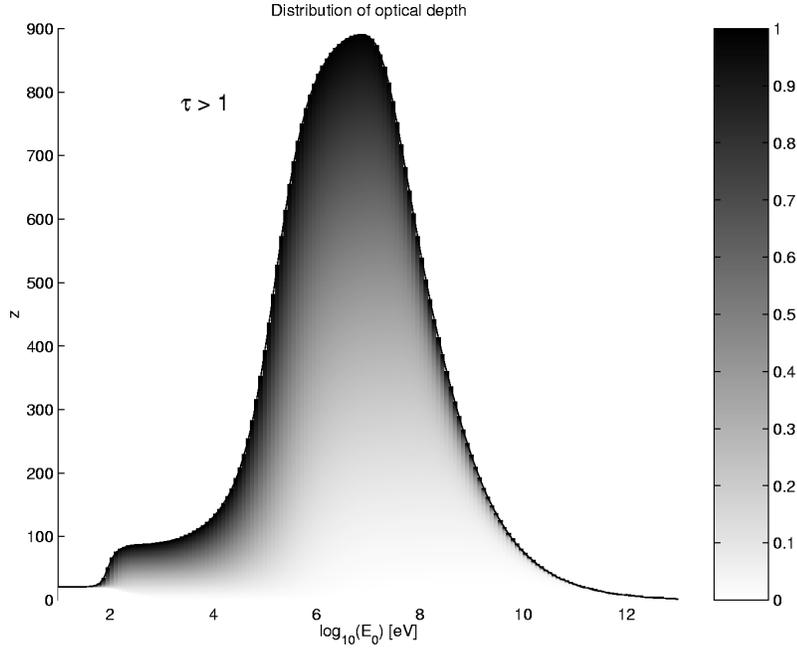}
  \end{center}
  \caption{Unit optical depth boundary using results in \,\cite{Zdziarski:1988ab}. Cosmological parameters are
  $\Omega_{TOT} = 1.0, \Omega_b = 0.044, h = 0.71, z_{ri} = 20, z_{rec} = 1300, T_0 = 2.725\,{\rm K}$ according to WMAP
    results \,\cite{Bennett:2003bz}.}
  \label{figure:taubound}
\end{figure}

The optical depth was evaluated using the formulae given in Section \,\ref{sec:opdepth}.
With the parameters in the preceding Section, the optical depth then 
takes the form shown in Figure \,\ref{figure:taubound}.

\section{GLAST}
The Gamma-ray Large Area Space Telescope (GLAST) due for launch in 2006 will improve measurements of the diffuse
gamma-ray background. GLAST contains two instruments, the Large Area Telescope (LAT) observing the gamma-ray
background and the GLAST Burst Monitor (GBM) searching for gamma-ray bursts.

A summary of the specifications of the LAT is shown in Table \ref{table:glast} along with
comparisons to EGRET. The GBM specifications are reported in Table \ref{table:gbm} with
comparisons to BATSE. For a fuller description of these characteristics, see Ref. \cite{GLAST:www}.

It is clear that GLAST measurements will significantly improve our knowledge of the diffuse gamma-ray background as well
as gamma-ray bursts and hence related phenomena such as PBH abundance and primordial density perturbation constraints.

 \begin{center}
    \begin{threeparttable}

    \begin{tabular}{|l|c|c|}
      \hline
      \bf{Quantity} & \bf{LAT (minimum spec.)} & \bf{EGRET} \\
      \hline
      Energy range & 20 MeV - 300 GeV & 20 MeV - 30 GeV \\
      \hline
      Peak effective area\tnote{a} & $> 8000 \, \rm cm^2$ & $1500 \,\rm cm^2$ \\
      \hline
      Field of view & $>$ 2 sr & 0.5 sr \\
      \hline
      Angular resolution\tnote{b} &  $< 3.5^\circ$ (100 MeV) &  $5.8^\circ$ (100 MeV) \\
      & $<0.15^\circ$ ($>$10 GeV) & \\ 
      \hline
      Energy resolution\tnote{c} & $< 10\%$ & $10\%$ \\
      \hline
      Deadtime per event & $<100 \,\rm \mu s$ & 100 ms \\
      \hline
      Source location determination\tnote{d} & $<0.5'$ & $15'$ \\
      \hline
      Point source sensitivity\tnote{e} & $<6 \times 10^{-9} \,\rm cm^{-2}\, s^{-1}$ & $\sim 10^{-7} \,\rm cm^{-2}\, s^{-1}$ \\
      \hline
    \end{tabular}
    \begin{tablenotes}
    \item[a] After background rejection
    \item[b] Single photon, $68\%$ containment, on-axis
    \item[c] 1-$\sigma$, on-axis
    \item[d] 1-$\sigma$ radius, flux $10^{-7} \,\rm cm^{-2}\, s^{-1}$ ($>$100 MeV), high $|b|$
    \item[e] $>$ 100 MeV, at high $|b|$, for exposure of one-year all sky survey, photon spectral index -2 
    \end{tablenotes}
     \caption{LAT specifications compared to EGRET. Reproduced from \cite{GLAST:www}.}
     \label{table:glast} 
    \end{threeparttable}
  \end{center}

 \begin{center}
    \begin{threeparttable}

    \begin{tabular}{|l|c|c|}
      \hline
      \bf{Quantity} & \bf{GBM (minimum spec.)} & \bf{BATSE} \\
      \hline
      Energy range & $<$ 10 keV - $>$ 25 MeV & 25 keV - 10 MeV \\
      \hline
      Field of view & all sky not occulted by the Earth & 4p sr \\
      \hline
      Energy resolution\tnote{a} & $<10\%$ & $<10\%$ \\
      \hline
      Deadtime per event & $<15\, \mu s$ & \\
      \hline
      Burst sensitivity\tnote{b} & $<0.5\,\rm cm^{-2}\, s^{-1}$ & $0.2 \,\rm cm^{-2}\, s^{-1}$ \\
      \hline
      Alert GRB location\tnote{c} & $\sim 15^\circ$ & $\sim 25^\circ$ \\
      \hline
      Final GRB location\tnote{d} & $\sim 3^\circ$ & $ 1.7^\circ$ \\
      \hline  
    \end{tabular}
    \begin{tablenotes}
    \item[a] 1-$\sigma$, 0.1 - 1 MeV
    \item[b] 50 - 300 keV
    \item[c] Calculated on-board; 1 second burst of 10 photons $\rm cm^{-2}\, s^{-1}$, 50 - 300 keV
    \item[d] Final ground computed locations; 1 second burst of 10 photons $\rm cm^{-2}\, s^{-1}$, 50 - 300 keV 
    \end{tablenotes}
     \caption{GBM specifications compared to BATSE. Reproduced from \cite{GLAST:www}}
     \label{table:gbm} 
    \end{threeparttable}
  \end{center}

\chapter{Simulation}

\section{\textsc{Pythia}}
The black hole emission was simulated using the particle physics event generator 
\textsc{Pythia} 6.212 \,\cite{Sjostrand:2000wi}. \textsc{Pythia} is a frequently used event generator
using the Lund model of string fragmentation for QCD particles. It produces 
results well within experimental constraints for known physics and includes
various theoretical models for hypothetical physics. Specifically, it can
simulate sparticle processes for various SUSY models. \textsc{Pythia} assumes R-parity
conservation. Only the lowest order sparticle pair production is included. 
SUSY decays of the top quark are included, but all other Standard Model particle decays are
unaltered. \textsc{Pythia} takes MSSM/mSUGRA parameters and calculates masses, cross 
sections etc. using analytical formulae.

Since \textsc{Pythia} is designed for beam collisions, and not direct 
decay of particles, some modifications had to be made, according to 
instructions 
from the author Torbj\"{o}rn Sj\"{o}strand. The changes made were in the
 function 
PYRESD. For details, see Appendix \ref{apx:pmod}.

\section{Convolution Methods}
We wish to perform the convolution of the direct emission spectra with the
 fragmentation functions. To do this, one could choose (at least) two 
different approaches: 
\begin{enumerate}
\item Evaluate the fragmentation functions fully for some initial-particle energies, fit a scaling-relation for all energies and then integrate (most likely numerically).
\item Generate a direct emission spectrum of particle energies and let them fragment and decay to find the final spectra.
\end{enumerate}
The benefit of the first option is that we obtain an expression for all temperatures $T$ of the black hole. One of the downsides is that there are as of now no good parameterisations of the fragmentation functions for general coloured particles \,\cite{Sjostrand:2003}. One could probably find fits, but this would require considerable computer time.

The benefit of the second option is that we do not need any knowledge of how the fragmentation functions behave. 
It is also a more direct representation of the actual physical process. We save computer time from this, but the 
downside is that we only get the instantaneous flux for a specific black hole temperature $T$. We need to 
perform a new simulation for each black hole temperature we are interested in. However, between mass 
thresholds, the instantaneous flux is not affected by any new emitted degrees of freedom, and so should behave 
fairly predictably.

For our purposes, the second option was chosen. This is also the choice made by MacGibbon-Webber 
\,\cite{MacGibbon:1990zk,MacGibbon:1991tj} and Barrau et al. \,\cite{Barrau:2001ev}.

\section{Rejection Method}

To perform the simulation we need to generate particle energies according to the direct (Hawking) emission spectrum. To achieve this, we use the \emph{rejection method} initially introduced by von Neumann in 1951. A variant of it consists of the following algorithm:
\newtheorem{rejalg}{Definition}
\begin{rejalg}[Rejection method]
We wish to generate random numbers according to the probability density $p(x)$ in the interval $[a,b]$.
Let $p_{m} \geq \max_{a \leq x \leq b} p(x)$. Let Z be the desired random variable.
\begin{enumerate}
\item Generate a random variable $X \in U(a,b)$
\item Generate a random variable $Y \in U(0,p_{m})$
\item If $Y > p(X)$, start over. 
\item Set Z=X, which is the generated random number (i.e. when $Y \leq p(X)$).
\end{enumerate}
\end{rejalg}

\newtheorem{rej}{Theorem}
\begin{rej}
The \emph{rejection method} produces the random variable $Z$ distributed according to $p(x)$ in $[a,b]$.
\end{rej}
Proof:
\begin{displaymath}
P(Z\leq r) = P(X\leq r | Y\leq p(X)) = \frac{\int_{-\infty}^{r}\int_{0}^{p(x)} \frac{1}{b-a}\frac{1}{p_{m}} dy dx}
{\int_{-\infty}^{\infty}\int_{0}^{p(x)} \frac{1}{b-a}\frac{1}{p_{m}} dy dx} =
\frac{\int_{-\infty}^{r} p(x) dx}
{\int_{-\infty}^{\infty} p(x) dx}
\end{displaymath}
To use the rejection method, we need to find the probability density for the directly emitted spectrum. Let $Q_{j}$ be a random variable equal to the total energy of an emitted particle of species $j$ with rest mass $\mu_{j}$. We can then write (assuming $q\geq \mu_{j}$)
\begin{equation}
P(Q_{j}\leq q) = \frac{\alpha_{j} \int_{\mu_{j}}^{q} \frac{d\dot{N}_{j}}{dQ} dQ}{\alpha_{j}\dot{N_{j}}} \, ,
\end{equation}
since this corresponds to the fraction of emitted particles with $Q_{j}\leq q$. We use the direct emission rate, Eq.(\ref{eq:hawking}).
This is the probability distribution, so the probability density is
\begin{equation}
f_{Q_{j}}(q) = \frac{d}{dq} P(Q_{j}\leq q) = \frac{\frac{d\dot{N_{j}}}{dq}(q)}{\dot{N_{j}}} \, ,
\end{equation}
which one might have seen straight away. 

\section{Simulation Algorithm}
\label{sec:sim}
To perform our simulation, we can now use the following algorithm for a black hole with temperature $T$:

\begin{enumerate}
\item Randomly select with probability $1/n$ a particle species $j$ from the emitted species $\{p_{i}\}_{1}^{n}$ $K$ times (this is to do the same amount of simulation on each species).
\item Find a particle energy $Q_{j}$ using the rejection method with probability density $f_{Q_{j}}$.
The limits $\mu_{j}/T\leq Q_{j}/T \leq 10$ can be suitable.
\item Create the particle in \textsc{Pythia}, let it fragment and decay the products.
\item Keep track of the different final products and store them for each initial particle species separately.
\item When finished generating particles, normalise the fluxes to the total flux, i.e. multiply the stored fluxes coming from species $j$ by $\alpha_{j}\dot{N_{j}}/\tilde{n}_{j}$, where $\tilde{n}_{j} \approx K/n$ is the number of $j$ particles created during the program run.
\end{enumerate}

\chapter{Results}

\section{Instantaneous Photon Flux}

\begin{figure}
  \begin{center}
    \includegraphics[width=0.75\textwidth]{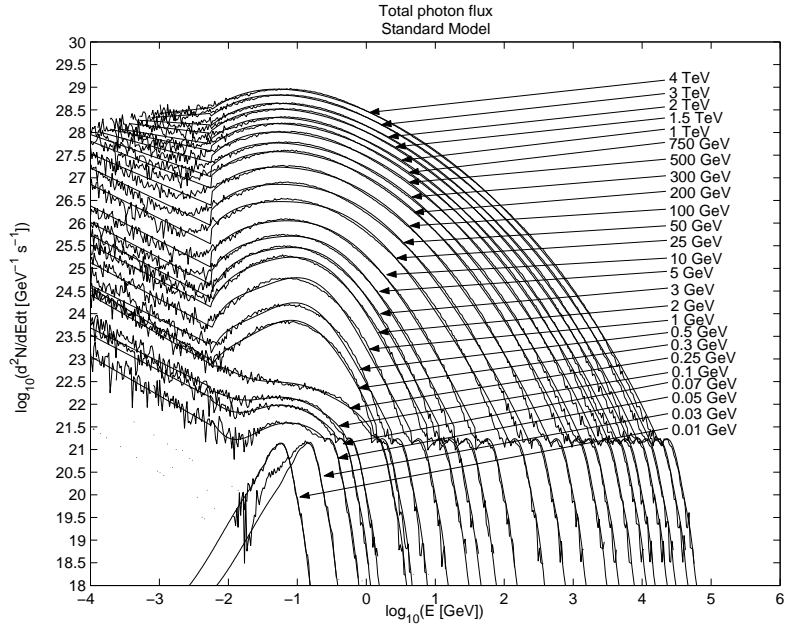}
  \end{center}
  \caption{Total photon flux from black holes in the Standard 
    Model. Jagged curves are simulation data and smooth curves
    the fitting. The table to the right gives the black hole temperature $T$.}
  \label{figure:SMinst}
\end{figure}

\begin{figure}
  \begin{center}
    \includegraphics[width=0.75\textwidth]{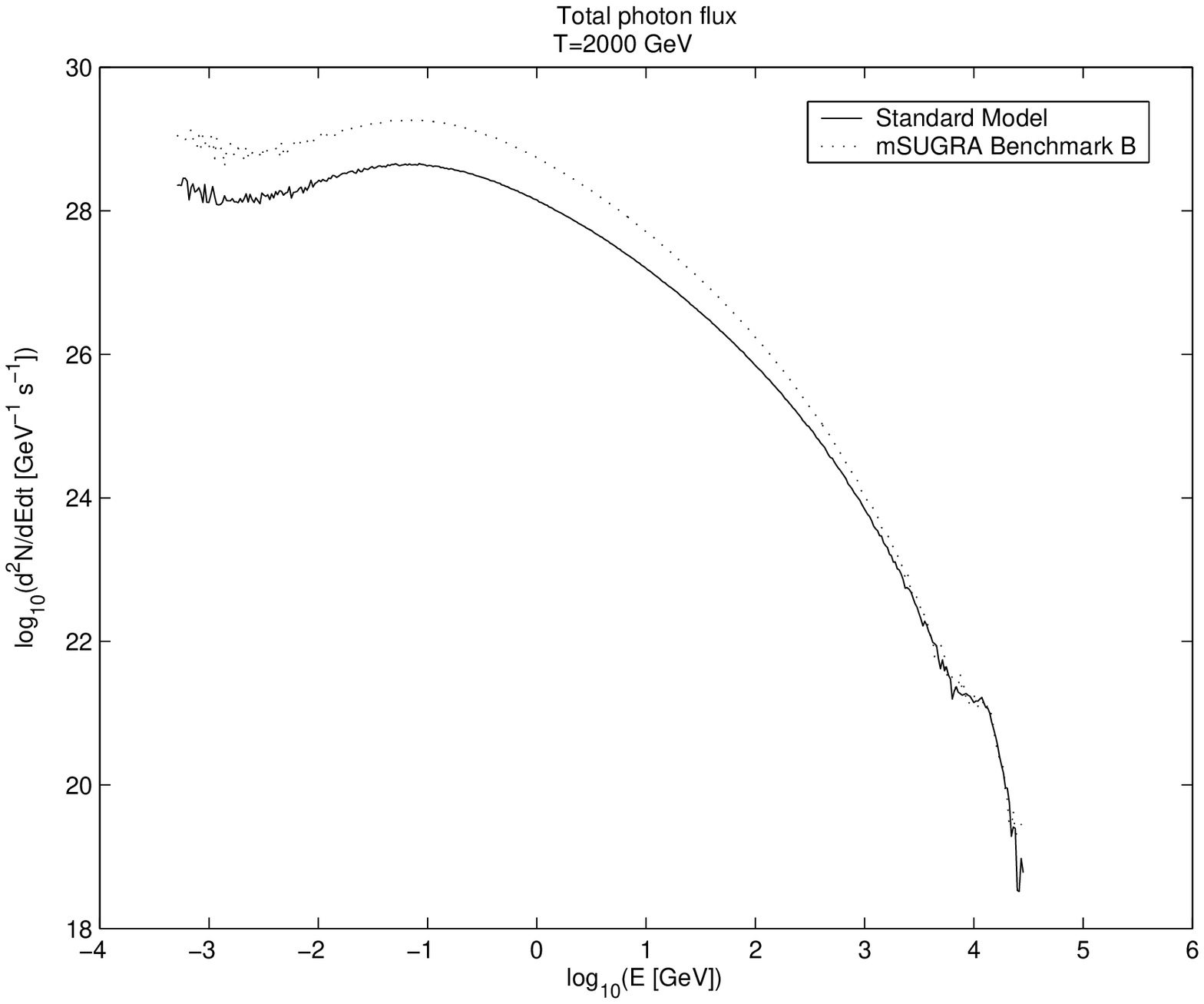}
  \end{center}
  \caption{Total photon flux from a T=2000 GeV black hole in the Standard 
    Model and mSUGRA Benchmark B.}
  \label{figure:SMB2000}
\end{figure}

The instantaneous photon flux was simulated for the Standard Model and mSUGRA Benchmark B (see Appendix \ref{apx:models})
according to the algorithm in Section \ref{sec:sim}. We limit the analysis to this mSUGRA benchmark model for reasons
explained further in Section \ref{sec:resdiff}.
The integrated
fluxes $\dot{N_{j}}$ were found by numerically integrating the direct emission spectra.
Numerical Recipes routines \cite{Press:2001nr} were used for numerical integration, random number
generation and splines fitting. The results from the simulations are shown in Figure \,\ref{figure:SMinst} - \,\ref{figure:SMB2000}.

The dominant peak at $E \approx 100\rm\, MeV$ comes from QCD fragmentation products, significantly pion decay
 ($\pi^0 \rightarrow 2\gamma$). The low-energy, log-linear part comes from electroweak particles, mainly
bremsstrahlung. The high-energy kink is the direct photon emission. 

Comparing our simulation results with those of MacGibbon and Webber \,\cite{MacGibbon:1990zk},
we find that for temperatures in the range $T \sim 10 - 100\,\rm GeV$ our results for the flux
around the peak region is of the order $20\%$ less, the difference decreasing with increasing energy. This
difference seems to come from the difference in QCD masses used (for instance, 
\,\cite{MacGibbon:1990zk} uses
$m_t = 50\,\rm GeV$, $m_g \approx 0.6 \,\rm GeV$ whereas we use $m_t = 175\,\rm GeV$, $m_g = 0\,\rm GeV$).
There seems also to be some difference in the treatment of the QCD threshold and direct pion emission, giving
a difference in flux around $T=0.3\,\rm GeV$ of the order $50\%$. This is within the limits of the 
uncertainty in particle model behaviour around $\Lambda_{QCD}$ \cite{MacGibbon:1990zk}.

\begin{figure}
  \begin{center}
    \includegraphics[width=0.75\textwidth]{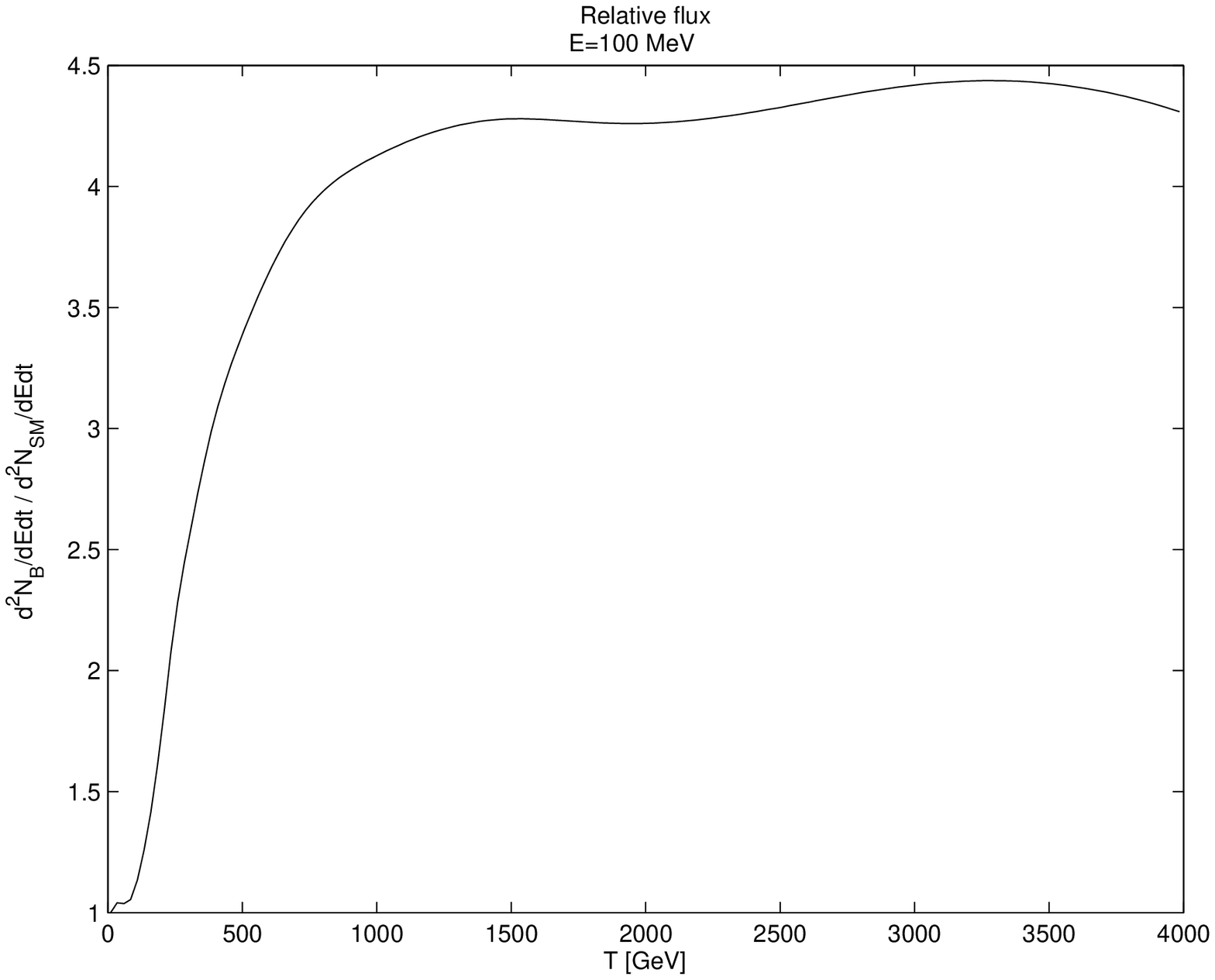}
  \end{center}
  \caption{The relative flux of a PBH in mSUGRA Benchmark B compared to the
    Standard Model. Note that the fluctuations at high temperature lie within
    uncertainties.}
  \label{figure:relflux}
\end{figure}

As can be seen in Figure \,\ref{figure:relflux}, the instantaneous flux from a PBH in mSUGRA Benchmark B is 
roughly a factor 4 higher than in the Standard Model
at peak energies, when all particles are emitted with relativistic energies. Comparing this with the
expected needed increase in flux at high energies to produce a significant effect on the DGB found
in Section \,\ref{sec:speff}, we see that it is too small by several orders of magnitude. We thus expect
that no observable effect should be present in the dominant part of the photon spectrum. 
This is explored further in the calculation of the diffuse flux. 

\begin{figure}
  \begin{center}
    \includegraphics[width=0.75\textwidth]{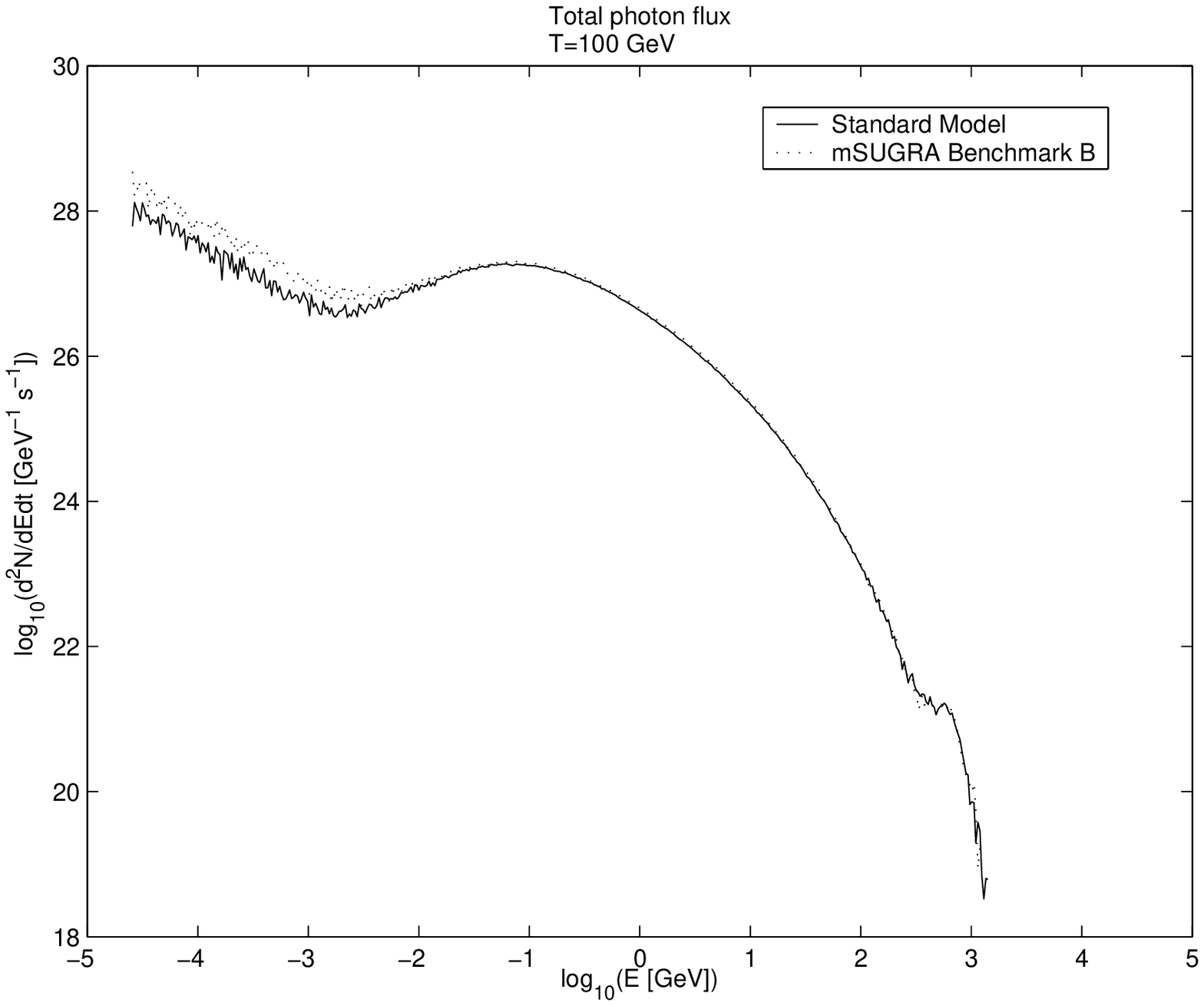}
  \end{center}
  \caption{Total photon flux from a T=100 GeV black hole in the Standard 
    Model and mSUGRA Benchmark B.}
  \label{figure:SMB100}
\end{figure}

\begin{figure}
  \begin{center}
    \includegraphics[width=0.75\textwidth]{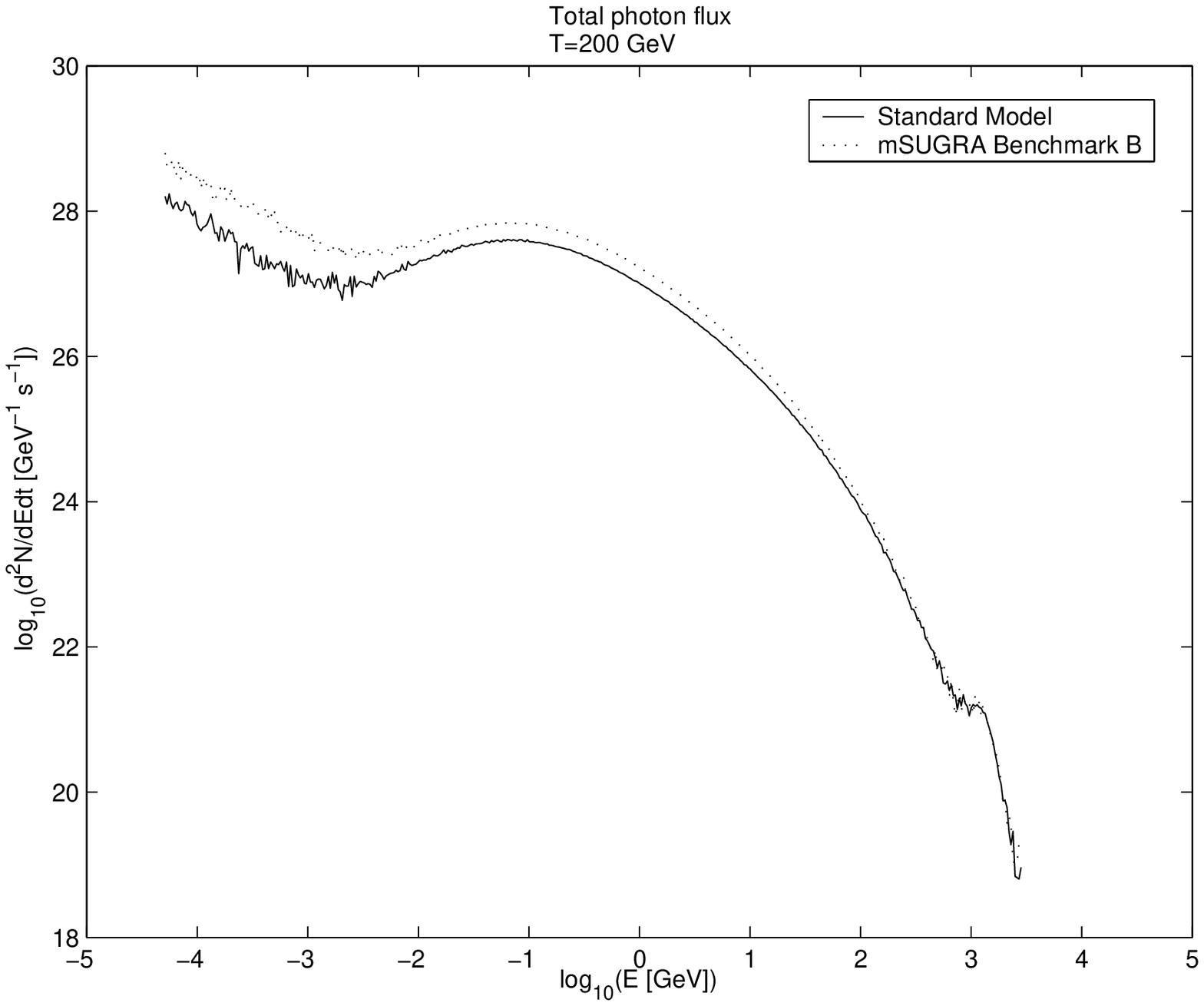}
  \end{center}
  \caption{Total photon flux from a T=200 GeV black hole in the Standard 
    Model and mSUGRA Benchmark B.}
  \label{figure:SMB200}
\end{figure}

\begin{figure}
  \begin{center}
    \includegraphics[width=0.75\textwidth]{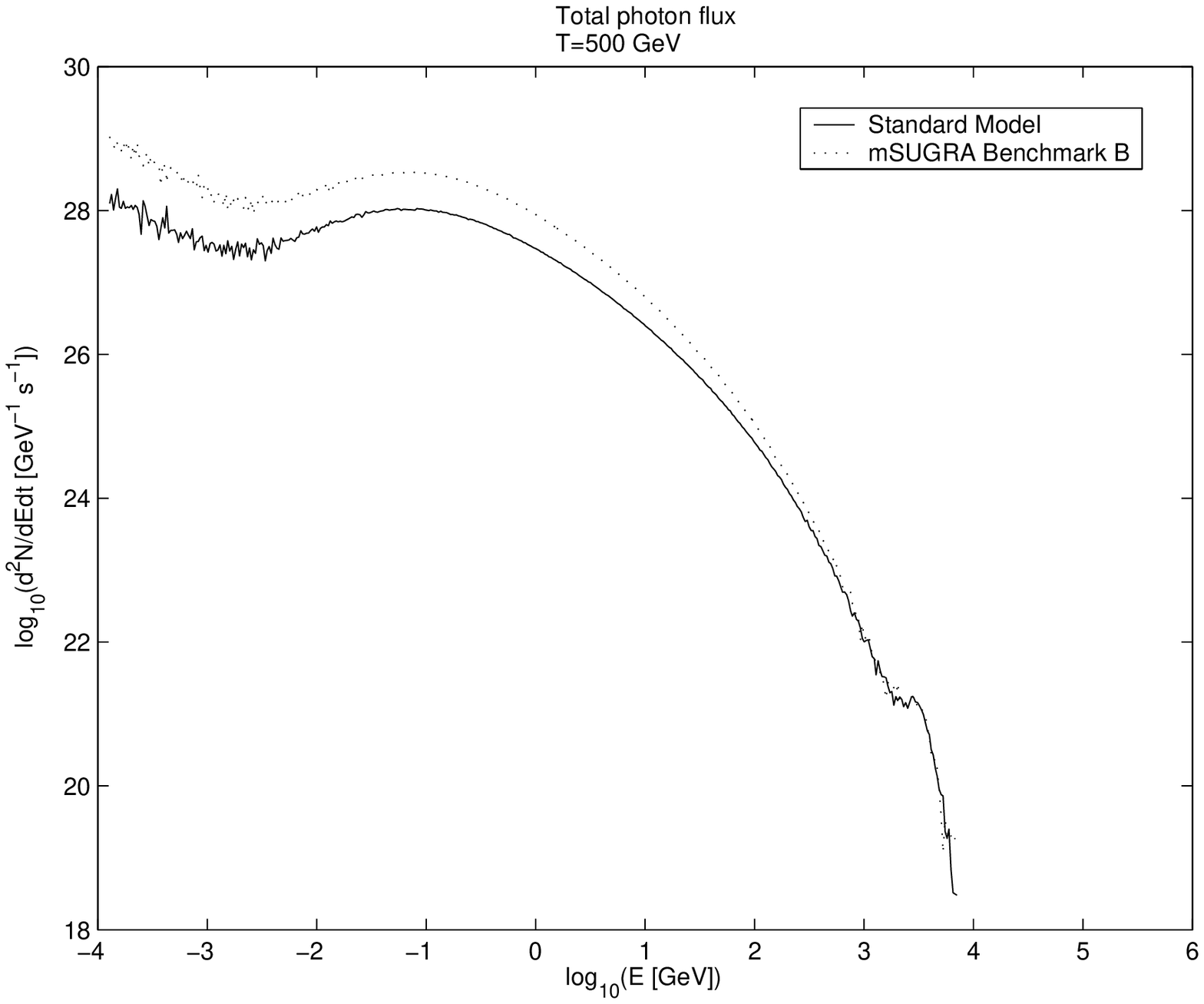}
  \end{center}
  \caption{Total photon flux from a T=500 GeV black hole in the Standard 
    Model and mSUGRA Benchmark B.}
  \label{figure:SMB500}
\end{figure}

\section{Integrated Photon Flux}
Since each Standard Model degree of freedom has a supersymmetric partner degree of freedom,
one might expect that the total flux from a PBH in a supersymmetric scenario would be 
roughly twice that of a Standard Model scenario (at least in the limit were 
all particles are emitted at relativistic energies). 
However, one also has to consider the difference in direct emission flux
for different
particle spins. A rough approximation in the limit that all emitted particles
are relativistic could then be
\begin{equation}
\frac{\dot{N}_{\gamma, SUSY}}{\dot{N}_{\gamma, SM}} \approx 1+
\frac{\sum_{j} \alpha_{j} \frac{\dot{N}_{\tilde{s}_{j}}} {\dot{N}_{s_{j}}}}
{\sum_{j} \alpha_{j}} \approx 3.8 - 3.9 \, ,
\end{equation}
where $\alpha_{j}$ denotes degrees of freedom of particle/sparticle species $j$, 
$\dot{N}_{\tilde{s}_{j}}$ is the total \emph{direct} flux, Eq. (\ref{eq:hawking}), of sparticle species $i$ with 
spin $\tilde{s}_{j}$ (similarly the denominator is the Standard Model particle flux).
We sum
of course only over emitted particles that might eventually produce photons.
Comparing this somewhat naive prediction in the relativistic limit,
with our results for mSUGRA Benchmark B, T=4000 GeV:
\begin{eqnarray}
\nonumber
  \dot{N}_{\gamma, SUSY} & \approx & 1.97\times10^{31} \,\,{\rm s^{-1}} \, ,\\
  \dot{N}_{\gamma, SM} & \approx & 5.65\times10^{30} \,\,{\rm s^{-1}} \, ,
\end{eqnarray}
one finds 
\begin{equation}
  \dot{N}_{\gamma, SUSY} \approx 3.5\dot{N}_{\gamma, SM} \, ,
\end{equation}
which given the roughness of the approximation is in reasonably good agreement.

\section{Photon Spectra from PBHs}

\subsection{Point Source Flux}
\begin{figure}
  \begin{center}
    \includegraphics[width=0.75\textwidth]{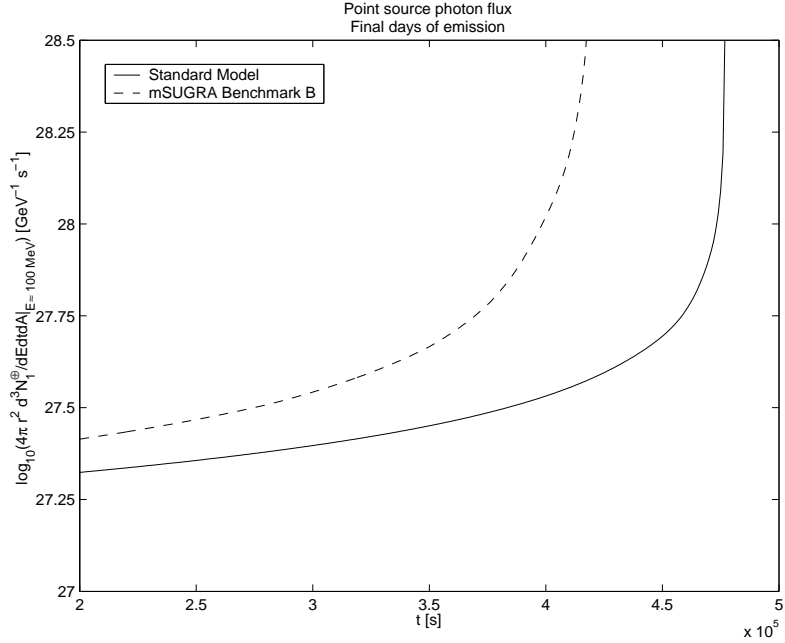}
  \end{center}
  \caption{Point source photon flux during the final days of PBH evaporation for the Standard Model and 
  mSUGRA Benchmark B. Note that the time scale is arbitrary.}
  \label{figure:finaldays}
\end{figure}

The likelihood that one should happen to locate and measure the photon flux from a single expiring PBH seems low.
On the other hand, if we were to observe the final emission stages of a PBH, we might easily impose
limits on allowed particle models since here we have a significant difference in flux for the peak energy, as
can be seen in Figure \ref{figure:finaldays}. Note that the time scale is arbitrary.
There is a clear quantitative and
qualitative difference between the Standard Model and mSUGRA Benchmark B curves.
Measuring this curve would allow conclusions regarding the
mass spectrum of the true particle model.

\subsection{Diffuse Flux}
\label{sec:resdiff}
We limit ourselves in the analysis to the mSUGRA Benchmark Model B. This is 
because Model B has the lowest mass spectrum of the benchmark models considered,
hence potentially the greatest effect on the photon spectra. For details on
the mSUGRA Benchmark Models, see Appendix \ref{apx:models} and Ref. \cite{Ellis:2001hv}.
Since one does not expect the effect of
sparticles to be highly significant, an upper limit such as this should 
suffice for the diffuse gamma-ray flux.

To find the diffuse photon flux, the numerical data for the instantaneous photon flux obtained using \textsc{Pythia}
was fitted and the integral (\ref{eq:djde}) numerically integrated. The initial mass function and cosmological 
parameters used are described in Section \ref{sec:model}.

The fitting of the instantaneous flux was done in a number of steps. The functional form chosen was
\begin{eqnarray}
  \log_{10}\left(\frac{d^2N_{\gamma}(E,T)/dEdt}{1 \,\rm GeV^{-1}\, s^{-1}} \right) & \approx &
  f_{lin}(E,T)\times W_{-\infty}^{-2.25}(E) +
  \nonumber \\
 &&  + f_{poly}(E,T) \times 
  W_{-2.25}^{\log_{10}\left(\frac{E_{max}}{1 \,\rm GeV}\right)}(E) \, ,
\end{eqnarray}
where $f_{lin}$ is a linear part, $f_{poly}$ a higher order polynomial and
\begin{equation}
  W_a^b(E) = 
 \theta\left[\log_{10}\left(\frac{E}{1\,\rm GeV} \right) - a \right]
  - \theta\left[\log_{10}\left(\frac{E}{1\,\rm GeV} \right) - b \right] \, ,
\end{equation}
$\theta$ being the Heaviside step function. The upper limit $\log_{10}(E/1\,\rm GeV) = -2.25$ is the approximate point where the
polynomial part starts to dominate over the linear part. $E_{max}$ is the approximate energy where the flux has dropped significantly.
This will coincide with the point where the direct photon emission dominates.
 The polynomials have the form
\begin{equation}
  f_{lin}(E,T) \approx 
  \sum_{n=0}^1 p_{n}(T)\left[\log_{10}\left(\frac{E}{1 \,\rm GeV}\right)\right]^n \, ,
\end{equation}
\begin{equation}
  f_{poly}(E,T) \approx 
  \sum_{n=0}^6 c_{n}(T)\left[\log_{10}\left(\frac{E}{1 \,\rm GeV}\right)\right]^n \, .
\end{equation}
For $T \apprge 2000$ the coefficients exhibit a limiting behaviour, approximately parameterisable as
\begin{equation}
\label{eq:pn}
p_{n} \approx q_{n1}\left(\frac{T}{1 \,\rm GeV}\right) + q_{n0} \, ,
\end{equation}
\begin{equation}
\label{eq:cn}
c_{n} \approx d_{n1}\ln\left(\frac{T}{1 \,\rm GeV}\right) + d_{n0} \, .
\end{equation}

\begin{table}
  \begin{center}
    \begin{tabular}{|l|ll|}
      \hline
       $n$ & $q^{SM}_{n1}$ & $q^{SM}_{n0}$ \\
       \hline
       0 & 0.00052 & 26.91986 \\
       1 & 0.00015 & -0.38854 \\
       \hline
    \end{tabular}
  \end{center}
  \caption{Coefficients $q^{SM}_{nm}$ for the Standard Model.}
  \label{table:qnSM}
\end{table}

\begin{table}
  \begin{center}
    \begin{tabular}{|l|ll|}
      \hline
       $n$ & $d^{SM}_{n1}$ & $d^{SM}_{n0}$ \\
       \hline
       0 &  0.47408 & 24.54293 \\
       1 &  0.03341 & -0.94904 \\
       2 &  0.02473 & -0.40625 \\
       3 &  0.01237 & -0.09398 \\
       4 & -0.00034 & -0.00562 \\
       5 & -0.00127 &  0.01257 \\
       6 &  0.00026 & -0.00241 \\
      \hline
    \end{tabular}
  \end{center}
  \caption{Coefficients $d^{SM}_{nm}$ for the Standard Model.}
  \label{table:dnSM}
\end{table}

\begin{table}
  \begin{center}
    \begin{tabular}{|l|ll|}
      \hline
       $n$ & $q^B_{n1}$ & $q^B_{n0}$ \\
       \hline
       0 &  0.00047 & 27.35710 \\
       1 &  0.00012 & -0.46435 \\
       \hline
    \end{tabular}
  \end{center}
  \caption{Coefficients $q^B_{nm}$ for mSUGRA Benchmark B.}
  \label{table:qnB}
\end{table}

\begin{table}
  \begin{center}
    \begin{tabular}{|l|ll|}
      \hline
       $n$ & $d^B_{n1}$ & $d^B_{n0}$ \\
       \hline
       0 & 0.49768 &  24.95461 \\
       1 & 0.05629 &  - 1.23295 \\
       2 & 0.01511 &  - 0.36052 \\
       3 & 0.00644 &  - 0.00381 \\
       4 & 0.00409 &  - 0.04291 \\
       5 & -0.00040 & - 0.00150 \\
       6 & -0.00016 & - 0.00225 \\
       \hline
    \end{tabular}
  \end{center}
  \caption{Coefficients $d^B_{nm}$ for mSUGRA Benchmark B.}
  \label{table:dnB}
\end{table}

When numerically integrating the diffuse photon flux, the coefficients $d_n$, $q_n$ were determined
using cubic splines on data from least squares fits on the numerical data for the instantaneous flux.
For temperatures $T>4000\,\rm GeV$ for which no numerical data existed, the Eqs. (\ref{eq:pn})-(\ref{eq:cn})
were used. This approach proved reliable up to $T \approx 5000 \,\rm GeV$. For higher temperatures,
the temperature was approximated to $T = 5000 \,\rm GeV$. This produces a negligible error in the
diffuse flux.

The value of $E_{max}$, the highest energy that gives a significant flux, was determined using cubic splines
on simulation data. Above $E_{max}$, the direct photon flux was still included.

Fitting of the instantaneous flux using Gauss-Seidel to various non-linear functions with a Boltzmann factor, 
a functional form one might expect would be successful, were also attempted but did not yield good accuracy.

\begin{figure}
  \begin{center}
    \includegraphics[width=0.75\textwidth]{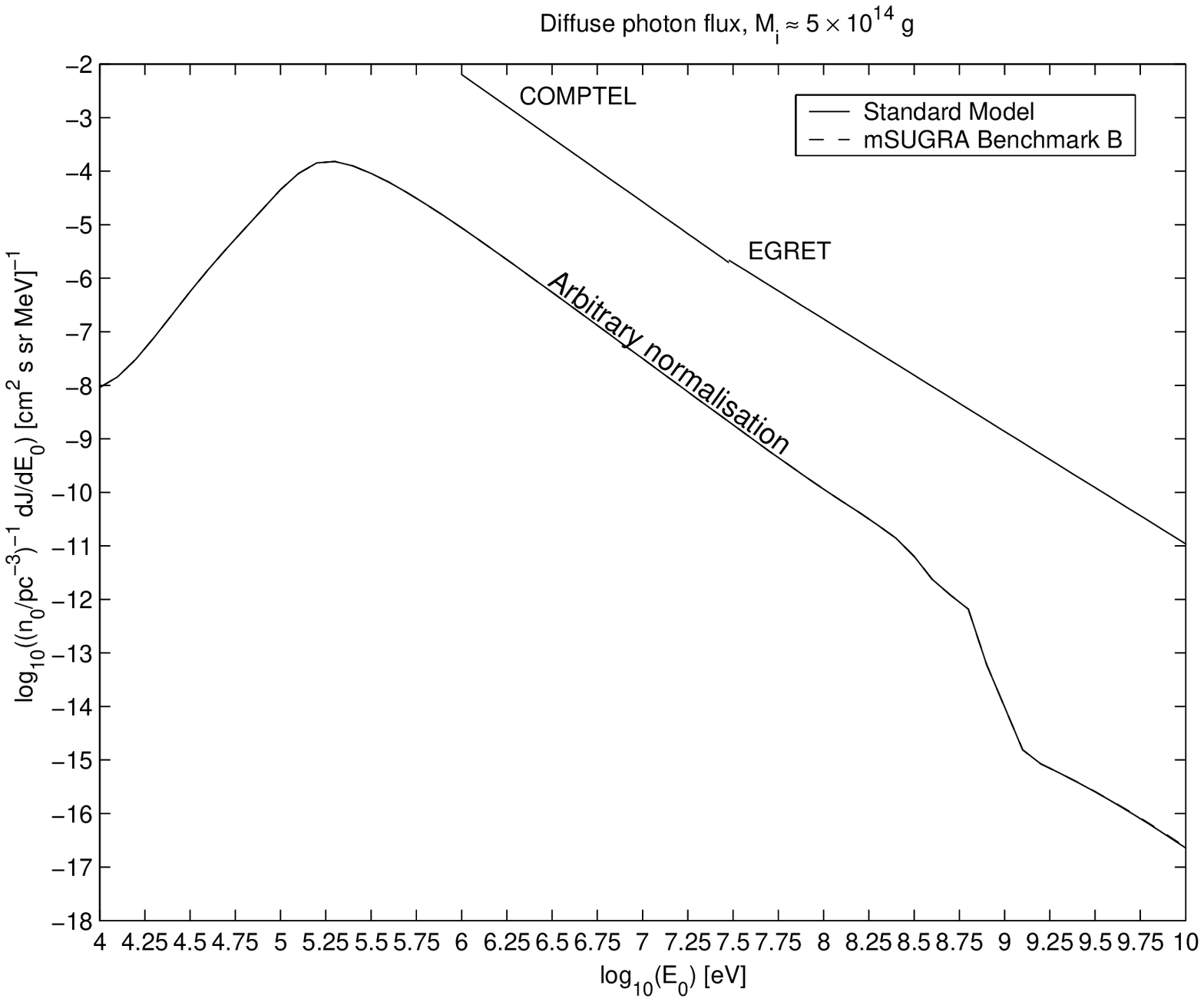}
  \end{center}
  \caption{Diffuse photon flux from a delta-function PBH mass spectrum (see Section \ref{sec:model}), 
    with $M_i = 5.125 \times 10^{14}\,\rm g$.
    This corresponds to $T_{RH} \approx 4 \times 10^8\,\rm GeV$, $z_i^{SM} \approx 5 \times 10^{27}$
    and $z_i^{B} \approx 6 \times 10^{27}$ (see Eq. (\ref{eq:zi})-(\ref{eq:mrh})).
    The normalisation is arbitrary, i.e. the flux is normalised to the present-day PBH number density $n_0$ as defined in
    Section \ref{sec:model}.
    The COMPTEL \,\cite{Kribs:1997ac} and EGRET \,\cite{Sreekumar:1998un} values are of course absolute flux values, not
    normalised to PBH density.
    The apparent discontinuities before and close to the drop at $E_0 \approx 1 \,\rm GeV$ are artefacts
    from the numerical treatment and should not be considered physical.}
  \label{figure:obs5e14}
\end{figure}

\begin{figure}
  \begin{center}
    \includegraphics[width=0.75\textwidth]{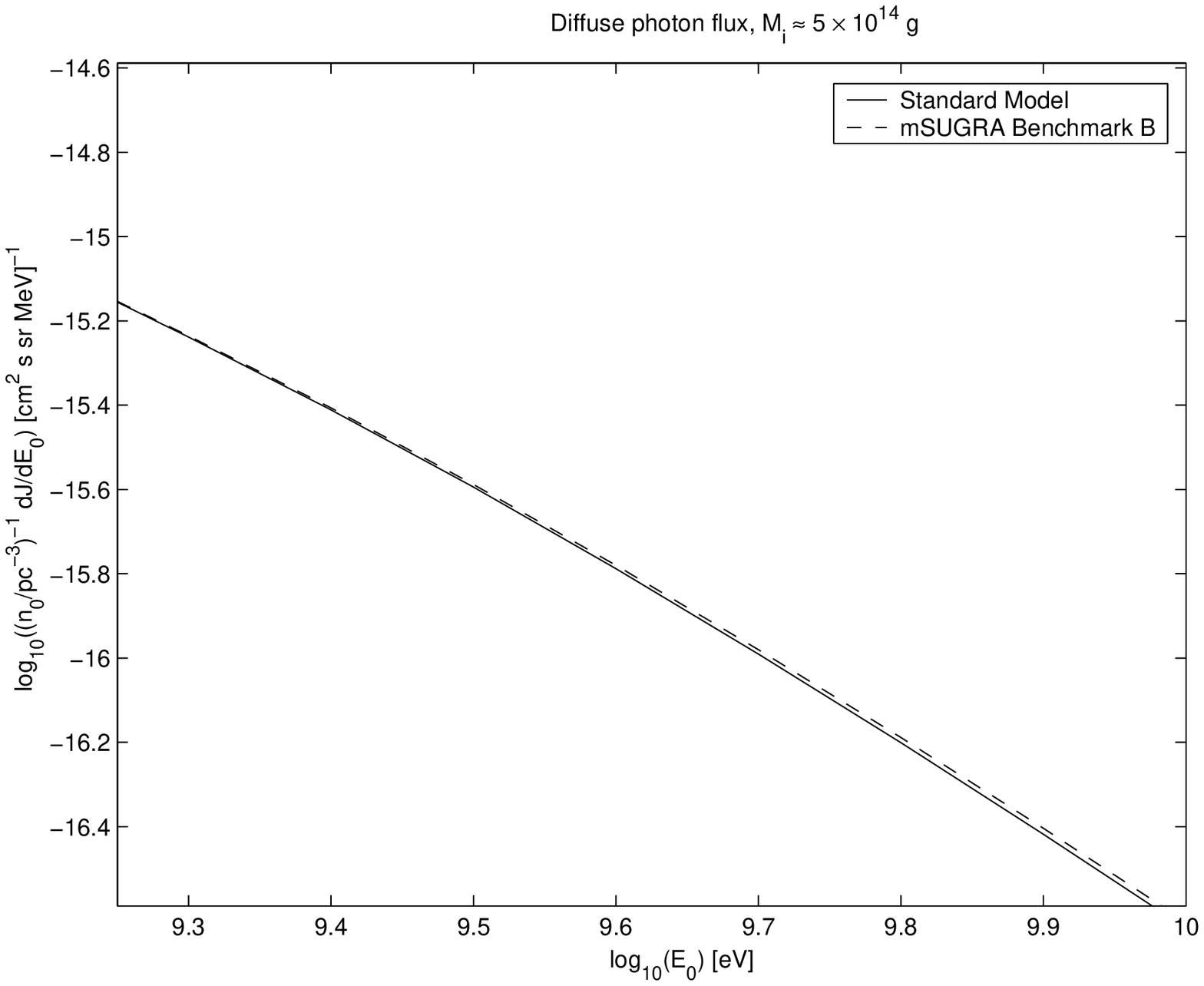}
  \end{center}
  \caption{Diffuse photon flux from a delta-function PBH mass spectrum (see Section \ref{sec:model}),
    with $M_i = 5.125 \times 10^{14}\,\rm g$.
    The normalisation is arbitrary, i.e. the flux is normalised to the present-day PBH number density $n_0$ as defined in
    Section \ref{sec:model}.
    Zoom of high-energy part of spectrum. The mSUGRA Benchmark B flux is of the
    order $1-5\%$ higher than that for the Standard Model.}
  \label{figure:obs5e14zm}
\end{figure}

\begin{figure}
  \begin{center}
    \includegraphics[width=0.75\textwidth]{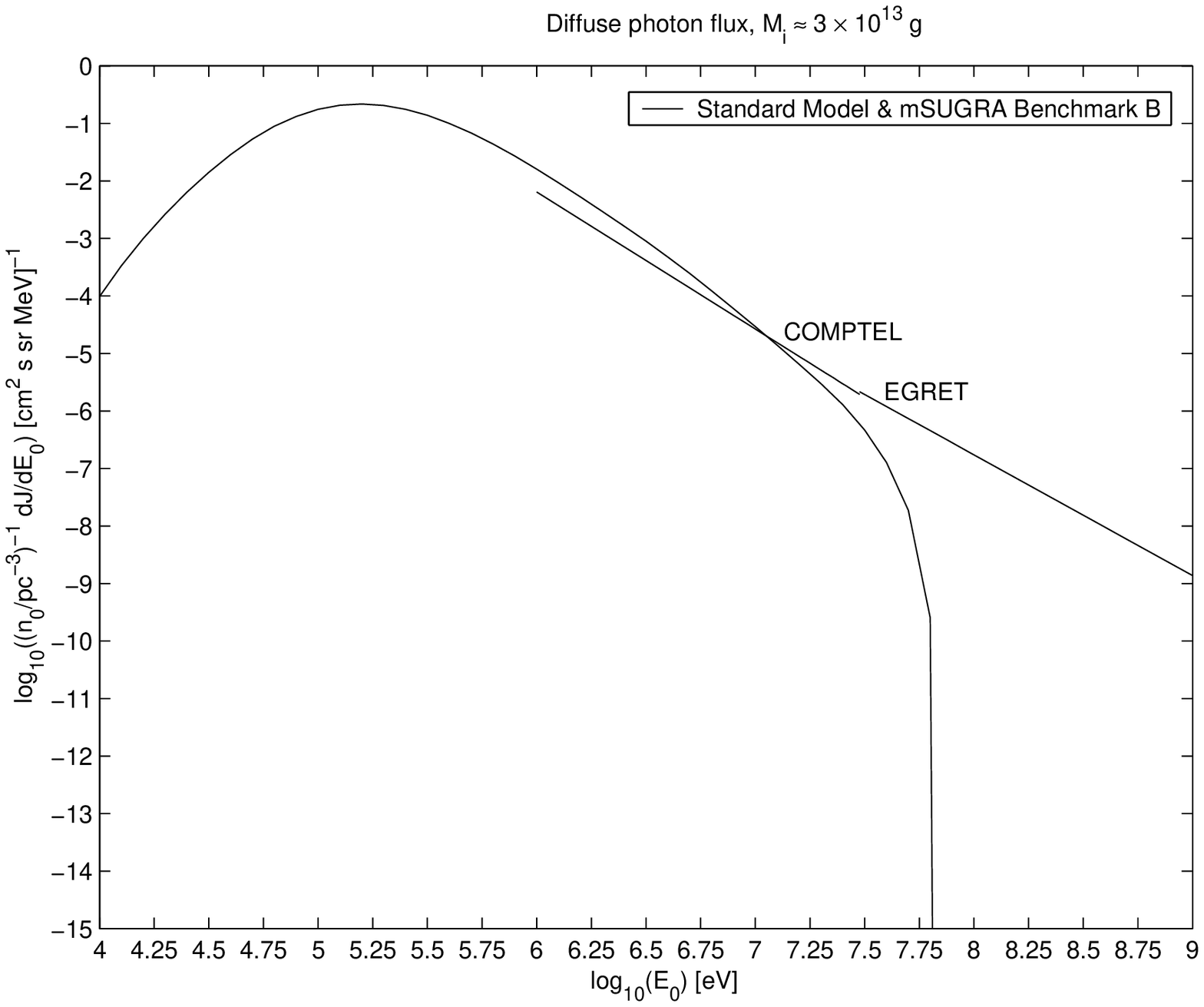}
  \end{center}
  \caption{Diffuse photon flux from a delta-function PBH mass spectrum (see Section \ref{sec:model}),
    with $M_i \approx 3  \times 10^{13}\,\rm g$.
    This corresponds to $T_{RH} \approx 2 \times 10^9\,\rm GeV$, $z_i^{SM} \approx 3 \times 10^{28}$
    and $z_i^{B} \approx 4 \times 10^{28}$ (see Eq. (\ref{eq:zi})-(\ref{eq:mrh})).
    The normalisation is arbitrary, i.e. the flux is normalised to the present-day PBH number density $n_0$ as defined in
    Section \ref{sec:model}.
    The COMPTEL \,\cite{Kribs:1997ac} and EGRET \,\cite{Sreekumar:1998un} values are of course absolute flux values, not
    normalised to PBH density.}
  \label{figure:obs3e13SMB}
\end{figure}

\begin{figure}
  \begin{center}
    \includegraphics[width=0.75\textwidth]{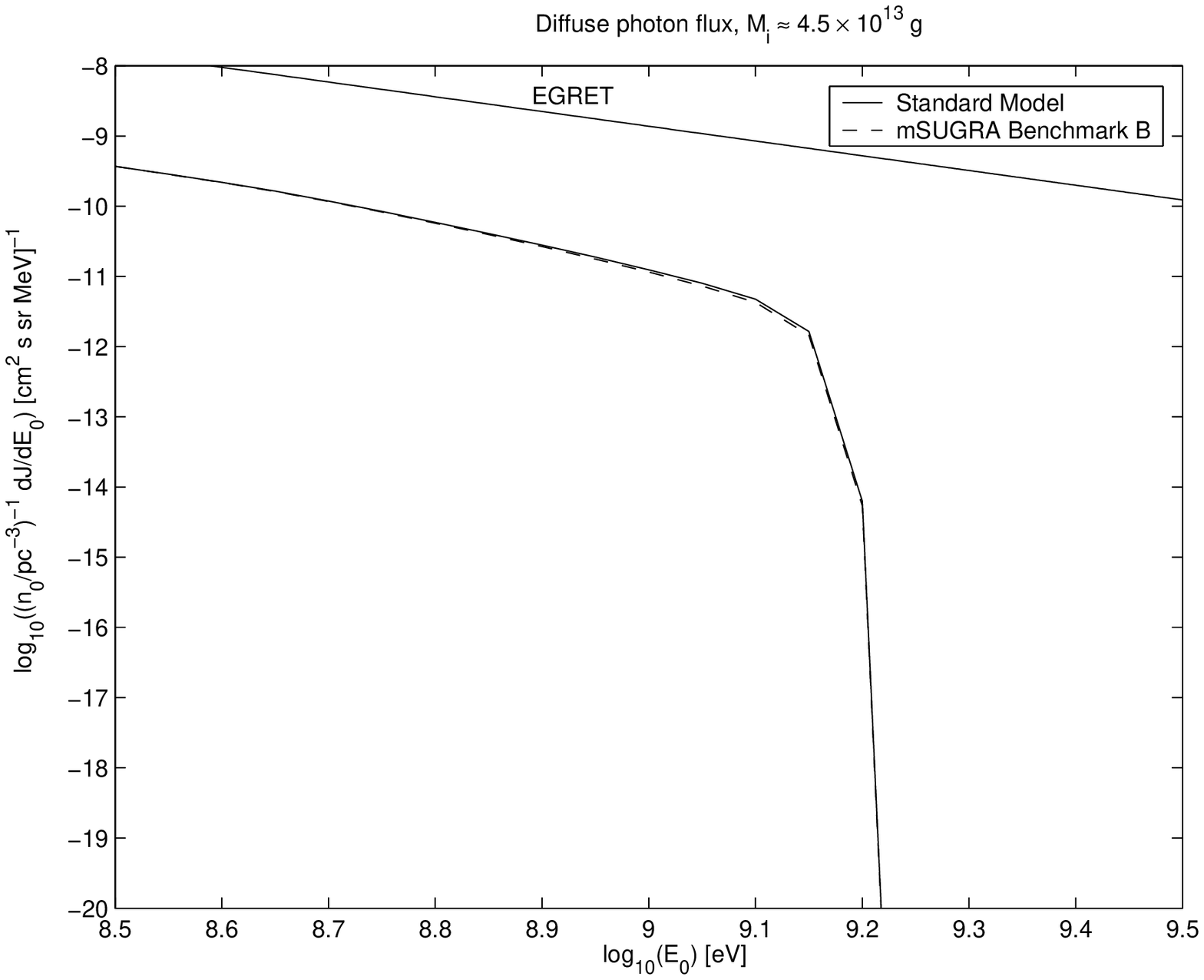}
  \end{center}
  \caption{Diffuse photon flux from a delta-function PBH mass spectrum (see Section \ref{sec:model}),
    with $M_i \approx 4.5  \times 10^{13}\,\rm g$.
    This corresponds to $T_{RH} \approx 1 \times 10^9\,\rm GeV$, $z_i^{SM} \approx 2 \times 10^{28}$
    and $z_i^{B} \approx 3 \times 10^{28}$ (see Eq. (\ref{eq:zi})-(\ref{eq:mrh})).
    The normalisation is arbitrary, i.e. the flux is normalised to the present-day PBH number density $n_0$ as defined in
    Section \ref{sec:model}.
    The EGRET \,\cite{Sreekumar:1998un} values are of course absolute flux values, not
    normalised to PBH density.}
  \label{figure:obs45e13SMB}
\end{figure}

\begin{figure}
  \begin{center}
    \includegraphics[width=0.75\textwidth]{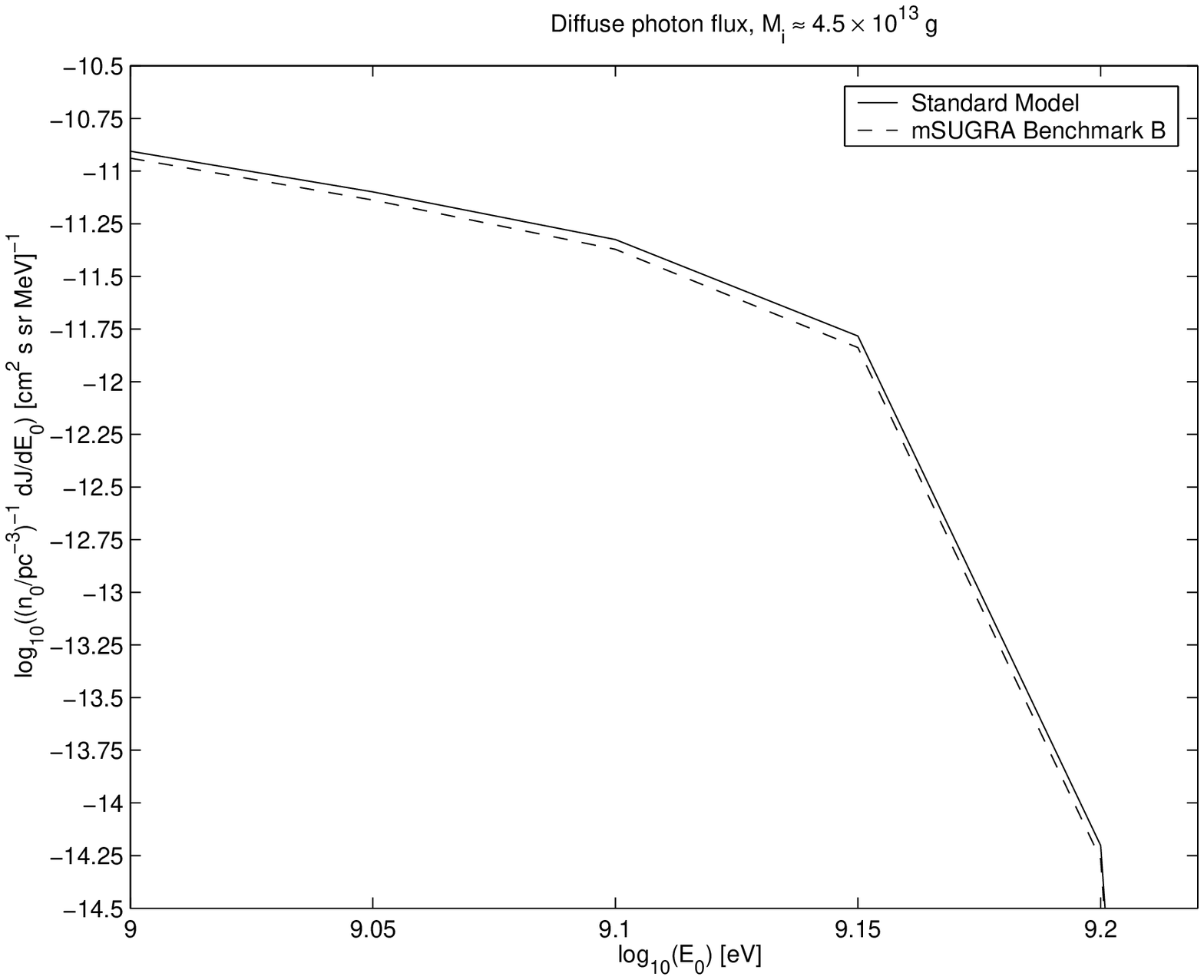}
  \end{center}
  \caption{Diffuse photon flux from a delta-function PBH mass spectrum (see Section \ref{sec:model}),
    with $M_i \approx 4.5  \times 10^{13}\,\rm g$.
    The normalisation is arbitrary, i.e. the flux is normalised to the present-day PBH number density $n_0$ as defined in
    Section \ref{sec:model}.
    Zoom of high-energy part of spectrum. The Standard Model flux is of the
    order $10\%$ higher than that for mSUGRA Benchmark B.}
  \label{figure:obs45e13SMBzm}
\end{figure}

\begin{figure}
  \begin{center}
    \includegraphics[width=0.75\textwidth]{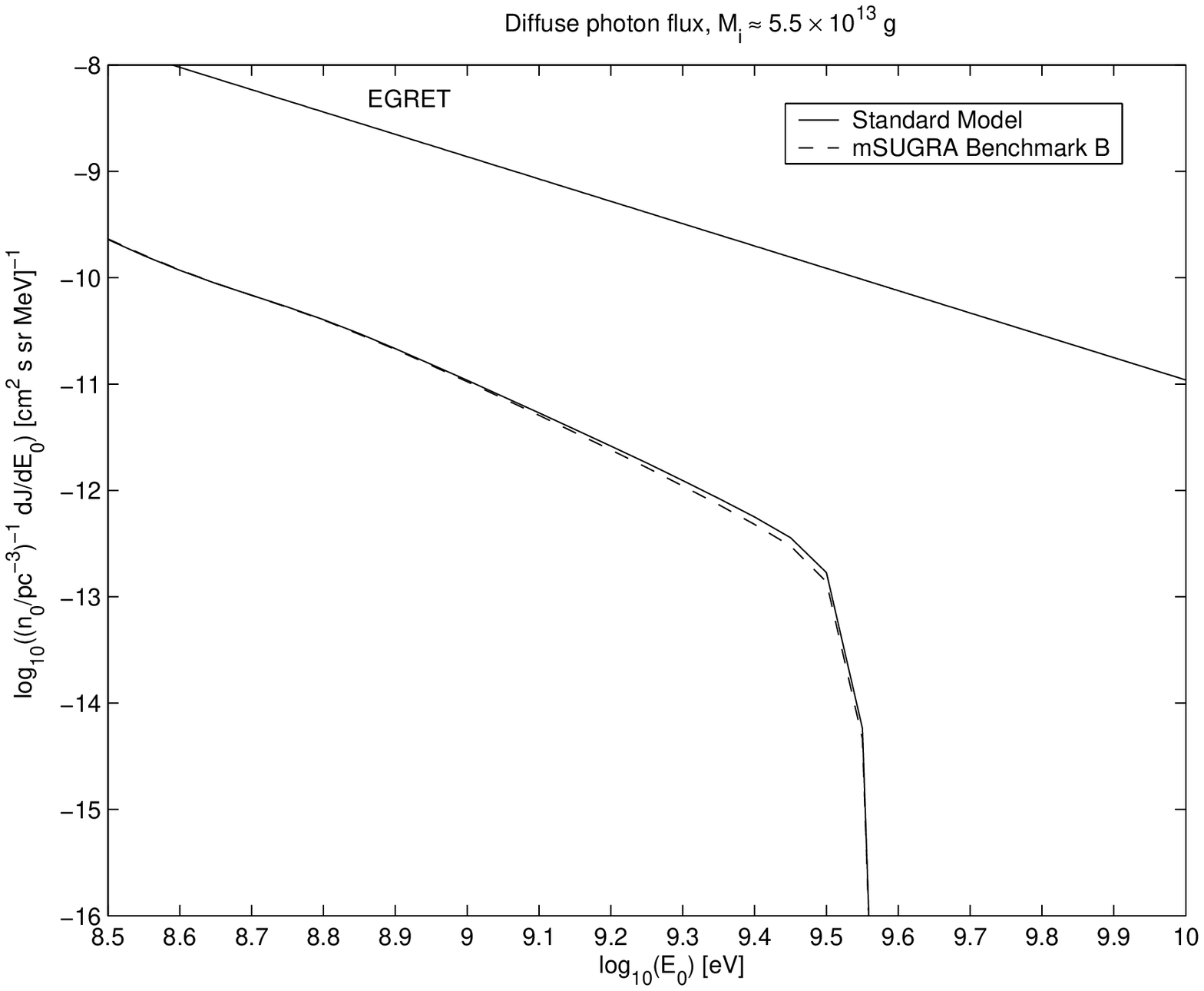}
  \end{center}
  \caption{Diffuse photon flux from a delta-function PBH mass spectrum (see Section \ref{sec:model}),
    with $M_i \approx 5.5  \times 10^{13}\,\rm g$.
    This corresponds to $T_{RH} \approx 1 \times 10^9\,\rm GeV$, $z_i \approx 2 \times 10^{28}$ (see Eq. (\ref{eq:zi})-(\ref{eq:mrh})).
    The normalisation is arbitrary, i.e. the flux is normalised to the present-day PBH number density $n_0$ as defined in
    Section \ref{sec:model}.
    The EGRET \,\cite{Sreekumar:1998un} values are of course absolute flux values, not
    normalised to PBH density.}
  \label{figure:obs55e13SMB}
\end{figure}

\begin{figure}
  \begin{center}
    \includegraphics[width=0.75\textwidth]{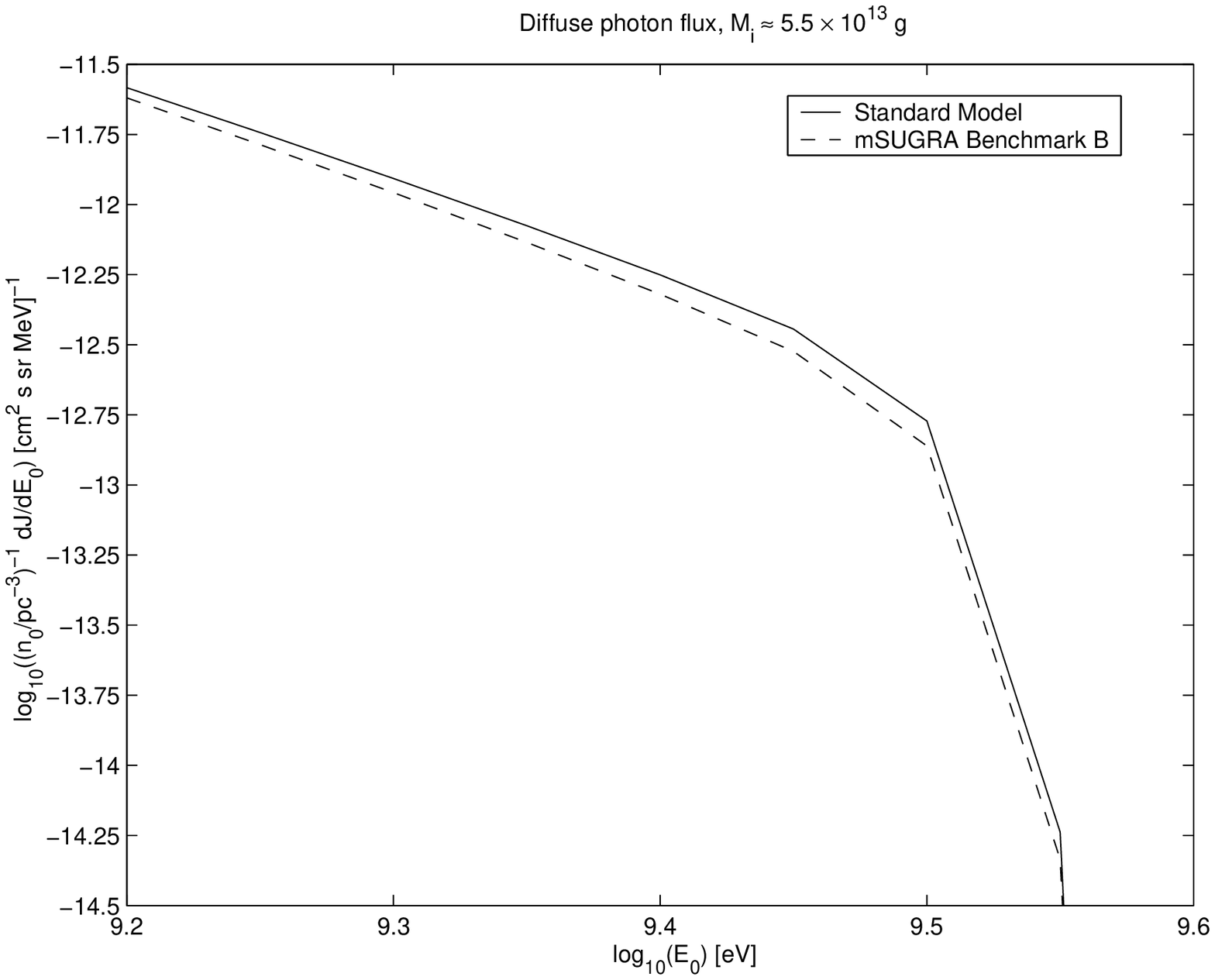}
  \end{center}
  \caption{Diffuse photon flux from a delta-function PBH mass spectrum (see Section \ref{sec:model}),
    with $M_i \approx 5.5  \times 10^{13}\,\rm g$.
    The normalisation is arbitrary, i.e. the flux is normalised to the present-day PBH number density $n_0$ as defined in
    Section \ref{sec:model}.
    Zoom of high-energy part of spectrum. The Standard Model flux is of the
    order $20\%$ higher than that for mSUGRA Benchmark B.}
  \label{figure:obs55e13SMBzm}
\end{figure}

\begin{figure}
  \begin{center}
    \includegraphics[width=0.75\textwidth]{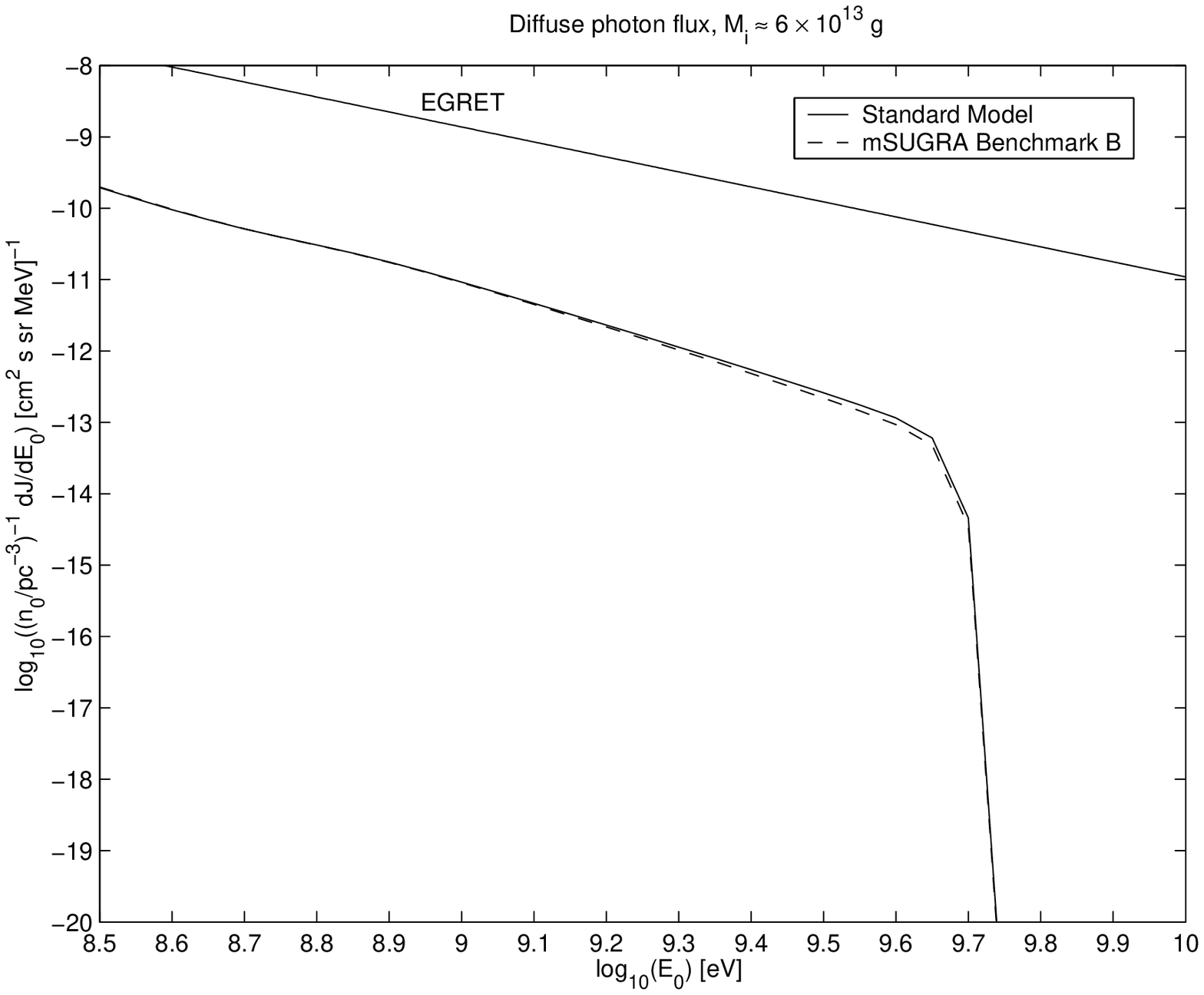}
  \end{center}
  \caption{Diffuse photon flux from a delta-function PBH mass spectrum (see Section \ref{sec:model}),
    with $M_i \approx 6  \times 10^{13}\,\rm g$.
    This corresponds to $T_{RH} \approx 1 \times 10^9\,\rm GeV$ and $z_i \approx 2 \times 10^{28}$ 
    (see Eq. (\ref{eq:zi})-(\ref{eq:mrh})).
    The normalisation is arbitrary, i.e. the flux is normalised to the present-day PBH number density $n_0$ as defined in
    Section \ref{sec:model}.
    The EGRET \,\cite{Sreekumar:1998un} values are of course absolute flux values, not
    normalised to PBH density.}
  \label{figure:obs6e13SMB}
\end{figure}

\begin{figure}
  \begin{center}
    \includegraphics[width=0.75\textwidth]{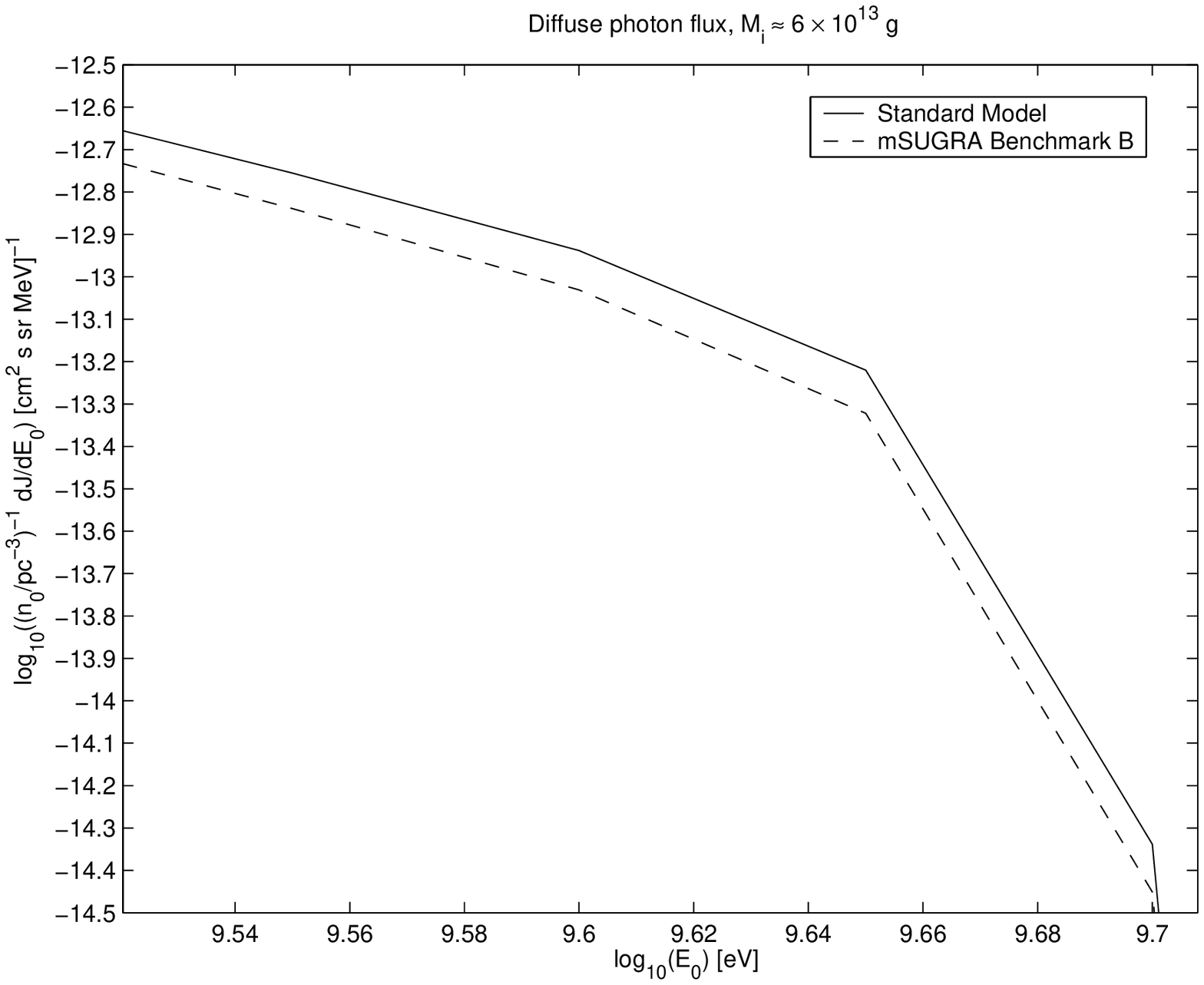}
  \end{center}
  \caption{Diffuse photon flux from a delta-function PBH mass spectrum (see Section \ref{sec:model}),
    with $M_i \approx 6  \times 10^{13}\,\rm g$.
    The normalisation is arbitrary, i.e. the flux is normalised to the present-day PBH number density $n_0$ as defined in
    Section \ref{sec:model}. Zoom of high-energy part of spectrum. The Standard Model flux is of the
    order $25-30\%$ higher than that for mSUGRA Benchmark B.}
  \label{figure:obs6e13SMBzm}
\end{figure}

\begin{figure}
  \begin{center}
    \includegraphics[width=0.75\textwidth]{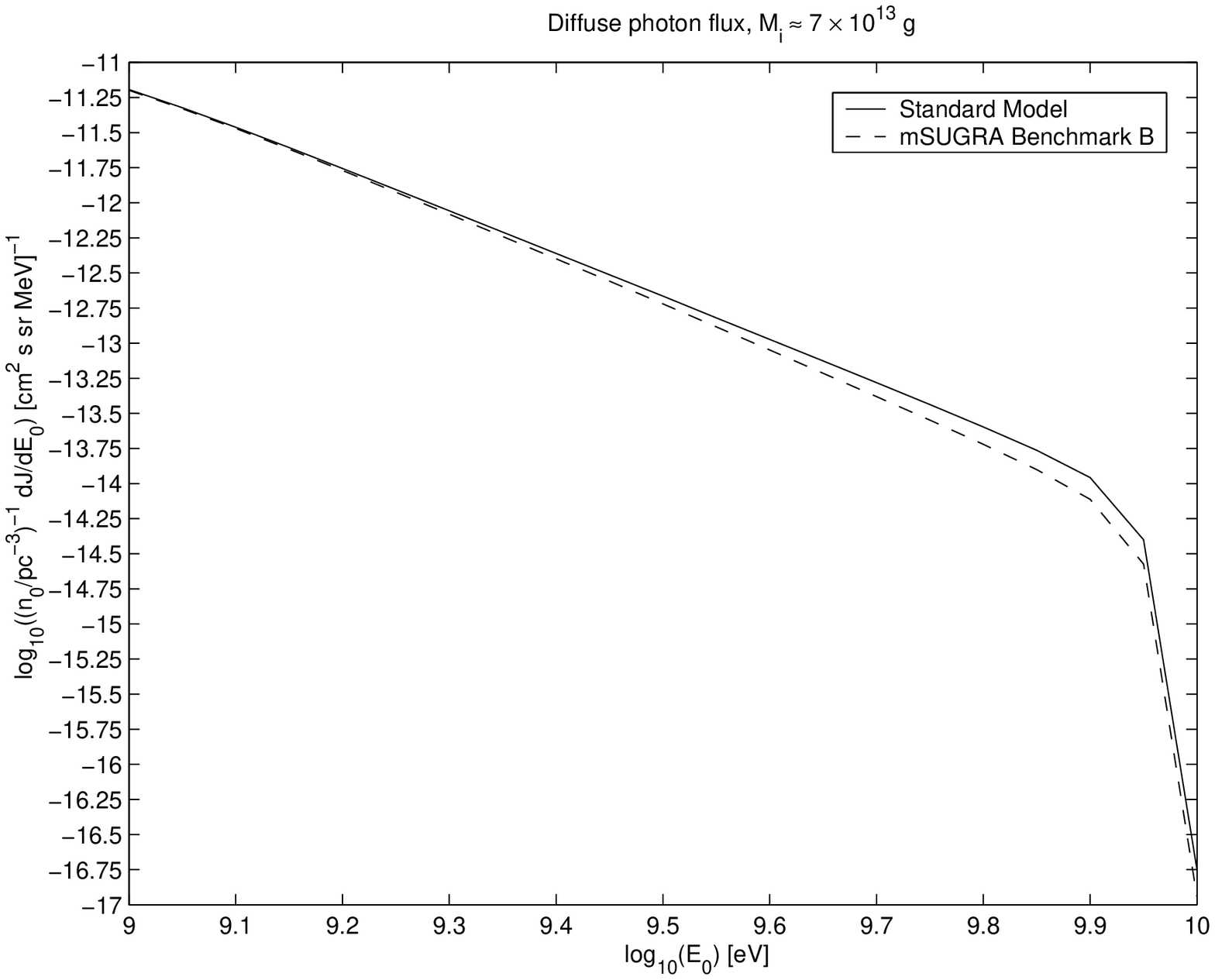}
  \end{center}
  \caption{Diffuse photon flux from a delta-function PBH mass spectrum (see Section \ref{sec:model}),
    with $M_i \approx 7 \times 10^{13}\,\rm g$.
    This corresponds to $T_{RH} \approx 1 \times 10^9\,\rm GeV$ and $z_i \approx 2 \times 10^{28}$ (see Eq. (\ref{eq:zi})-(\ref{eq:mrh})).
    The normalisation is arbitrary, i.e. the flux is normalised to the present-day PBH number density $n_0$ as defined in
    Section \ref{sec:model}.
    The Standard Model flux is of the order $40-50\%$ higher than that for mSUGRA Benchmark B.}
  \label{figure:obs7e13SMB}
\end{figure}

\begin{figure}
  \begin{center}
    \includegraphics[width=0.75\textwidth]{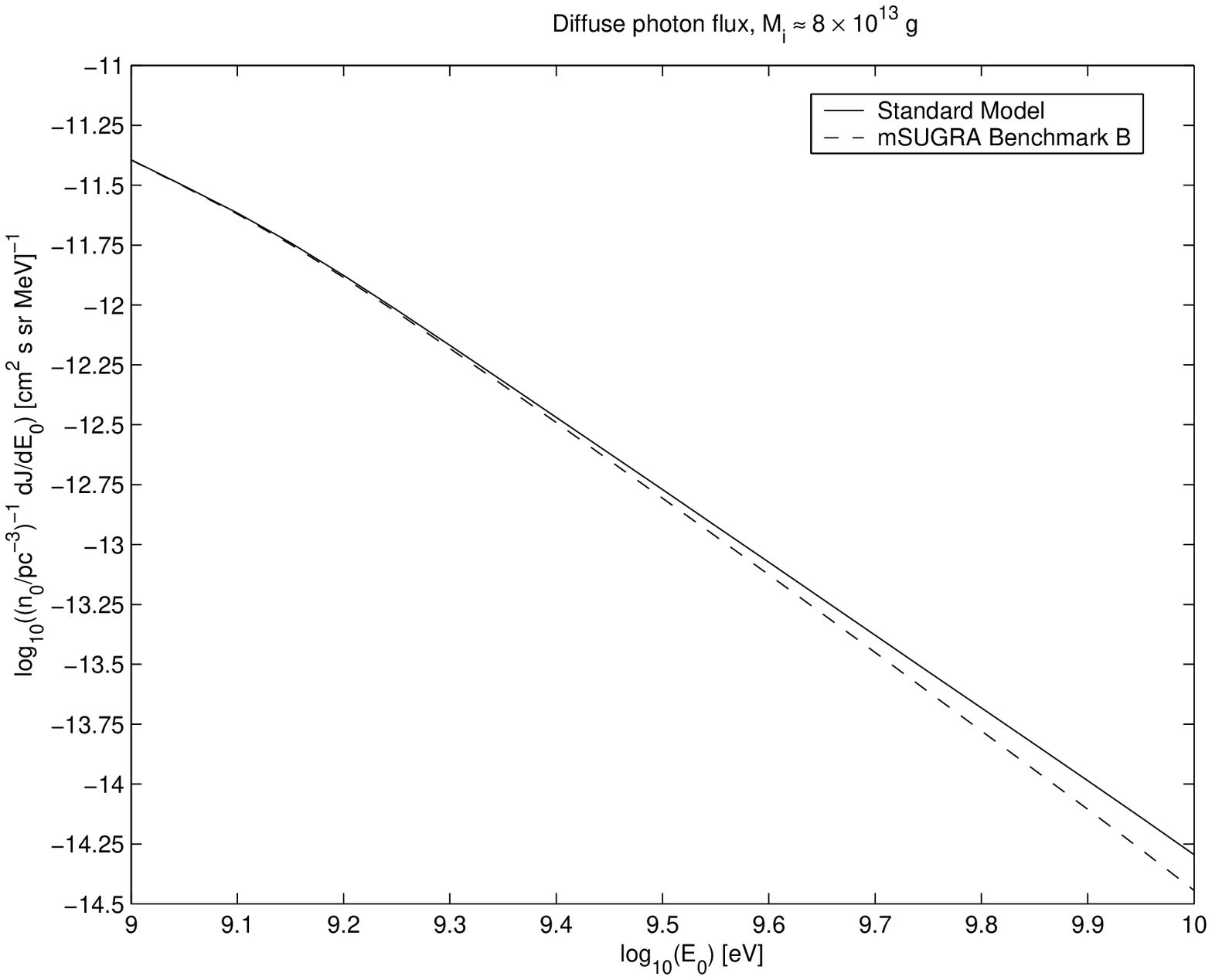}
  \end{center}
  \caption{Diffuse photon flux from a delta-function PBH mass spectrum (see Section \ref{sec:model}),
    with $M_i \approx 8 \times 10^{13}\,\rm g$.
    This corresponds to $T_{RH} \approx 1 \times 10^9\,\rm GeV$ and $z_i \approx 2 \times 10^{28}$ (see Eq. (\ref{eq:zi})-(\ref{eq:mrh})).
    The normalisation is arbitrary, i.e. the flux is normalised to the present-day PBH number density $n_0$ as defined in
    Section \ref{sec:model}.
    The Standard Model flux is of the
    order $25-30\%$ higher than that for mSUGRA Benchmark B.}
  \label{figure:obs8e13SMB}
\end{figure}

The results from the numerical integration is shown in Figs. \,\ref{figure:obs5e14} - \,\ref{figure:obs8e13SMB}.
Note that the Figures have an arbitrary normalisation, given by the present-day PBH number density $n_0$ as 
defined in Section \ref{sec:model}.
The peak in the diffuse spectrum is located at roughly $E_{0,peak} \approx 500\,\rm keV$. 
This peak comes from the dominant photon production due to QCD fragmentation (primarily pion decay), with
energy at emission $E_{peak} \approx 100 \,{\rm MeV}$.
The high-energy part comes mainly from direct photon emission.

As can be seen in Fig. \ref{figure:obs5e14},
the effect from sparticles in the dominant part of the spectrum is completely negligible. 
For $M_i \approx 5 \times 10^{14} \,\rm g$, there is a $\sim 1-5\%$ higher flux for
mSUGRA Benchmark B for high 
energy photons (Figure \,\ref{figure:obs5e14zm}), completely within numerical uncertainties. 
Comparing the diffuse flux curve to that of MacGibbon and Carr \cite{MacGibbon:1991ca}, one finds that
we do not have a bump at $E_0 \approx 100\,\rm MeV$, but rather a drop at $E_0 \approx 500-1000 \,\rm MeV$.
This difference seems to depend on the difference in initial mass function (they assume a Carr
initial mass function) and cosmological parameters. The bump, or drop, comes from the
optical depth. The flux at $E_0 \apprge 1 \,\rm GeV$ comes from the high-temperature
stage of emission with $z \approx 0$. Note that the apparent discontinuities close to the
drop are artefacts from the numerical treatment, and should not be taken as physical.

Figure \ref{figure:obs3e13SMB} shows that also for $M_i \approx 3\times 10^{13} \,\rm g$
the supersymmetric contributions are negligible. Furthermore, the optical depth effectively cuts of all photons
with $E_0 \apprge 60\,\rm MeV$. 

Turning to other initial masses (Figs. \ref{figure:obs45e13SMB}-\ref{figure:obs8e13SMB}),
one sees that in the high-energy part of the spectrum,
a bit counter-intuitive perhaps, the Standard
Model flux is greater. In this part of the spectrum, it turns out that for a range of initial masses
($4 \times 10^{13} \,{\rm g} \apprle M_i \apprle 1 \times 10^{14} \,{\rm g}$)
the faster increase in black hole
temperature associated with a SUSY particle spectrum is the dominating effect (over difference in
instantaneous flux). This is due to the fact that the SUSY black hole spends less time
``at each temperature''. The lower the sparticle mass spectrum is, the broader is the photon energy range in which
this effect will be present. For a higher mass spectrum than in mSUGRA Benchmark B, one would then expect
a similar effect, but ``beginning'' at higher photon energy.
The magnitude of the difference will be dominated by the optical depth, since
the effect comes from the late stages of PBH emission (previous emission needs to be attenuated not to
dominate the late stage emission).
Because of this, PBHs that reach the final stages of
emission close to when the universe becomes thin to the photon energy range in question will show the greatest
difference, since previous emission is highly attenuated. The results show that the effect occurs for 
$E_0 \apprge 1 \,\rm GeV$, and the greatest difference is for $M_i \approx 7 \times 10^{13} \,\rm g$ where
the Standard Model flux is $\sim 40-50\%$ greater than that for mSUGRA Benchmark B. For smaller initial masses, more
of the interesting photon energy range is cut off by the optical depth. For larger initial masses the effect
is still present, but quickly becomes dominated by previous emission, turning over to 
a Standard Model flux $\sim 1-5\%$ greater than the mSUGRA Benchmark B flux for $M_i \approx
5 \times 10^{14} \,\rm g$. The total effect on the observed spectrum will of course be dependent on the shape
of the PBH initial mass function, but should in general be much less than $50\%$. It should be pointed out that
the treatment of the function $\Phi$ in the mass loss rate was not very exact close to mass thresholds (step function),
which might affect the results slightly, although not significantly. 

Additionally, normalising the spectrum in Figure \ref{figure:obs5e14} to that measured
by COMPTEL \,\cite{Kribs:1997ac} and EGRET \,\cite{Sreekumar:1998un}, one sees that the diffuse PBH 
photon flux at energies $E_0 \apprge 1 \,\rm GeV$ is some orders of magnitude less than the
measured one. Hence, we do not expect to discriminate between the
Standard Model and MSSM particle models by measuring the diffuse photon flux from PBHs, at least if PBHs with initial
mass $M_i \approx 5 \times 10^{14} \, \rm g$ dominate.

\section{Other Particles}
The instantaneous flux of other astrophysically stable particles show the same general behaviour
as the photon flux as regards differences between the Standard Model and mSUGRA. This was explicitly
checked from the simulation data obtained. We will not consider these particles further in this work.

\chapter{Conclusions}
Emission from PBHs remains an interesting testing ground for high-energy physics and the conditions in the early Universe.
In these considerations, the effects of supersymmetric particle models on the gamma-ray flux from PBHs were studied.

The numerical data for the instantaneous photon flux from PBHs found agrees quite well with previous works, although some
differences exist due to some difference in model assumptions. The instantaneous flux in mSUGRA models was found to be roughly
a factor 4 greater than in the Standard Model, for peak photon energies in the regime were all emitted particles are
relativistic.

The instantaneous flux of other astrophysically stable particles was found to exhibit the same general behaviour as the
photon flux.

For the point source flux, the shape of the flux at peak energies as a function of time will be quantitatively and qualitatively
different for the Standard Model and mSUGRA, allowing limits to be imposed on the true particle model's mass spectrum.
Albeit a far-fetched prospect to be able to detect
the emission from a PBH in its final year to days, it would provide valuable information on a variety of physics,
particularly the mass spectrum of the true particle model.

The results show that no observable difference should be present in the dominant part of the diffuse gamma-ray
flux. In the high-energy range $E_0 \apprge 1 \,\rm GeV$, a PBH in the Standard Model produces a diffuse flux
that is typically much less than $\sim 50\%$ greater than for a PBH in mSUGRA Benchmark B.
The exact difference will be dependent on the
shape of the PBH initial mass function. For other SUSY models, the photon energy range where this difference
is present will start at
higher or lower energies for heavier and lighter particle mass spectra respectively. In the context of DGB measurements, 
particularly the upcoming GLAST project, it seems unlikely although perhaps not impossible to see such an effect. A detailed
study using a specific initial mass function is needed to settle the question definitely. 

The effect of possible photospheres would be to attenuate the low-energy (QED photosphere) and high-energy (QCD photosphere)
flux. A detailed analysis is outside the scope of this work.

\appendix
\chapter{mSUGRA Benchmark Models}
\label{apx:models}
\section{B}
The parameters of mSUGRA Benchmark B according to Ref. \cite{Ellis:2001hv}
is shown in Table \,\ref{table:pb}. The resultant mass spectrum of non-Standard
Model particles is shown in Table \,\ref{table:mb}.

\begin{table}
  \begin{center}
    \begin{tabular}{|ll|}
      \hline
       Parameter & Value \\
       \hline
       $m_{1/2}$ & 255 GeV \\
       $m_0$ & 102 GeV \\
       $\tan(\beta)$ & 10 \\
       sgn$(\mu)$ & +1\\
       $A$ & 0 \\
       \hline
    \end{tabular}
  \end{center}
  \caption{Parameters for mSUGRA Benchmark B}
  \label{table:pb}
\end{table}

\begin{table}
  \begin{center}
    \begin{tabular}{|ll|}
      \hline
       Particle & Mass [GeV] \\
       \hline
       $\tilde{\chi}_{1}^0$ & 101.73 \\
       $h^{0}$ & 111.64 \\
       $\tilde{\mu}^{\pm}_R$ & 148.49 \\
       $\tilde{e}^{\pm}_R$ & 148.49 \\
       $\tilde{\tau}_1^{\pm}$ & 148.89 \\
       $\tilde{\chi}_1^{\pm}$ & 190.46 \\
       $\tilde{\chi}_{2}^0$ & 191.24 \\
       $\tilde{\nu}_{\mu,L}, \bar{\tilde{\nu}}_{\mu,R}$ & 200.14 \\
       $\tilde{\nu}_{e,L}, \bar{\tilde{\nu}}_{e,R}$ & 200.14 \\
       $\tilde{\nu}_{\tau,L}, \bar{\tilde{\nu}}_{\tau,R}$ & 205.58 \\
       $\tilde{\mu}_L^{\pm}$ & 215.41 \\ 
       $\tilde{e}_L^{\pm}$ & 215.41 \\
       $\tilde{\tau}_2^{\pm}$ & 223.96 \\
       $\tilde{\chi}_{3}^0$ & 366.75 \\
       $\tilde{\chi}_{4}^0$ & 386.69 \\
       $\tilde{\chi}_2^{\pm}$   & 387.30 \\
       $A^0$ & 408.54 \\
       $H^0$ & 408.80 \\
       $H^{\pm}$ & 416.11 \\
       $\tilde{t}_1, \bar{\tilde{t}}_1$ & 451.26 \\
       $\tilde{b}_1, \bar{\tilde{b}}_1$ & 557.60 \\
       $\tilde{s}_R, \bar{\tilde{s}}_L$ & 581.95 \\
       $\tilde{d}_R, \bar{\tilde{d}}_L$ & 581.95 \\
       $\tilde{b}_2, \bar{\tilde{b}}_2$ & 582.97 \\
       $\tilde{c}_R, \bar{\tilde{c}}_L$ & 583.11 \\
       $\tilde{u}_R, \bar{\tilde{u}}_L$ & 583.11 \\
       $\tilde{c}_L, \bar{\tilde{c}}_R$ & 601.13 \\
       $\tilde{u}_L, \bar{\tilde{u}}_R$ & 601.13 \\
       $\tilde{s}_L, \bar{\tilde{s}}_R$ & 605.79 \\
       $\tilde{d}_L, \bar{\tilde{d}}_R$ & 605.79 \\
       $\tilde{t}_2, \bar{\tilde{t}}_2$ & 631.22 \\
       $\tilde{g}$ & 681.31 \\
       \hline
    \end{tabular}
  \end{center}
  \caption{Sparticle/non-standard particle mass spectrum in mSUGRA Benchmark B}
  \label{table:mb}
\end{table}
\section{A, C-M}
For parameters, see Ref. \,\cite{Ellis:2001hv}. The mass spectra can easily be calculated using e.g. \textsc{Pythia}
by giving the parameters as input.

\chapter{\textsc{Pythia} modifications}
\label{apx:pmod}
The changes made in \textsc{Pythia} 6.212 were in the function PYRESD, and follow below.
\newline
\newline
Line 13058:
\begin{verbatim}
  C...Order incoming partons and outgoing resonances.
  C***old      IF(JTMAX.EQ.2.AND.ISUB.NE.0.AND.MSTP(47).GE.1.AND.
         IF(JTMAX.EQ.2.AND.MSTP(47).GE.1.AND. ! new line
\end{verbatim}
Line 13190 and 13210:
\begin{verbatim}
  C...Store incoming and outgoing momenta, with random rotation to
  C...avoid accidental zeroes in HA expressions.
  C        IF(ISUB.NE.0) THEN ! removed (made comment)
             DO 470 I=IMIN,IMAX
           (... etc ...)
     500     CONTINUE
  C        ENDIF ! removed (made comment)
\end{verbatim}
Line 13233 and 13246:
\begin{verbatim}
  C...Calculate four-products.
  C        IF(ISUB.NE.0) THEN ! removed (made comment)
             DO 540 I=1,2
           (... etc ...)
     560     CONTINUE
  C        ENDIF ! removed (made comment)
         ENDIF
\end{verbatim}

\begin{acknowledgments}
I would like to thank my supervisor Lars Bergstr\"om for helpful discussions
and encouragement.

Anne Green has been helpful with background material, discussions and computing
and typesetting issues, as well as proofreading. For that I am grateful.

Joakim Edsj\"o has provided valuable help with how to use \textsc{Pythia}, as well as
discussion of the simulation aspects. I appreciate it.

Several other people in the Field and Particle Theory Group at Stockholm
University have helped out with computer problems, etc. 

Thanks also go to Torbj\"{o}rn Sj\"{o}strand (Lund University) for being very helpful with how
 to modify \textsc{Pythia} for my purposes and related questions.

Aur\'elien Barrau and Ga\"elle Boudoul (ISN, Grenoble) have been helpful
 with references and comments on the simulation treatment. 

Jane MacGibbon has given helpful comments on general PBH aspects and simulation.

I would also like to thank H\aa kan Snellman (KTH) for valuable advise on
my choice of thesis project and studies in general.

Rickard Armiento (KTH) graciously provided the very useful \LaTeX\ document class used to
typeset this thesis.

The scanning pen C-Pen 200 from C Technologies has been a valuable and extensively used tool in
organising references.

\end{acknowledgments}


\begin{thebibliography}{10}

\bibitem{Hawking:1974rv}
S.~W. Hawking,
\newblock Nature {\bf 248}, 30 (1974).

\bibitem{Hawking:1975sw}
S.~W. Hawking,
\newblock Commun. Math. Phys. {\bf 43}, 199 (1975).

\bibitem{GLAST:www}
Glast nasa homepage,
\newblock http://glast.gsfc.nasa.gov/.

\bibitem{MacGibbon:1990zk}
J.~H. MacGibbon and B.~R. Webber,
\newblock Phys. Rev. {\bf D41}, 3052 (1990).

\bibitem{MacGibbon:1991tj}
J.~H. MacGibbon,
\newblock Phys. Rev. {\bf D44}, 376 (1991).

\bibitem{Halzen:1991uw}
F.~Halzen, E.~Zas, J.~H. MacGibbon, and T.~C. Weekes,
\newblock Nature {\bf 353}, 807 (1991).

\bibitem{Oliensis:1984ih}
J.~Oliensis and C.~T. Hill,
\newblock Phys. Lett. {\bf B143}, 92 (1984).

\bibitem{Ellis:2001hv}
J.~R. Ellis, J.~L. Feng, A.~Ferstl, K.~T. Matchev, and K.~A. Olive,
\newblock Eur. Phys. J. {\bf C24}, 311 (2002), astro-ph/0110225.

\bibitem{Carr:1974nx}
B.~J. Carr and S.~W. Hawking,
\newblock Mon. Not. Roy. Astron. Soc. {\bf 168}, 399 (1974).

\bibitem{Carr:1975qj}
B.~J. Carr,
\newblock Astrophys. J. {\bf 201}, 1 (1975).

\bibitem{Hawking:1971}
S.~W. Hawking,
\newblock Month. Not. Roy. Astron. Soc. {\bf 152}, 75 (1971).

\bibitem{Hawking:1982ga}
S.~W. Hawking, I.~G. Moss, and J.~M. Stewart,
\newblock Phys. Rev. {\bf D26}, 2681 (1982).

\bibitem{La:1989st}
D.~La and P.~J. Steinhardt,
\newblock Phys. Lett. {\bf B220}, 375 (1989).

\bibitem{Crawford:1982yz}
M.~Crawford and D.~N. Schramm,
\newblock Nature {\bf 298}, 538 (1982).

\bibitem{Hawking:1989bb}
S.~W. Hawking,
\newblock Phys. Lett. {\bf B231}, 237 (1989).

\bibitem{Polnarev:1991bb}
A.~G. Polnarev and R.~Zemboricz,
\newblock Phys. Rev. {\bf D43}, 1106 (1991).

\bibitem{Canuto:1978bb}
V.~Canuto,
\newblock Mon. Not. Roy. Astron. Soc. {\bf 184}, 721 (1978).

\bibitem{Bergstrom:1999bk}
L.~Bergstr{\"o}m and A.~Goobar,
\newblock {\em Cosmology and Particle Astrophysics} (Wiley, 1999).

\bibitem{Choptuik:1993jv}
M.~W. Choptuik,
\newblock Phys. Rev. Lett. {\bf 70}, 9 (1993).

\bibitem{Evans:1994pj}
C.~R. Evans and J.~S. Coleman,
\newblock Phys. Rev. Lett. {\bf 72}, 1782 (1994), gr-qc/9402041.

\bibitem{Niemeyer:1998mt}
J.~C. Niemeyer and K.~Jedamzik,
\newblock Phys. Rev. Lett. {\bf 80}, 5481 (1998), astro-ph/9709072.

\bibitem{Niemeyer:1999ak}
J.~C. Niemeyer and K.~Jedamzik,
\newblock Phys. Rev. {\bf D59}, 124013 (1999), astro-ph/9901292.

\bibitem{Kolb:1990bk}
E.~Kolb and M.~Turner,
\newblock {\em The Early Universe} (Addison-Wesley, 1990).

\bibitem{Kim:1999iv}
H.~I. Kim, C.~H. Lee, and J.~H. MacGibbon,
\newblock Phys. Rev. {\bf D59}, 063004 (1999), astro-ph/9901030.

\bibitem{Kim:1996hr}
H.~I. Kim and C.~H. Lee,
\newblock Phys. Rev. {\bf D54}, 6001 (1996).

\bibitem{Press:1974iz}
W.~H. Press and P.~Schechter,
\newblock Astrophys. J. {\bf 187}, 425 (1974).

\bibitem{Carr:1993aq}
B.~J. Carr and J.~E. Lidsey,
\newblock Phys. Rev. {\bf D48}, 543 (1993).

\bibitem{Green:1999xm}
A.~M. Green and A.~R. Liddle,
\newblock Phys. Rev. {\bf D60}, 063509 (1999), astro-ph/9901268.

\bibitem{Green:2001kw}
A.~M. Green,
\newblock Phys. Rev. {\bf D65}, 027301 (2002), astro-ph/0105253.

\bibitem{Green:1997sz}
A.~M. Green and A.~R. Liddle,
\newblock Phys. Rev. {\bf D56}, 6166 (1997), astro-ph/9704251.

\bibitem{Bunn:1996py}
E.~F. Bunn, A.~R. Liddle, and M.~J. White,
\newblock Phys. Rev. {\bf D54}, 5917 (1996), astro-ph/9607038.

\bibitem{Bringmann:2001yp}
T.~Bringmann, C.~Kiefer, and D.~Polarski,
\newblock Phys. Rev. {\bf D65}, 024008 (2002), astro-ph/0109404.

\bibitem{Blais:2002gw}
D.~Blais, T.~Bringmann, C.~Kiefer, and D.~Polarski,
\newblock Phys. Rev. {\bf D67}, 024024 (2003), astro-ph/0206262.

\bibitem{Hawking:1977wz}
S.~W. Hawking,
\newblock Sci. Am. {\bf 236}, 34 (1977).

\bibitem{Page:1977um}
D.~N. Page,
\newblock Phys. Rev. {\bf D16}, 2402 (1977).

\bibitem{Gibbons:1975kk}
G.~W. Gibbons,
\newblock Commun. Math. Phys. {\bf 44}, 245 (1975).

\bibitem{Barrau:2001ev}
A.~Barrau {\em et~al.},
\newblock (2001), astro-ph/0112486.

\bibitem{Teukolsky:1973ha}
S.~A. Teukolsky,
\newblock Astrophys. J. {\bf 185}, 635 (1973).

\bibitem{Teukolsky:1974ab}
S.~A. Teukolsky and W.~H. Press,
\newblock Astrophys. J. {\bf 193}, 443 (1974).

\bibitem{Page:1976phd}
D.~N. Page,
\newblock {\em Accretion into and emission from black holes},
\newblock PhD thesis, Caltech, 1976.

\bibitem{Page:1976df}
D.~N. Page,
\newblock Phys. Rev. {\bf D13}, 198 (1976).

\bibitem{Page:1976ki}
D.~N. Page,
\newblock Phys. Rev. {\bf D14}, 3260 (1976).

\bibitem{Simkins:1986phd}
R.~D. Simkins,
\newblock {\em Massive scalar particle emission from Schwarzschild black
  holes},
\newblock PhD thesis, Penn State, 1986.

\bibitem{Parikh:1999mf}
M.~K. Parikh and F.~Wilczek,
\newblock Phys. Rev. Lett. {\bf 85}, 5042 (2000), hep-th/9907001.

\bibitem{Alexeyev:2002}
S.~Alexeyev, A.~Barrau, O.~Khovanskaya, and M.~Sazhin,
\newblock gr-qc/0201069.

\bibitem{Heckler:1997jv}
A.~F. Heckler,
\newblock Phys. Rev. Lett. {\bf 78}, 3430 (1997), astro-ph/9702027.

\bibitem{Heckler:1997ab}
A.~F. Heckler,
\newblock Phys. Rev. {\bf D55}, 480 (1997).

\bibitem{Cline:1998xk}
J.~M. Cline, M.~Mostoslavsky, and G.~Servant,
\newblock Phys. Rev. {\bf D59}, 063009 (1999), hep-ph/9810439.

\bibitem{Daghigh:2001gy}
R.~G. Daghigh and J.~I. Kapusta,
\newblock Phys. Rev. {\bf D65}, 064028 (2002), gr-qc/0109090.

\bibitem{Page:1976ab}
D.~Page and S.~Hawking,
\newblock Astrophys. J. {\bf 206}, 1 (1976).

\bibitem{Naselsky:1978}
P.~Naselsky,
\newblock Sov. Astron. Lett. {\bf 4}, 209 (1978).

\bibitem{Zeldovich:1976}
Y.~Zel'dovich and A.~Starobinsky,
\newblock JETP Lett. {\bf 24}, 571 (1976).

\bibitem{Barrow:1980hh}
J.~D. Barrow,
\newblock Surveys High Energ. Phys. {\bf 1}, 183 (1980).

\bibitem{Lindley:1980b}
D.~Lindley,
\newblock Mon. Not. Roy. Astron. Soc. {\bf 196}, 317 (1980).

\bibitem{Miyama:1978}
S.~Miyama and K.~Sato,
\newblock Prog. Theo. Phys. {\bf 59}, 1012 (1978).

\bibitem{Novikov:1979a}
I.~D. Novikov, A.~G. Polnarev, A.~A. Starobinsky, and Y.~B. Zeldovich,
\newblock Astron. Astrophys. {\bf 80}, 104 (1979).

\bibitem{Rothman:1981}
T.~Rothman and R.~Matzner,
\newblock Astrophys. Space. Sci. {\bf 75}, 2291 (1981).

\bibitem{Carr:2000my}
B.~J. Carr,
\newblock (2000), astro-ph/0102390.

\bibitem{MacGibbon:1991ca}
J.~MacGibbon and B.~Carr,
\newblock Astrophys. J. {\bf 371}, 447 (1991).

\bibitem{Carr:1998mc}
B.~J. Carr and J.~H. MacGibbon,
\newblock Phys. Rep. {\bf 307}, 141 (1998).

\bibitem{Sreekumar:1998un}
P.~Sreekumar {\em et~al.},
\newblock Astrophys. J. {\bf 494}, 523 (1998), astro-ph/9709257.

\bibitem{Kribs:1997ac}
G.~D. Kribs and I.~Z. Rothstein,
\newblock Phys. Rev. {\bf D55}, 4435 (1997), hep-ph/9610468.

\bibitem{Carr:1994ar}
B.~J. Carr, J.~H. Gilbert, and J.~E. Lidsey,
\newblock Phys. Rev. {\bf D50}, 4853 (1994), astro-ph/9405027.

\bibitem{Kribs:1999bs}
G.~D. Kribs, A.~K. Leibovich, and I.~Z. Rothstein,
\newblock Phys. Rev. {\bf D60}, 103510 (1999), astro-ph/9904021.

\bibitem{Bunn:1997da}
E.~F. Bunn and M.~J. White,
\newblock Astrophys. J. {\bf 480}, 6 (1997), astro-ph/9607060.

\bibitem{Liddle:1998nt}
A.~R. Liddle and A.~M. Green,
\newblock Phys. Rept. {\bf 307}, 125 (1998), gr-qc/9804034.

\bibitem{Bennett:2003bz}
C.~L. Bennett {\em et~al.},
\newblock (2003), astro-ph/0302207.

\bibitem{Jungman:1996df}
G.~Jungman, M.~Kamionkowski, and K.~Griest,
\newblock Phys. Rept. {\bf 267}, 195 (1996), hep-ph/9506380.

\bibitem{Cooper:2001bk}
F.~Cooper, A.~Khare, and U.~Sukhatme,
\newblock {\em Supersymmetry in Quantum Mechanics} (World Scientific, 2001).

\bibitem{Wess:1992bk}
J.~Wess and J.~Bagger,
\newblock {\em Supersymmetry and Supergravity}, 2nd ed. (Princeton University
  Press, 1992).

\bibitem{Zdziarski:1988ab}
A.~A. Zdziarski and R.~Svensson,
\newblock Astrophys. J. {\bf 344}, 551 (1989).

\bibitem{Weinberg:1972bk}
S.~Weinberg,
\newblock {\em Gravitation and Cosmology} (Wiley, 1972).

\bibitem{Sjostrand:2000wi}
T.~Sj{\"o}strand {\em et~al.},
\newblock Comput. Phys. Commun. {\bf 135}, 238 (2001), hep-ph/0010017.

\bibitem{Sjostrand:2003}
T.~Sj{\"o}strand,
\newblock private communication, 2003.

\bibitem{Press:2001nr}
W.~H. Press, S.~A. Teukolsky, W.~T. Vetterling, and B.~P. Flannery,
\newblock {\em Numerical Recipes in Fortran 77: The Art of Scientific
  Computing}, 2nd ed. (Cambridge University Press, 2001).

\end{thebibliography}
\end{document}